\documentclass[twocolumn]{aastex62}
\usepackage{natbib}
\usepackage{color} 
\usepackage{aas_macros} 
\usepackage{ulem}
\usepackage{hyperref} 

\received{2020}
\revised{}
\accepted{}

\begin{document}

\title{Constraints on the assembly history of the Milky Way's smooth, diffuse stellar halo from the metallicity-dependent, radially-dominated velocity anisotropy profiles probed with K giants and BHB stars using LAMOST, SDSS/SEGUE, and {\it Gaia}}

\email{sarahbird@nao.cas.cn, xuexx@nao.cas.cn, liuchao@nao.cas.cn,} 
\email{jtshen@sjtu.edu.cn, cflynn@swin.edu.au, ycq@shao.ac.cn,}
\email{gzhao@nao.cas.cn, hjtian@lamost.org}

\author{Sarah A. Bird} \affil{Key Laboratory of Optical Astronomy, National Astronomical Observatories, Chinese Academy of Sciences, 20A Datun Road, Beijing 100101, People's Republic of China} 
\affil{China Three Gorges University, Yichang 443002, People's Republic of China} \affil{Center for Astronomy and Space Sciences, China Three Gorges University, Yichang 443002, People's Republic of China} 
\affil{Shanghai Astronomical Observatory, 80 Nandan Road, Shanghai 200030, People's Republic of China}

\author{Xiang-Xiang Xue} \affil{Key Laboratory of Optical Astronomy, National Astronomical
  Observatories, Chinese Academy of Sciences, 20A Datun Road, Beijing 100101, People's Republic of China} \affil{School of Astronomy and Space Science, University of Chinese Academy of Sciences, Beijing 100049, People's Republic of China}

\author{Chao Liu} \affil{Key Laboratory of Optical Astronomy, National Astronomical
  Observatories, Chinese Academy of Sciences, 20A Datun Road, Beijing 100101, People's Republic of China} \affil{School of Astronomy and Space Science, University of Chinese Academy of Sciences, Beijing 100049, People's Republic of China}

\author{Juntai Shen} 
\affil{Department of Astronomy, School of Physics and Astronomy, Shanghai Jiao Tong University, 800 Dongchuan Road, Shanghai 200240, People's Republic of China}
\affil{Key Laboratory for Particle Astrophysics and Cosmology (MOE) / Shanghai Key Laboratory for Particle Physics and Cosmology, Shanghai 200240, People's Republic of China}

\author{Chris Flynn} \affil{Centre for Astrophysics and Supercomputing, Swinburne
  University of Technology, Post Office Box 218, Hawthorn, VIC 3122,
  Australia}

\author{Chengqun Yang} \affil{Shanghai Astronomical Observatory, 80
  Nandan Road, Shanghai 200030, People's Republic of China}

\author{Gang Zhao}  \affil{Key Laboratory of Optical Astronomy, National Astronomical
  Observatories, Chinese Academy of Sciences, 20A Datun Road, Beijing 100101, People's Republic of China} 

\author{Hai-Jun Tian} \affil{China Three Gorges University, Yichang 443002, People's Republic of China} \affil{Center for Astronomy and Space Sciences, China Three Gorges University, Yichang 443002, People's Republic of China}

\begin{abstract}

We analyze the anisotropy profile of the Milky Way's smooth, diffuse stellar halo using 
SDSS/SEGUE blue horizontal branch stars and SDSS/SEGUE and LAMOST K giants. 
These intrinsically luminous stars allow us to probe the halo to approximately 
100 kpc from the Galactic center. Line-of-sight velocities, distances, metallicities, and proper motions 
are available for all stars via SDSS/SEGUE, LAMOST, and {\it Gaia}, and we use these data to construct 
a full 7D set consisting of positions, space motions, and metallicity. We remove substructure from our samples using integrals 
of motion based on the method of Xue et al. We find radially dominated kinematic profiles 
with nearly constant anisotropy within 20 kpc, beyond which the anisotropy profile gently declines although remains radially dominated to the furthest extents of our sample. 
Independent of star type or substructure removal, the anisotropy depends on metallicity, 
such that the orbits of the stars become less radial with decreasing metallicity. 
For $-1.7<$ [Fe/H] $<-1$, the smooth, diffuse halo anisotropy profile begins to decline at Galactocentric distances $\sim20$ kpc, from $\beta\sim0.9$ to 0.7 for K giants and from $\beta\sim0.8$ to 0.1 for blue horizontal branch stars.
For [Fe/H] $<-1.7$, the smooth, diffuse halo anisotropy remains constant along all distances with $0.2<\beta<0.7$ depending on the metallicity range probed, although independent on star type.
These samples are ideal 
for estimating the total Galactic mass as they represent the virialized stellar halo system.

\end{abstract}


\section{Introduction} \label{sec:intro}

Halo stars \citep{GaiaCollaborationBrown2018} preserve clues to the formation and evolution of galaxies as well as to their merger histories.
Due to their long orbital timescales, the relics of past events are preserved in both configuration and velocity space, and the chemistry and age of the stars additionally clarify the picture.
Even when substructure has become too diffuse to be detectable in configuration space, the member stars retain information of their common origin locked in their velocities, energies, angular momenta, and orbital properties; and for deriving these, accurate distances and 3D velocities are necessary, the limiting reagent of these being the velocities as only very small samples beyond a few kpc have been observed, although this predicament is now changing.

A number of processes contribute stars to the stellar halo \citep[e.g.,][]{Zolotov2009,Tissera2013,Cooper2015,Pillepich2015,Amorisco2017atlas,Di_Matteo2019}: infalling satellites, stars formed from gas within the halo, stars displaced from the disk. With each large survey, new substructure and remnant satellites are uncovered, as reviewed by \citet{Newberg2016} and \citet{Helmi2020}. With the large amount of substructure in the halo, only recently are statistically significant halo star samples being collected to help detail these past events buried in space and time.

The halo stars are also probes to the galactic dark matter distribution. The density and velocity of the relaxed halo star population are used in the Jeans equation to estimate the Galactic mass. Ideally for this estimation, substructure should be removed as a lemma of the Jeans equation is a relaxed sample in equilibrium \citep[e.g.,][]{Binney2008}.

As a step forward, we present a large sample of halo stars with full 7D positions, velocities, and metallicity, from which substructure have been removed. We analyze their velocities and metallicity as a step toward a future mass estimation.

When considering the 3D velocities and orbital parameters, a convenient term for spherical or near-spherical systems is the velocity anisotropy of the system. In a Galactocentric spherical coordinate system anisotropy is defined \citep{Binney1980,Binney2008} as
\begin{equation}
\beta = 1 - (\sigma_\theta^2+\sigma_\phi^2)/(2\sigma_r^2),
\end{equation}
where $\sigma$ is the velocity dispersion within the velocity components  of distance $r_\mathrm{gc}$ from the Galactic Center, polar angle, and azimuthal angle, $(r,\theta,\phi)$.
This single parameter characterizes the orbits of the stars, whether being radially dominated ($\beta>0$), isotropic ($\beta=0$), or tangentially dominated ($\beta<0$). 

Violent relaxation models and simulations suggest that, for Milky Way type galaxies, the anisotropy profile is smooth and slowly rising rising with increasing distance from the galactic centers \citep{Diemand2005,Abadi2006,Sales2007.379.1464,Rashkov2013}. 
\citet{Kafle2012} (analyzing the simulation suite of \citet{Bullock2005}) and \citet{Loebman2018} (analyzing simulations of \citet{Bullock2005}, \citet{Christensen2012}, and \citet[Making Galaxies in a Cosmological Context (MaGICC)]{Stinson2013}) typically find smooth, rising, radial $\beta$ profiles. 
Additionally the suite of idealized, collisionless N-body simulations of minor mergers and a particle-tagging technique by \citet{Amorisco2017atlas} are characterized by radial $\beta$ profiles. 

Within a few kpc of the Sun, the anisotropy is observationally determined to be radial, with values of $\beta=0.5-0.7$ \citep{Morrison1990,Chiba1998,Kepley2007,Smith2009,Bond2010,Evans2016,Posti2018}.

The stochastic nature of galaxy formation in $\Lambda$CDM cosmologies leads to various processes and events which contribute to the stellar halo, each mechanism having the possibility of contributing populations of halo stars unique in configuration space, kinematics, and chemistry \citep[e.g.,][]{Amorisco2017atlas,Di_Matteo2019}.
When anisotropy is analyzed with more detail, the picture is more complex. 
Trends in kinematic statistics 
have been explained by two or more halo components \citep[e.g.,][]{Majewski1992,Carney1996,Wilhelm1996,Kinman2007,Deason2011.411,Carollo2007,Carollo2010,Nissen2010,Beers2012,Schuster2012,An2013,An2015,Zuo2017}.
Different components to the stellar halo have also been analyzed in simulations \citep[e.g.,][]{Bekki2001,Brook2004,Abadi2006,Zolotov2009,Cooper2010,Oser2010,Font2011,McCarthy2012,Pillepich2015}. 
\citet{Deason2011.411}, \citet{Kafle2013,Kafle2017googly}, and \citet{Hattori2013} find observational evidence that $\beta$ changes with metallicity.

More than a few kpc from the Sun, the picture of $\beta$ has not been clear. A range of different profiles estimated from line-of-sight velocities has been proposed \citep{Sommer-Larsen1994,Sommer-Larsen1997,Sirko2004,Thom2005,Deason2011.411,Deason2012,Kafle2012,Kafle2014,King2015,Williams2015}. \citet{Sommer-Larsen1994} found the anisotropy profile transitioned from radial to tangentially dominated and suggested such a scenario could be achieved by the specific accretion history of infalling subsystems. \citet{Kafle2012} found a dip in the anisotropy profile attributing this a possible unaccounted feature in the Galactic potential or to the transition between two components of the stellar halo (such as proposed by the forementioned references). \citet{King2015} found a deeper, more extended tangential dip in the anisotropy profile and suggested the cause to be Sagittarius Stream members or other streams within their halo sample.

\citet{Flynn1996} assigned stars with the profile found by \citet{Sommer-Larsen1994} and tested the long-time stability of the kinematics within a Milky Way-like potential, finding the kinematics robust over many billion years. On the other hand, \citet{Bird2015mwhalo} performed a similar test for the profile found by \citet{Kafle2012} and found that if such a profile exists in the Milky Way, the kinematics are in a state of transition and the dip in the profile smooths over several hundred million years. \citet{Loebman2018} made a detailed analysis of $\beta$ in simulations, with careful attention to when $\beta$ reaches tangential (negative) values, when the otherwise smooth-radial profile exhibits dips, and the length of time these features remain. 
They showed that beta dips can be indicators of satellite passings or infall. These are seen as localized departures (dips) from the general trend of the $\beta$ profile, such features lasting several hundred million years. \citet{Loebman2018} found that large troughs (extended decreases in the anisotropy profile), as those seen by \citet{Sommer-Larsen1994} and \citet{King2015} can be produced by ancient major mergers.

Indeed, pencil beam surveys have shown different $\beta$ values in different sky locations 
\citep{Deason2013beta,Cunningham2016,Cunningham2019b}. These authors
used proper motions from multi-epoch $HST$ fields and ground based spectroscopic analysis of main sequence stars and were able to reach distances to 30 kpc. Although many fields showed $\beta>0.5$, several fields showed kinematics which were isotropic $\beta\sim0$.

Now with $Gaia$ and large stellar spectroscopic surveys such as LAMOST and SDSS becoming available, the direct measurement of the 3D velocity dispersions and $\beta$ over large distances is possible. 
Recently anisotropy has played a vital role in the study of \citet{Belokurov2018} who found a highly radial component to the velocity anisotropy derived from halo main sequence stars with $-1.7<$[Fe/H]$<-1.0$ within the region probed ($1-10$ kpc from the Sun and distance from the disk $|Z|<9$ kpc).
They explained this as evidencing an
early event in our Galaxy's history, the last significant merger $\sim10$ Gyr ago. The large radial anisotropy of this remnant merger even spawned a name for the progenitor, the ``$Gaia$-Sausage.'' 
\citet{Myeong2018action} found a clear difference in the kinematic and spatial distribution of relatively metal-rich versus metal-poor halo stars.
\citet{Bird2019beta} and \citet{Lancaster2019} using halo K giants and BHB stars, respectively, out to larger distances found this highly radial anisotropy for $-1.7<$[Fe/H]$<-1.0$ extends to Galactocentric distances of $\sim25$ kpc. These studies similarly found the radial anisotropy deceases with decreasing metallicity. The highly radial [Fe/H]$>-2$ halo component was also seen by \citet{Wegg2019} using the proper motions of RR Lyrae stars combined with models to derive the 3D velocity dispersions.

Such an early merger event has evidenced itself in energy and angular momentum space \citep{Koppelman2018,Helmi2018} and has been dubbed the ``blob'' and named ``$Gaia$-Enceladus.'' \citet{Deason2018.862} showed the high radially anisotropic halo stars share apocentric distances of $\sim20$ kpc. \citet{Haywood2018} analyzed the chemodynamics and linked the results to an early accretion event. 
\citet{Kruijssen2019} compared the age and metallicity of Milky Way globular clusters to simulations and found evidence for a similar early merger, naming it the ``Kraken.'' As the $Gaia$ data is further analyzed, a plethora of studies have uncovered more evidence which can be explained by an early large-dwarf accretion event as well as details of new streams and substructure \citep[e.g.,][]{Myeong2018.475,Myeong2018Sausageglobs,Myeong2019,Fernandez-Alvar2019,Fernandez-Alvar2019richtail,Simion2019,Iorio2019,Mackereth2018,Mackereth2019,Yuan_Zhen2019}. The number of works is growing, both theoretical and observational, which supports that most of the mass in Milky Way-type halos comes from the last (or last few) significant accretion event(s) \citep[e.g.,][]{Brook2003,Robertson2005,Bullock2005,Font2006_646,Johnston2008,Cooper2010,Deason2013,Deason2016,Fiorentino2015,Harmsen2017,Bell2017,Amorisco2017atlas,Fattahi2019,D-Souza2018,Monachesi2019,Elias2020}. 

\citet{Deason2011.416} at $r_\mathrm{gc}\sim27$ kpc found a break in the stellar halo density profile and explained this as caused by the last significant merger. \citet{Lancaster2019} found near the same break in the density profile, that the highly radial, metal rich component of their BHB star's anisotropy drops from radially dominated to isotropic. \citet{Bird2019beta} use K giants and found the anisotropy profile drops to lower radial values, but the drop was reduced by removing a large portion of the Sagittarius Stream. \citet{Cunningham2019b} found variations in $\beta$ between different fields from their pencil beam survey, but an overall view of $\beta$ rising with increasing Galactocentric distance. Concluding from these works, the overall picture of the Galactic anisotropy profile and the influence of streams and substructure has yet to be clarified. When considering the analysis of varying $\beta$ in simulations by \citet{Loebman2018} and \citet{Cunningham2019b} (using two galaxies in the {\it Latte} suite of the Feedback In Realistic Environments (FIRE) simulation project \citep{Wetzel2016,Hopkins2018,Sanderson2020}, these  seemingly differing results may be explained by the field observed and the substructure present. The question remains as to whether the Milky Way anisotropy profile at large distance is similar to the highly radial and slowly rising profile found in simulations \citep[e.g.,][]{Diemand2005,Abadi2006,Sales2007.379.1464,Rashkov2013}, or if the profile deviates to isotropic or tangential values.

We here make a more complete study, covering $2<r_\mathrm{gc}<110$ kpc, and comparing $\beta$ from two different star-types, halo K giants and BHB stars, and a range of metallicities ($-3<$ [Fe/H] $<-1$, depending on the sample) and test the effects on $\beta$ from removing substructure using a complete substructure finding method.

In Section \ref{sec:sample}, 
we present our sample and the effects on the kinematics due to substructure removal.
In Section \ref{sec:results}, we show
that the orbital families are predominantly radial at all radii
probed through the anisotropy parameter $\beta$. We find that the
amplitude of $\beta$ is a function of metallicity, which, for the more metal poor halo star sample, substantially
reduces relative to the metal rich halo stars, but still radial. In
Section \ref{sec:conclusion}, we discuss our results in terms of
the literature covering observational work and $N$-body/hydrodynamical simulations of the
formation and evolution of galaxy halos, and finally draw our conclusions.

\section{Data}
\label{sec:sample}

\subsection{Halo Samples}
  \label{sec:combined}

We analyze three stellar halo samples. Two samples we select from the BHB and K-giant catalogs of \citet{Xue2011} and \citet{Xue2014}, respectively. The detailed methods used for selecting the samples from SDSS/SEGUE and measuring distances are found in their respective papers. Our third sample consists of LAMOST DR5 K giants, similar to that used by \citet{Bird2019beta} and \citet{Yang2019a}. We select LAMOST K giants using the method of \citet{Liu2014} and determine distances using the method of \citet{Xue2014}, which is the same as \citet{Bird2019beta}. For K giants which have been observed multiple times within the LAMOST survey, we include in our final sample the observation which produces the smallest distance uncertainty.
Line-of-sight velocities and stellar parameters, i.e. metallicity [Fe/H], effective temperature $T_\mathrm{eff}$, and surface gravity $\log(g)$, are taken from the published catalogs. Our selection criteria for our halo star samples are summarized in Table \ref{table:data_selection} and we further elaborate upon in this Section \ref{sec:sample}.
We match our samples to {\it Gaia} DR2 (search radius $<1''$) to obtain proper motions. Distances are provided for the BHB stars by \citet{Xue2008} and \citet{Xue2011} and for the SDSS/SEGUE K giants by \citet{Xue2014}. We include a summary of our halo star samples in Table \ref{table:data}.

\begin{table}
\caption{Selection criteria. Beginning with the BHB sample from SDSS/SEGUE \citep{Xue2008,Xue2011} and K giants from SDSS/SEGUE \citep{Xue2014} and LAMOST \citep{Bird2019beta}, we retain stars within our halo sample (see Table \ref{table:data}) using these criteria.
}
\label{table:data_selection}
\begin{center}
\begin{tabular}{c}
\hline
Required criteria\\
\hline
\hline
{\it Gaia} DR2 match (search radius $<1''$)\\
$|Z|>2$ kpc\\
$[$Fe$/$H$]$ $<-1$\\
$|(V_r,V_\theta,V_\phi)| < 500$ km s$^{-1}$\\ 
$\delta(V_r,V_\theta,V_\phi)$ $<$ $(100,150,150)$ km s$^{-1}$\\ 
$E<0$ (km s$^{-1})^2$\\
semi-major axis $a<300$ kpc\\
\hline
Substructure defined\\
\hline
\hline
friend's of friend group $>6$ members\\
\hline
\end{tabular}
\end{center}
\end{table}

\begin{table*}
\caption{Sample summary. After applying the selection criteria summarized in Table \ref{table:data_selection}, we retain the number of stars $N$star listed below.}
\label{table:data}
\begin{center}
\begin{tabular}{ccccc}
\hline
Sample & $N$star & Stellar type & Distance method & Reference\\
\hline
LAMOST DR5 & 12728 & K-giant & \citet{Xue2014} & \citet{Liu2014,Bird2019beta}\\
\hline
SDSS/SEGUE & 5248 & K-giant & \citet{Xue2014} & \citet{Xue2014}\\
\hline
SDSS & 3982 & BHB & \citet{Xue2008,Xue2011} & \citet{Xue2011}\\
\hline
\end{tabular}
\end{center}
\end{table*}

We impose criteria in metallicity and position to select clean samples of halo BHB stars and K giants with minimal disk contamination. 
We select stars with $|Z|>2$ kpc (height above the Galactic disk mid-plane) 
and [Fe/H] $<-1$. 
Note that these are less stringent as those used for the LAMOST DR5 K giant sample of \citet{Bird2019beta}, and allow us to analyze the nearby stellar halo. Less than one percent of the SDSS BHB catalog lack metallicities or have spurious values; we exclude these from our sample.

A comparison of the 498
common stars between our LAMOST and SDSS K giant samples shows good agreement but with a systematic bias in the line-of-sight velocities of 
$8$ km s$^{-1}$, which we add to the LAMOST velocities
(systematic biases in velocity have been
calculated by comparing star samples from LAMOST with SDSS/SEGUE 
\citep{Tian2015,Yang2019a} and with {\it Gaia} \citep{Schonrich2017}). 
The mean $v_\mathrm{los}$ uncertainties are 9 km s$^{-1}$ (LAMOST KG), 2 km s$^{-1}$ (SDSS KG), and 4 km s$^{-1}$ (SDSS BHB). The proper motion uncertainties generally range between 0.01 to 0.5 mas yr$^{-1}$.

We derive Galactocentric Cartesian coordinates and velocities and Galactocentric spherical velocities for the stars in the same manner as \citet{Bird2019beta}:
Galactocentric Cartesian coordinates $(X,Y,Z)$ and velocities
$(U,V,W)$ follow the conventions in {\tt astropy} \citep{astropy2018} and Galactocentric spherical velocities ($V_r, V_\theta,V_\phi$) follow Eq. $2-7$ of \citet{Bird2019beta}. We propagate (1) uncertainties in the distances
of the stars, (2) line-of-sight radial velocity
uncertainties, 
(3) two {\it Gaia} DR2
proper motion error estimates, (4) and the {\it Gaia} DR2 proper motion covariance. These four uncertainties are propagated via 1000 Monte Carlo runs per star with the {\tt python} package {\tt pyia} \citep{Price-Whelan2018} to derive standard error for the uncertainties in the Galactocentric Cartesian and spherical coordinate systems.

We adopt the following to improve the purity of our halo K-giant and BHB star sample
and minimize the contamination from stars which may have faulty stellar parameters or may be incorrectly classified. 
As later described in Section \ref{sec:sub}, we use a Milky Way-like potential for calculating integrals of motion in order to remove substructure; we additionally use the integrals of motion of the stars derived from this process to flag contaminants. Stars with extremely high energy may have faulty stellar parameters or may be stars incorrectly classified. 
Blue stragglers and red clump stars are candidate contaminants to our BHB and K giant samples, respectively. These contaminants are intrinsically dimmer. If the distance methods for BHB stars and K giants are used for these dimmer stars, their distances will be greatly overestimated. The overestimated distances and likely larger proper motions (due to the star's closer than estimated proximity to the Sun) effectively increase the tangential velocity component ($V_\mathrm{tan} \propto d_\mathrm{helio} \times \mu_\mathrm{tot}$), the angular momentum, the energy, and thus the size of the orbit. 
To eliminate egregious outliers, we require $|(V_r,V_\theta,V_\phi)| < 500$ km s$^{-1}$, $\delta(V_r,V_\theta,V_\phi)$ $<$ $(100,150,150)$ km s$^{-1}$, $E<0$ (km s$^{-1})^2$, and semi-major axis $a<300$ kpc. 
A similar elimination of high velocity stars due to blue straggler contamination within their BHB sample has recently been discussed by \citet{Lancaster2019}. They show that these high velocity stars tend to occupy the same regions as blue straggler contaminants in color-color and Balmer line shape space and accordingly cull these stars using color and spectral classifications, removing all stars with SDSS colors satisfying
$u - g < 1.15$ and $g - r > -0.07$ as well as stars satisfying $u - g < 1.15$ and $c_\gamma < 0.925$, where $c_\gamma$ describes the shape of the H$\gamma$ line. We do not apply these criteria from \citet{Lancaster2019} as the initial BHB and K giant samples have minimal contamination from blue stragglers by applying Balmer line cuts \citep{Xue2008} and from red clump stars by keeping only stars above the horizontal branch \citep{Xue2014,Bird2019beta}.

\begin{figure*}[htb]
\begin{tabular}{ccc}
\includegraphics[width=.65\columnwidth]{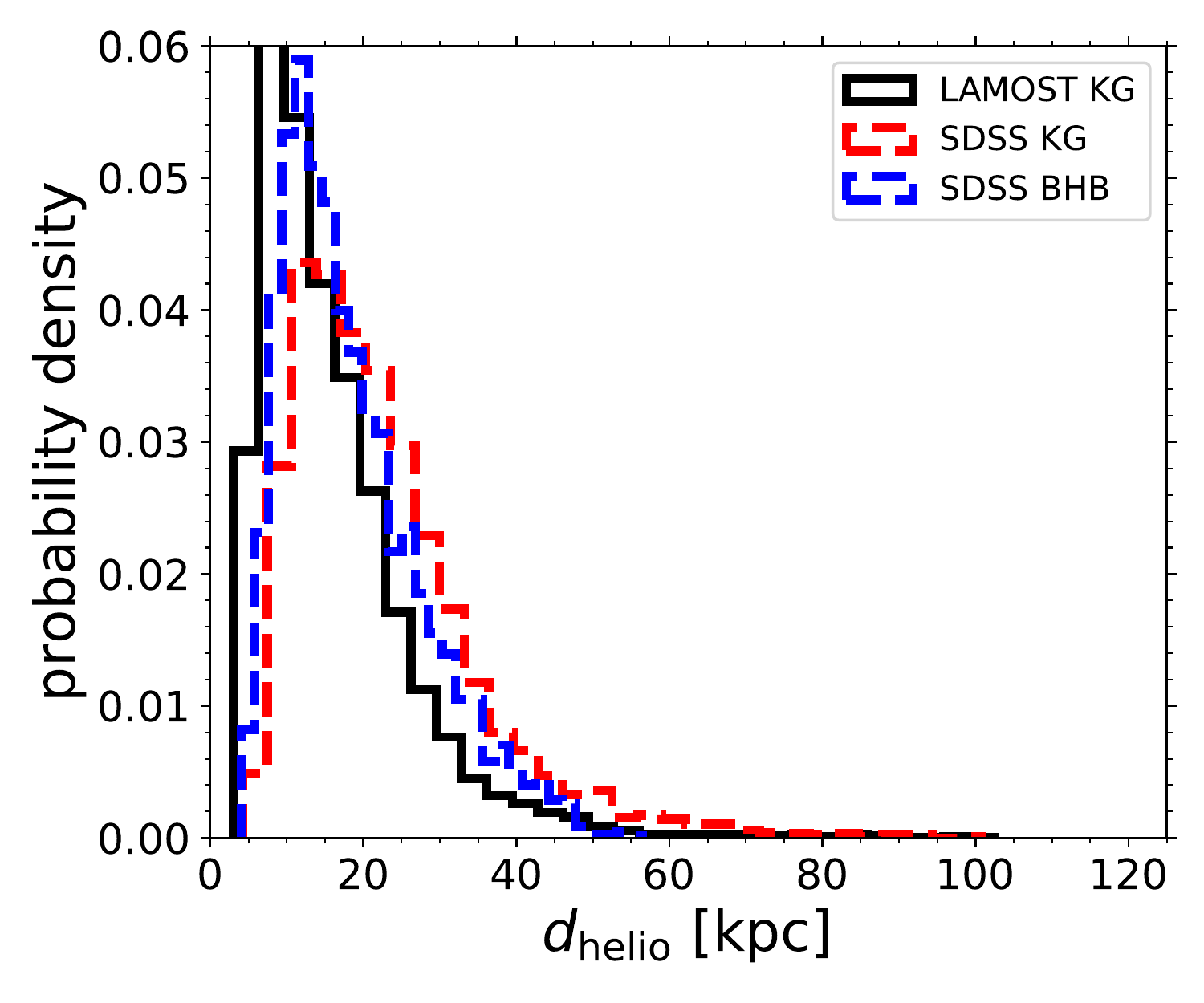}&
\includegraphics[width=.65\columnwidth]{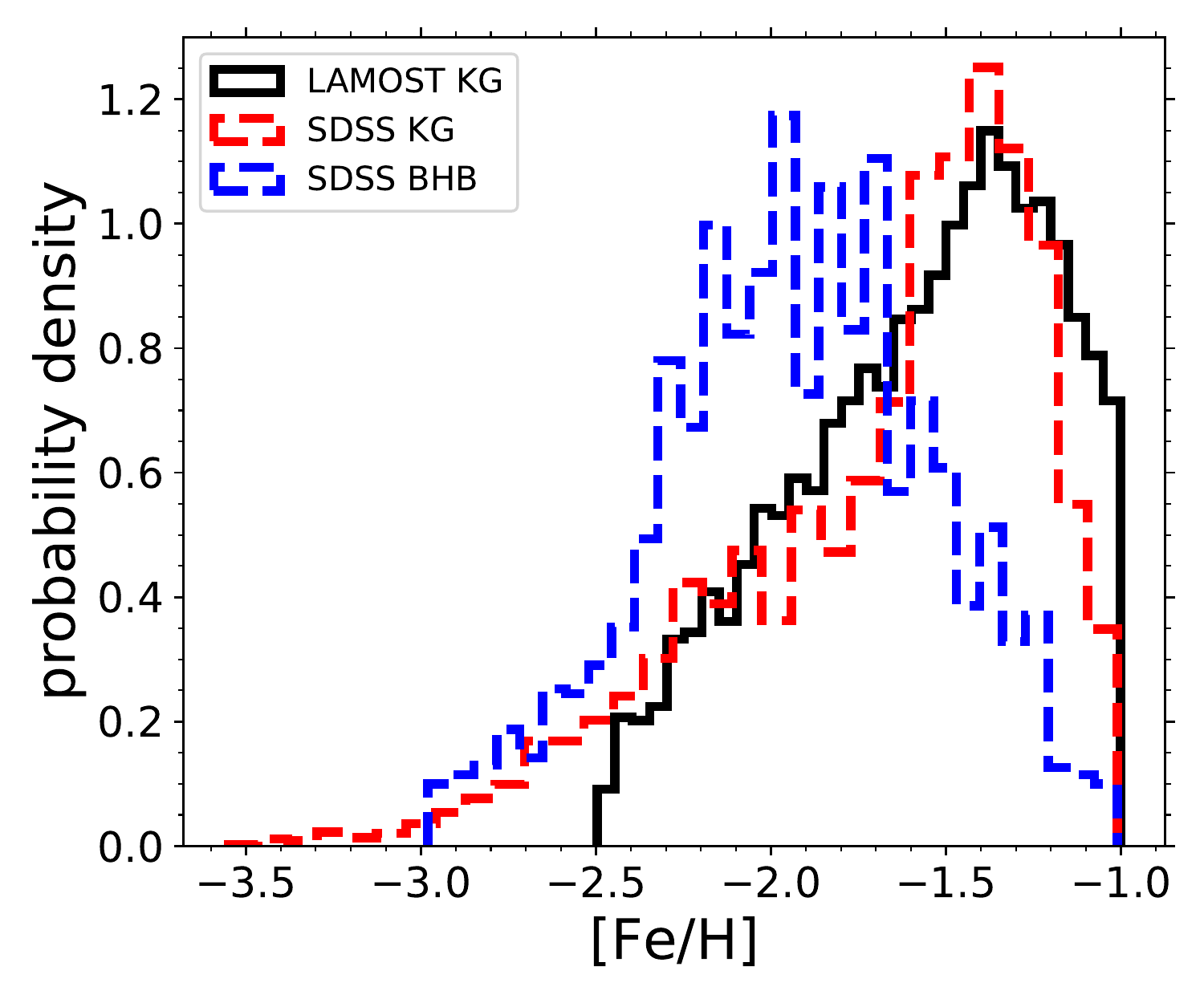}&
\includegraphics[width=.65\columnwidth]{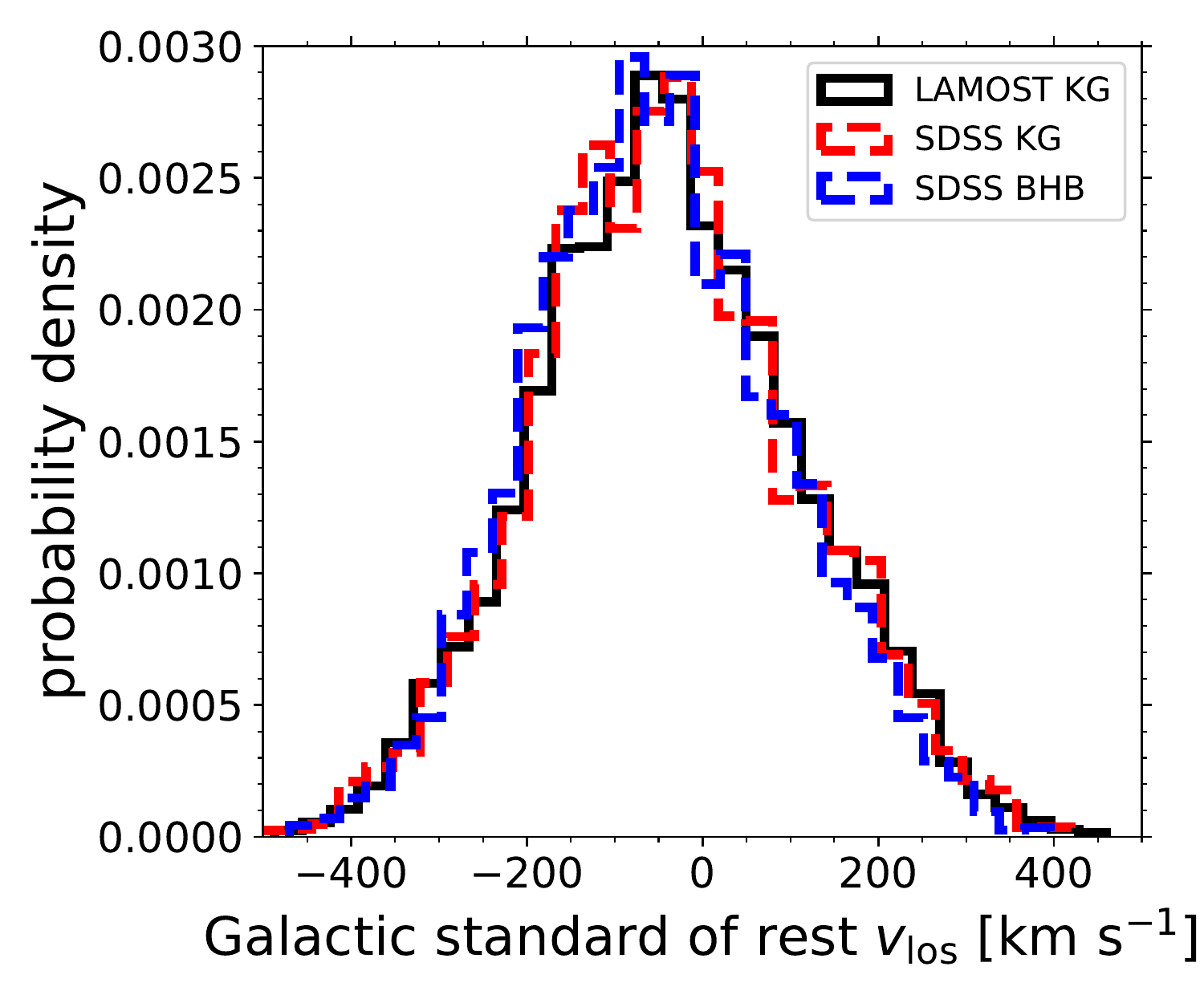}\\
\includegraphics[width=.65\columnwidth]{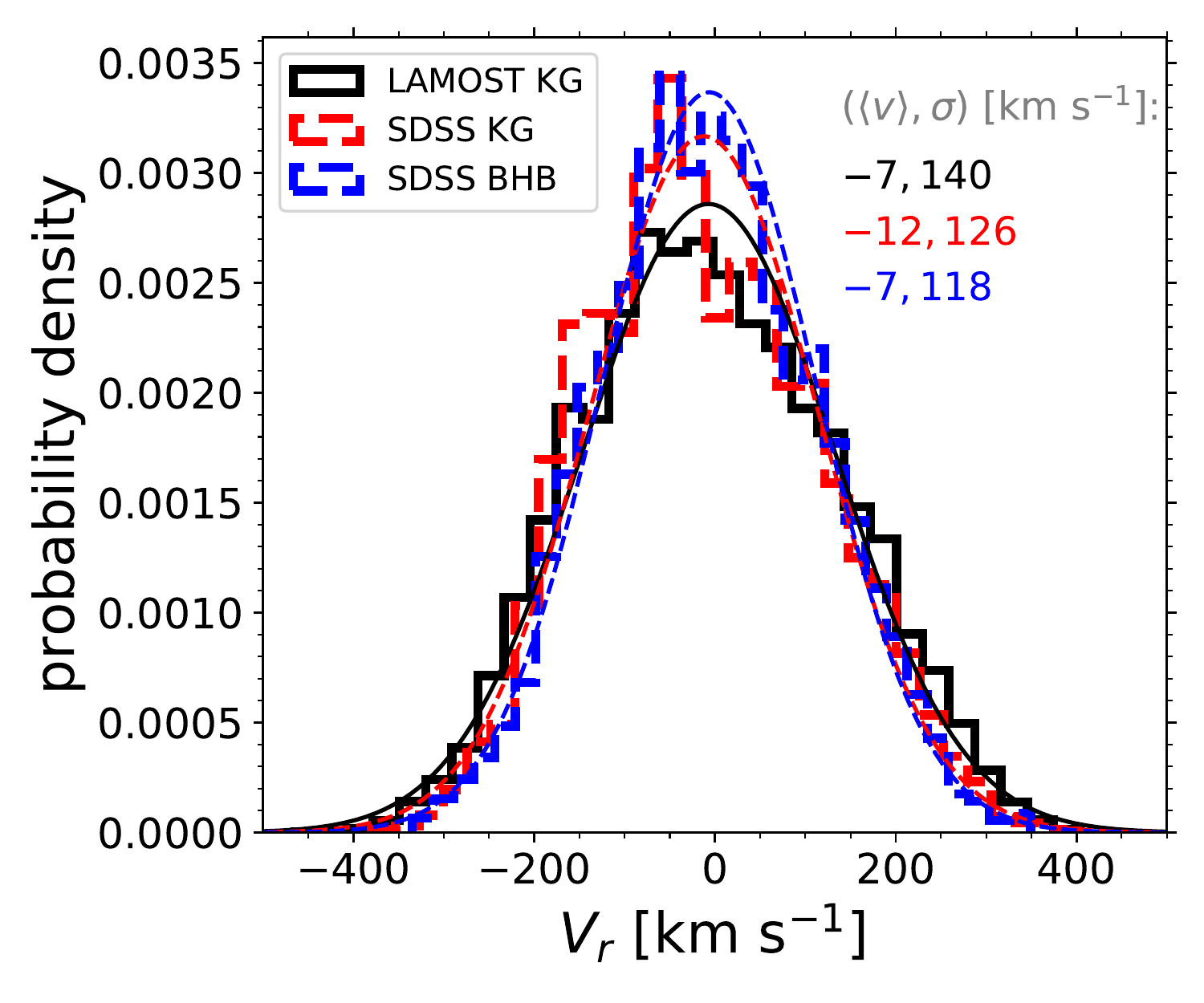}&
\includegraphics[width=.65\columnwidth]{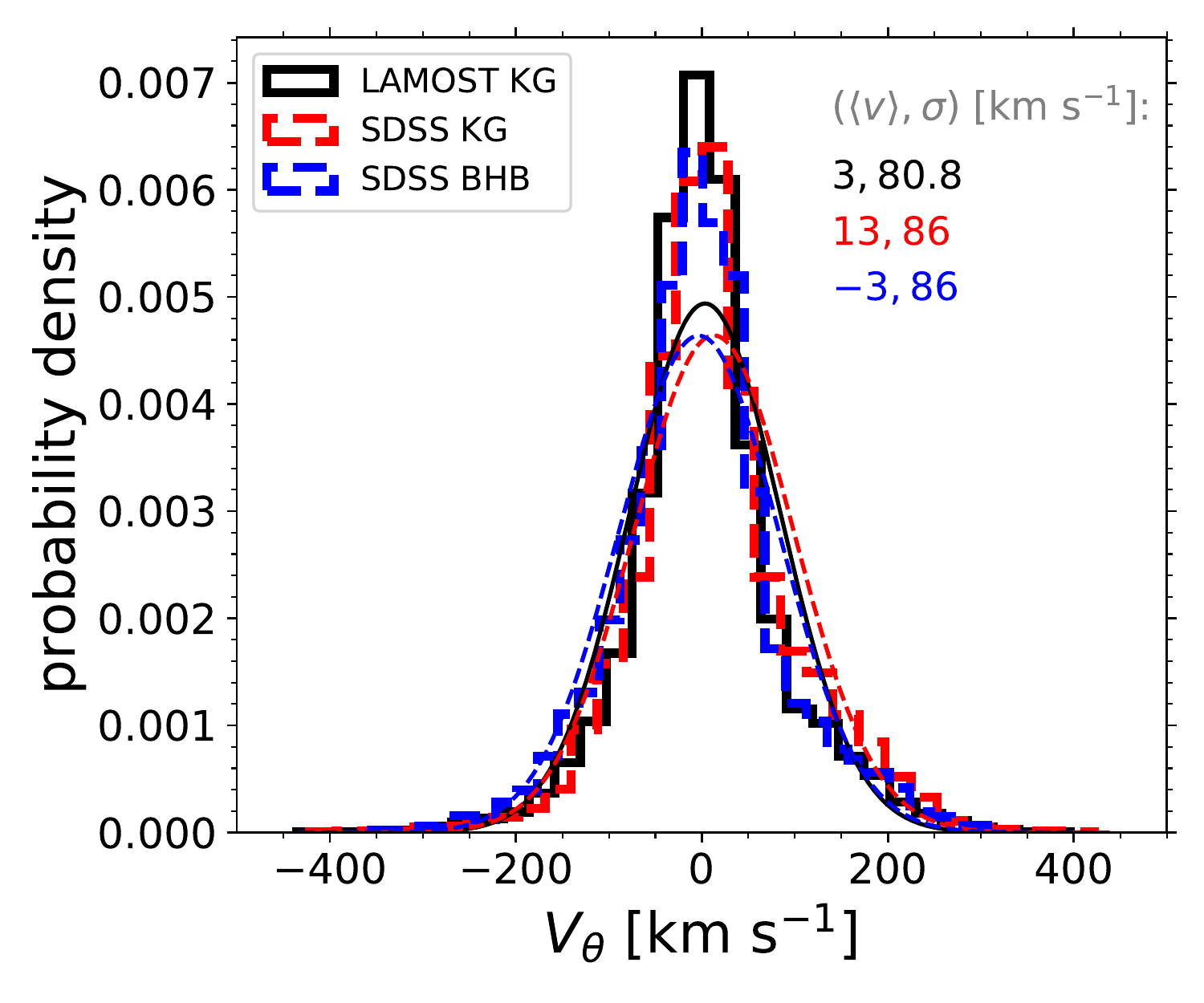}&
\includegraphics[width=.65\columnwidth]{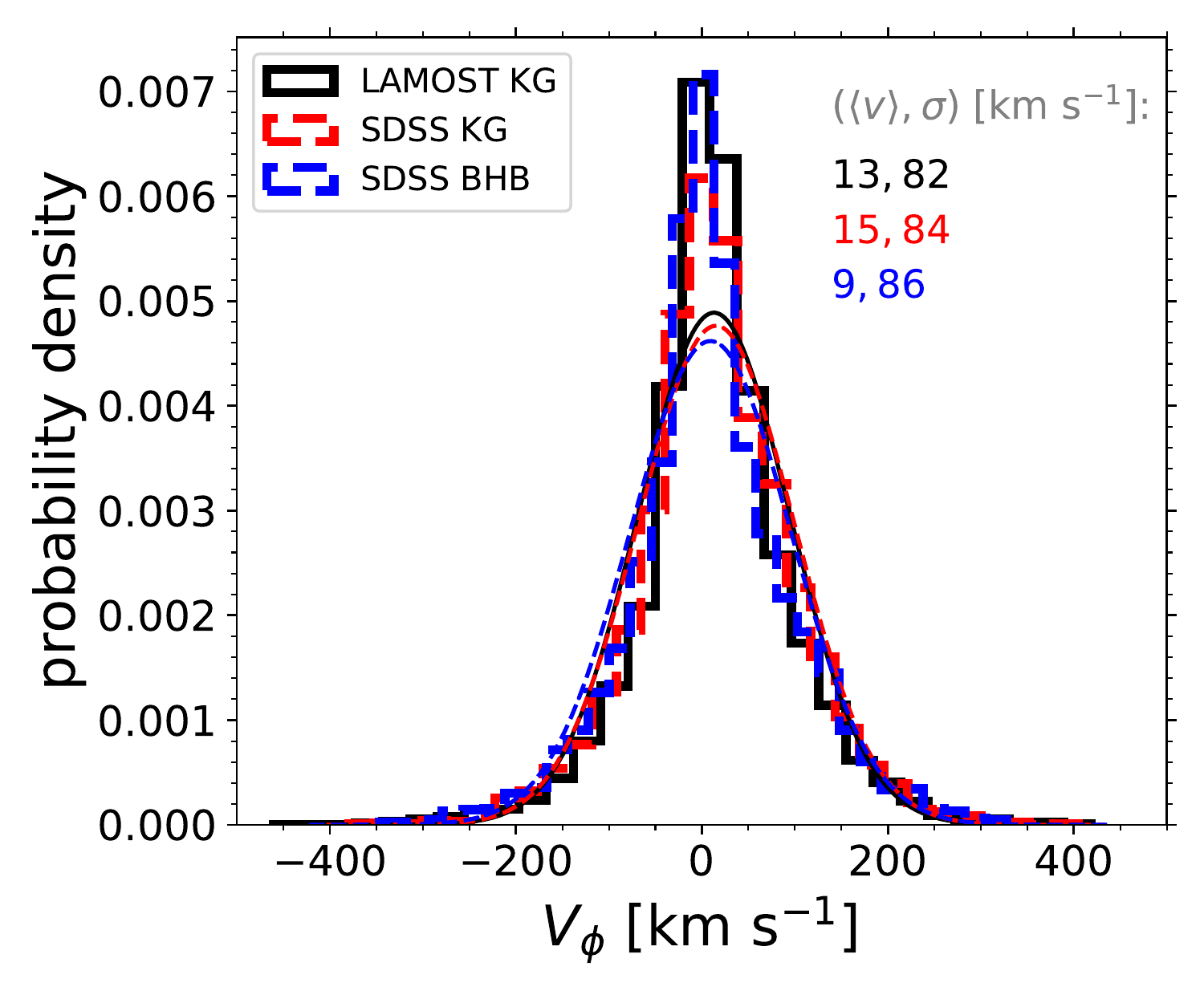}
\end{tabular}
\caption{Probability density distributions comparing the LAMOST and
  SDSS/SEGUE sample halo K giant and SDSS sample BHB $d_\mathrm{helio}$, [Fe/H], and heliocentric $v_\mathrm{los}$ in the Galactic standard of rest frame of reference in the upper panels (left, middle, and right panels, respectively) 
and 3D Galactocentric spherical velocities in the lower panels. The respective counts $N$ are
  normalized by (bin height)$\times$(bin width) to form a
  probability density. Only metallicity [Fe/H] differs much between the sample distributions.
In the lower panels, the thin lines (colors respective to the upper left panel legend) are Gaussian fits (mean and dispersion in the upper right panel corner).
}
\label{lamostsegue}
\end{figure*}

\begin{figure}
\includegraphics[width=\columnwidth]{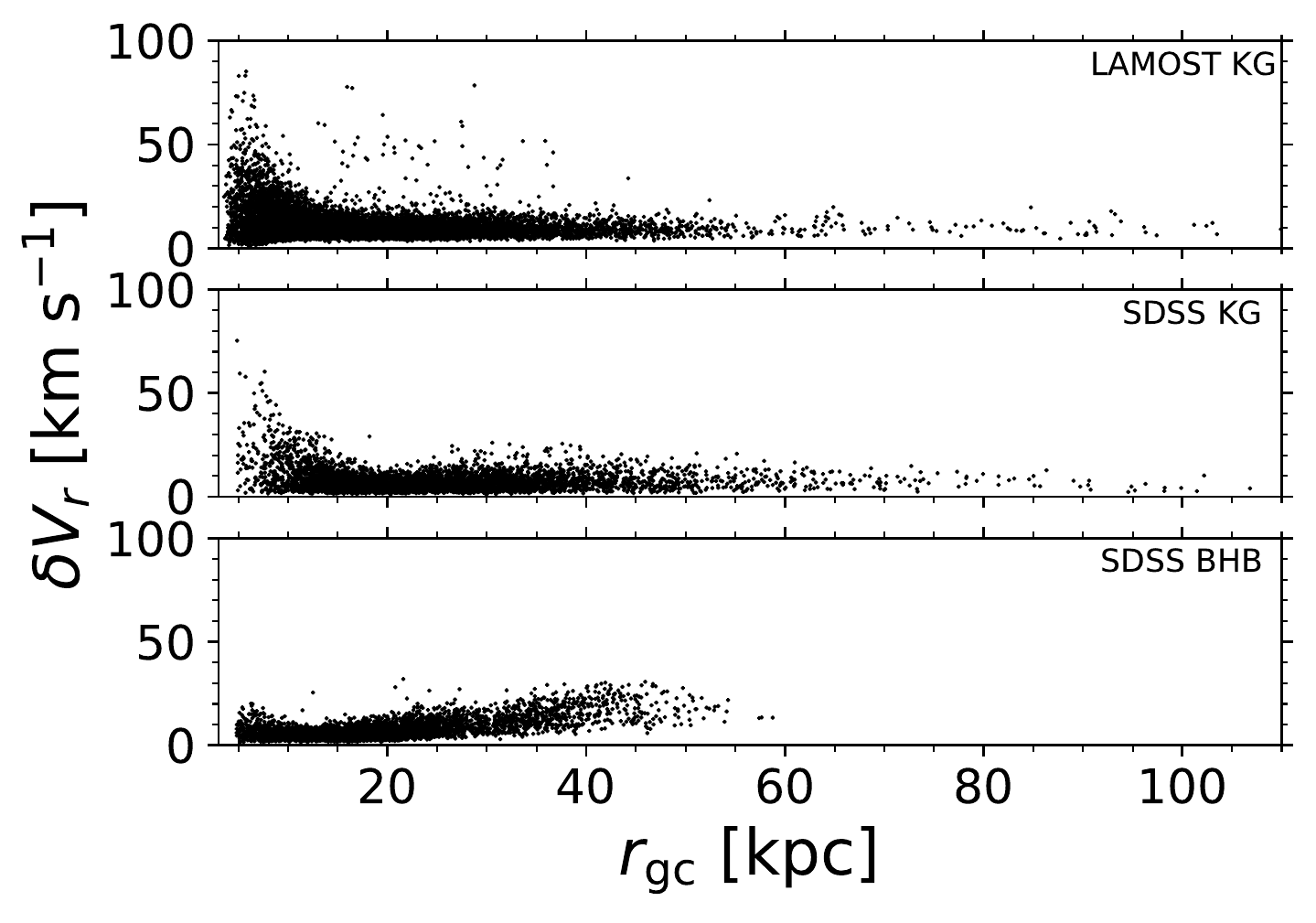}\\
\includegraphics[width=\columnwidth]{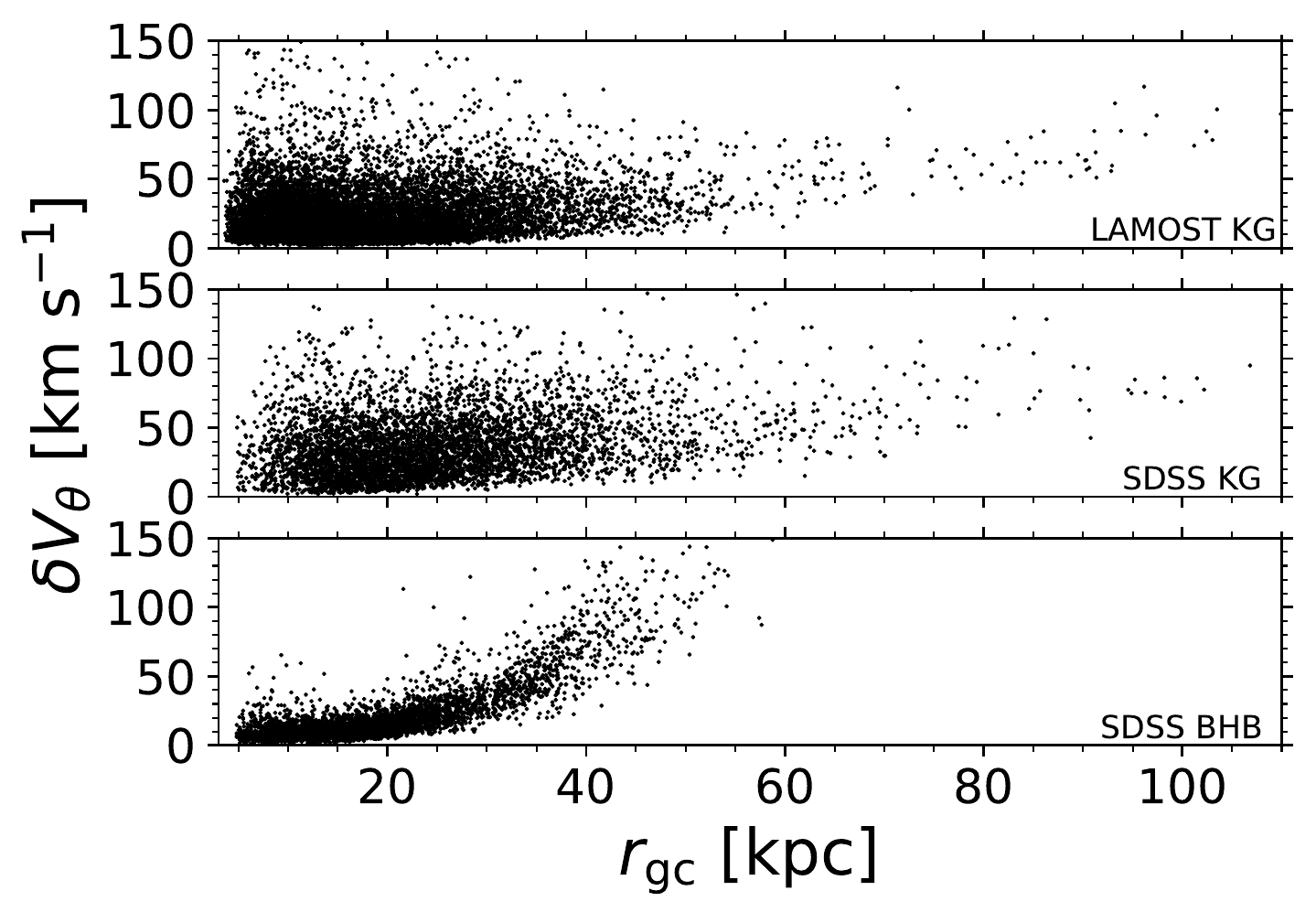}\\
\includegraphics[width=\columnwidth]{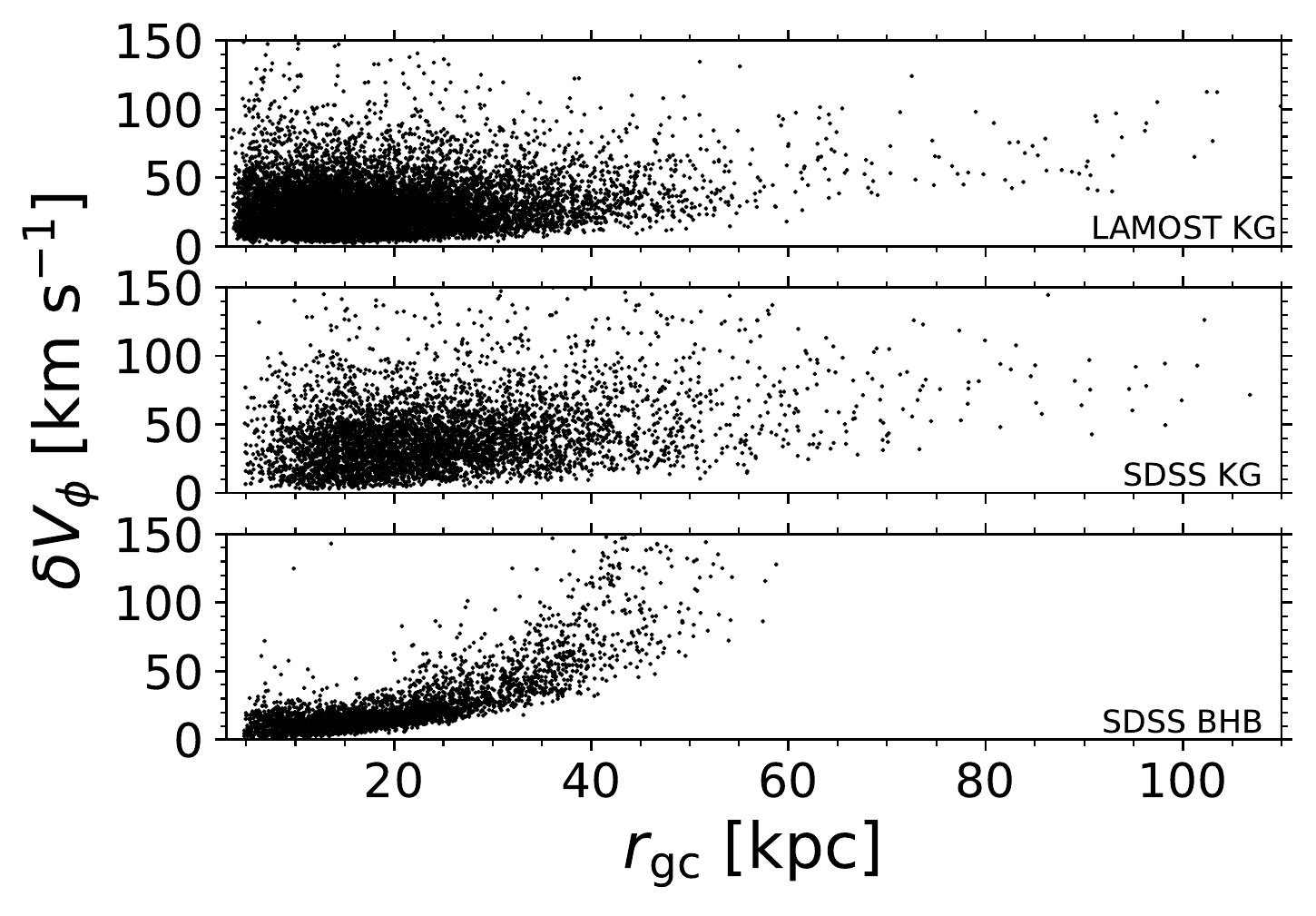}
  \caption{Uncertainties in Galactocentric spherical velocities ($\delta V_r$, $\delta V_\theta$, $\delta V_\phi$)
versus Galactocentric distance $r_\mathrm{gc}$ 
for our sample K-giant and BHB halo stars from LAMOST and SDSS.  
}
\label{fig:vlos_err}
\end{figure}

We plot the probability density distributions of distance, metallicity, and line-of-sight velocity for the K giant samples from LAMOST and SDSS and for the SDSS BHB sample in the upper panel of
Figure \ref{lamostsegue}. 
We plot the probability density distributions and Gaussian fits of spherical Galactocentric velocities $(V_r,V_\theta,V_\phi)$ for the samples and give mean velocity and the dispersions in the lower panels.
We show the
distribution of the spherical Galactocentric velocity uncertainties ($\delta V_r$, $\delta V_\theta$, $\delta V_\phi$) versus Galactocentric distance $r_\mathrm{gc}$ in Figure \ref{fig:vlos_err}. Note the larger spread in $\delta V_r$ for $r_\mathrm{gc}<20$ kpc as compared to the smaller spread for $r_\mathrm{gc}>20$ kpc. The larger spread is due to the larger contribution of proper motion which is required to calculate $V_r$ in specific directions in space within a few kpc of the Sun (e.g., stars located within a few kpc directly above the Sun). The proper motion uncertainties are larger compared to the small uncertainties in the line-of-sight velocities and thus, when propagated, result in larger $\delta V_r$.
The two K giant samples are qualitatively comparable and we combine these for our analysis of anisotropy; after applying all our selection criteria described in this section, we remove the 498 stars-in-common from the LAMOST K-giant sample, leaving a total sample of 17478 LAMOST/SDSS halo K giants.

The relative distance uncertainties for the K giants 
are shown 
in Figure 1 of \citet{Bird2019beta} and Figure 4 of \citet{Yang2019a} for LAMOST DR5 K giants and Figure 9 of \citet{Xue2014} for SEGUE K giants.
The propagated uncertainties in $r$, $g-r$, and [Fe/H] yield a median distance precision of 16 percent or better for the K giants. SDSS BHB stars have 5 percent distance precision \citep{Xue2008,Xue2011}.
\citet{Bird2019beta} find that the \citet{Xue2014} distance method for LAMOST K giants estimates distances which are closer by 10 percent on average compared with distances estimated by \citet{Bailer-Jones2018}. Correcting for the bias lead to a slight reduction in their velocity anisotropy estimates.
\citet{Yang2019b} use the same distance estimates as we use for LAMOST and SDSS K giants and SDSS BHB stars and make a comparison with $Gaia$ parallaxes. They find no distance bias for BHB stars and a 15 percent bias for K giants, such that the distances using the method of \citet{Xue2014} are closer compared to $Gaia$ based distances. They additionally find a decrease with $Gaia$ $G$ magnitude such that the distance bias diminishes for magnitudes $G>14$. Less than 10 percent of our sample halo stars have $G<14$.
Other distance methods have been applied to different samples (with overlaps to our sample), e.g., the methods of \cite{Schonrich2019}, \citet{Sanders2018}, \citet{Bailer-Jones2018}, \citet{Carlin2015}, \citet{Liu2014}, and \citet{Deason2011.416}.
We select the methods of \citet{Xue2014} and \citet{Xue2008,Xue2011} for K giants and BHB stars, respectively, as these methods are tailored to our selected star-types as well as to more metal poor stars such as our selected halo stars. We do not apply any correction for systematic uncertainties to our distances, although in the Appendix we investigate the uncertainty in our results due systematic distance uncertainties.
Distance biases of 10 percent introduce uncertainty on the anisotropy measurements of the halo sample
of $\Delta\beta\sim0.1-0.2$ depending on $r_\mathrm{gc}$, and are discussed further in
the Appendix.
Adopting distances scaled by $\pm10$\% instead of the
    \citet{Xue2008,Xue2011} scale for BHB stars reduces/increases, respectively, the anisotropy parameter $\beta$ (see  
the Appendix and Figure \ref{fig:distance_systematics_beta}).

\subsection{Substructure Removal: \citet[in preparation]{Xue2019} Method}
  \label{sec:sub}

Dynamical properties of stars which remain constant over many orbits of the galaxy can be used to characterize and classify them into substructures. These constants are called integrals of motion. These remain constant under the assumption that the Galactic potential is stationary and the stars are not perturbed along their orbits. This is not true for galaxies as a whole during their formation and evolution, but for halo stars orbiting the galaxy, this approximation is an adequate simplification. Stars which share similar integrals of motion have similar orbits. Stars which clump in integral-of-motion phase space are likely candidates of sharing the same origin. Despite clumping in integral-of-motion space, the stars may be spread along different points in their orbits, covering a large spatial volume, thus making any hypothesis of common origins through just the use of spatial coordinates difficult.

We remove substructure from our halo samples using the method of \citet[in preparation]{Xue2019} who identify substructure through clustering in integral-of-motion space, namely with
eccentricity $e=(r_\mathrm{apo}-r_\mathrm{peri})/(r_\mathrm{apo}+r_\mathrm{peri})$, semi-major axis $a=(r_\mathrm{apo}+r_\mathrm{peri}$)/2,  $(l,b)_\mathrm{orbit}$ (the direction of the orbital plane's polar axis relative to the defined Galactic coordinate plane), and the direction of apocenter $l_\mathrm{apo}$ (the angle between apocenter and the projection of the Galactocentric $X$-axis on the orbital plane). 
\citet[in preparation]{Xue2019} use 6D positions and velocities to calculate these five integrals of motion $\hat{O}=(e,a,l_\mathrm{orbit},b_\mathrm{orbit},l_\mathrm{apo})$ within a Galactic potential (a \citet{Hernquist1990} bulge and exponential disk for the stellar components and a spherical NFW profile \citep{Navarro1995.275.720,Navarro1996,Navarro1997} for dark matter) and search for stars clustering in these properties using the friends-of-friends algorithm. Groups with $\ge 6$ member stars are defined as substructure and we remove these to define the smooth, diffuse halo. The results of this method have also recently been used to locate and analyze Sagittarius member stars by \citet{Yang2019b}.

Substructure identification through clustering in the integral-of-motion space is a highly effective method and has previously been explored by many studies using both observations
\citep[e.g.,][]{Helmi1999Nature,Helmi2006,Helmi2017,Helmi2018,Chiba2000,Re_Fiorentin2005,Re_Fiorentin2015,Klement2008,Klement2009,Klement2011,Kepley2007,Morrison2009,Smith2009,Yuan_Zhen2018,Li_Hefan2019,Koppelman2019} and
simulations \citep[e.g.,][]{Helmi2000,Knebe2005,Meza2005,Penarrubia2006,Font2006_646,Brown2005,Choi2007,Gomez2010,Jean-Baptiste2017}. Two excellent reviews are \citet{Klement2010} and \citet{Smith2016}.
Now with the large samples of halo stars with proper motions from {\it Gaia} and line-of-sight velocities from LAMOST and SDSS, a much more extensive search for substructure can be made in integral-of-motion 
space. Previously in \citet{Bird2019beta} we used $E$-$L$ space to select the most prominent subset of the largest substructure, that of the Sagittarius Stream. We compare results in Section \ref{sec:beta1beta2}.
A full description of the substructure removal method will be the topic of \citet[in preparation]{Xue2019}. 

The removal of substructure reduces our halo samples from 12728 LAMOST K giants, 5248 SDSS K giants, and 3982 SDSS BHB stars, to 7684, 3277, and 2526, respectively.

\subsection{Effects of Substructure Removal on Stellar Halo Kinematics}
\label{sec:sub-rem}

\begin{figure*}
\includegraphics[width=1.8\columnwidth]{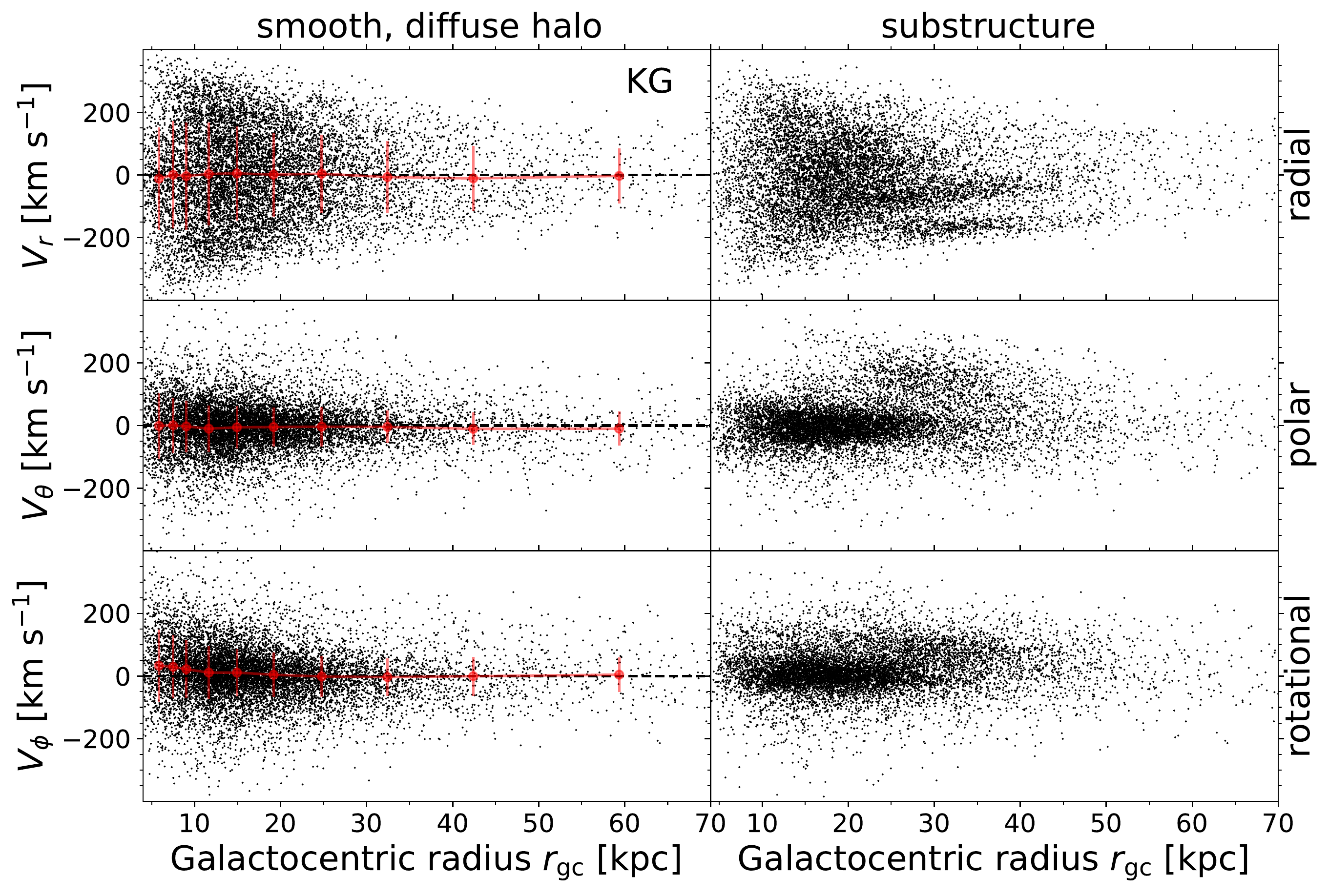}
\includegraphics[width=1.8\columnwidth]{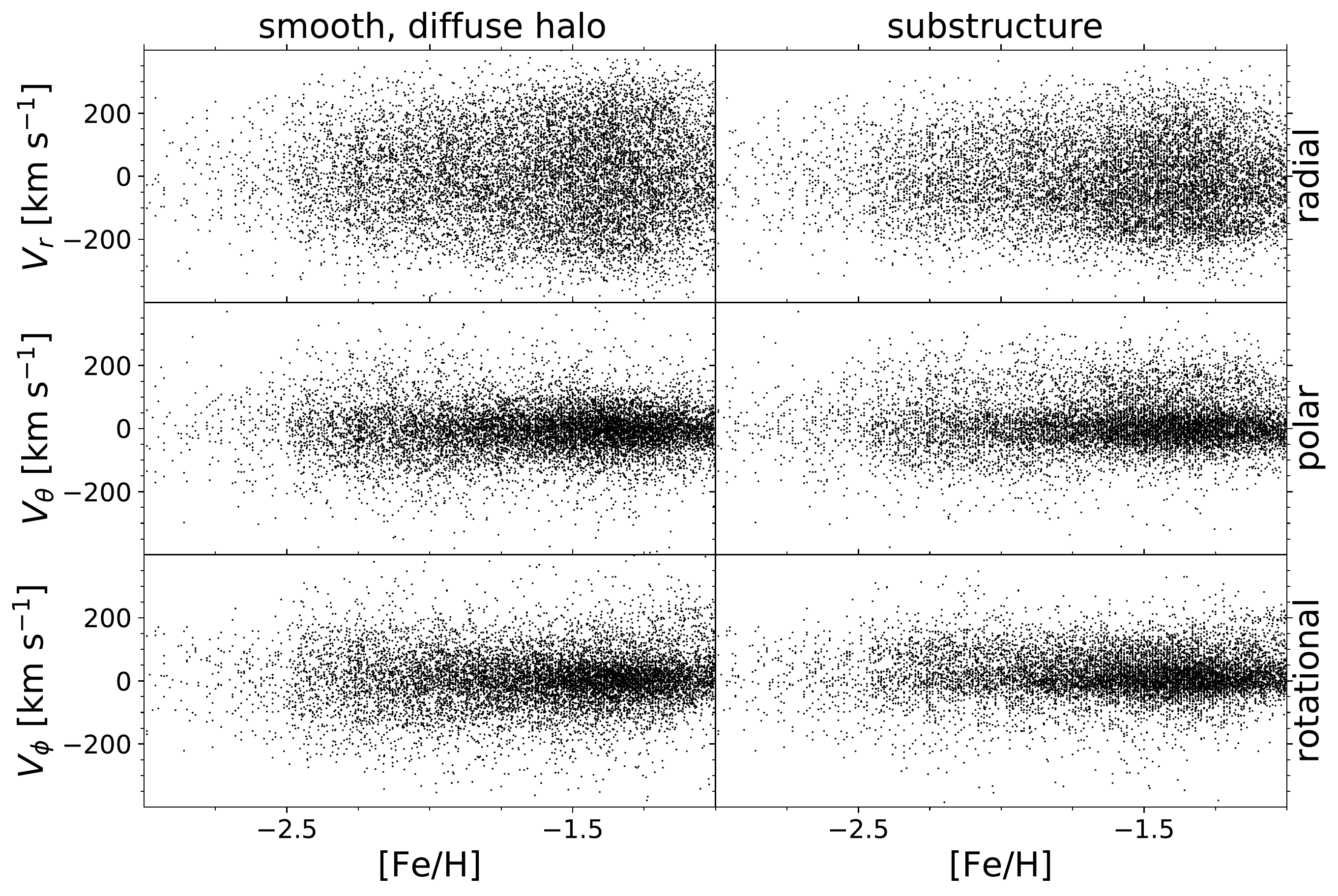}
  \caption{Space velocities in spherical coordinates
    $(r,\theta,\phi)$ for our sample of 17478 
halo 
    LAMOST and SDSS K giants for which we have LAMOST/SDSS line-of-sight velocities and
    metallicities, and {\it Gaia} proper motions (common stars are removed from LAMOST). 
The velocities are shown vs. metallicity [Fe/H] in the lower six panels, and vs. Galactocentric distance $r_\mathrm{gc}$ in the upper six panels.
Left column panels are the smooth, diffuse halo stars and the right are the removed substructure (halo stars are selected as described in Section \ref{sec:combined} and substructure is identified using the method of \citet[in preparation]{Xue2019}, as described in Section \ref{sec:sub}). Red bars on the markers indicate the respective velocity dispersion calculated in bins of $r_\mathrm{gc}$ and red markers indicate the mean velocity. The dispersion profile is further elaborated upon in Figure \ref{fig:profs} and within the text.
} 
  \label{fig:fehvel-kg}
\end{figure*}

\begin{figure*}
\includegraphics[width=1.8\columnwidth]{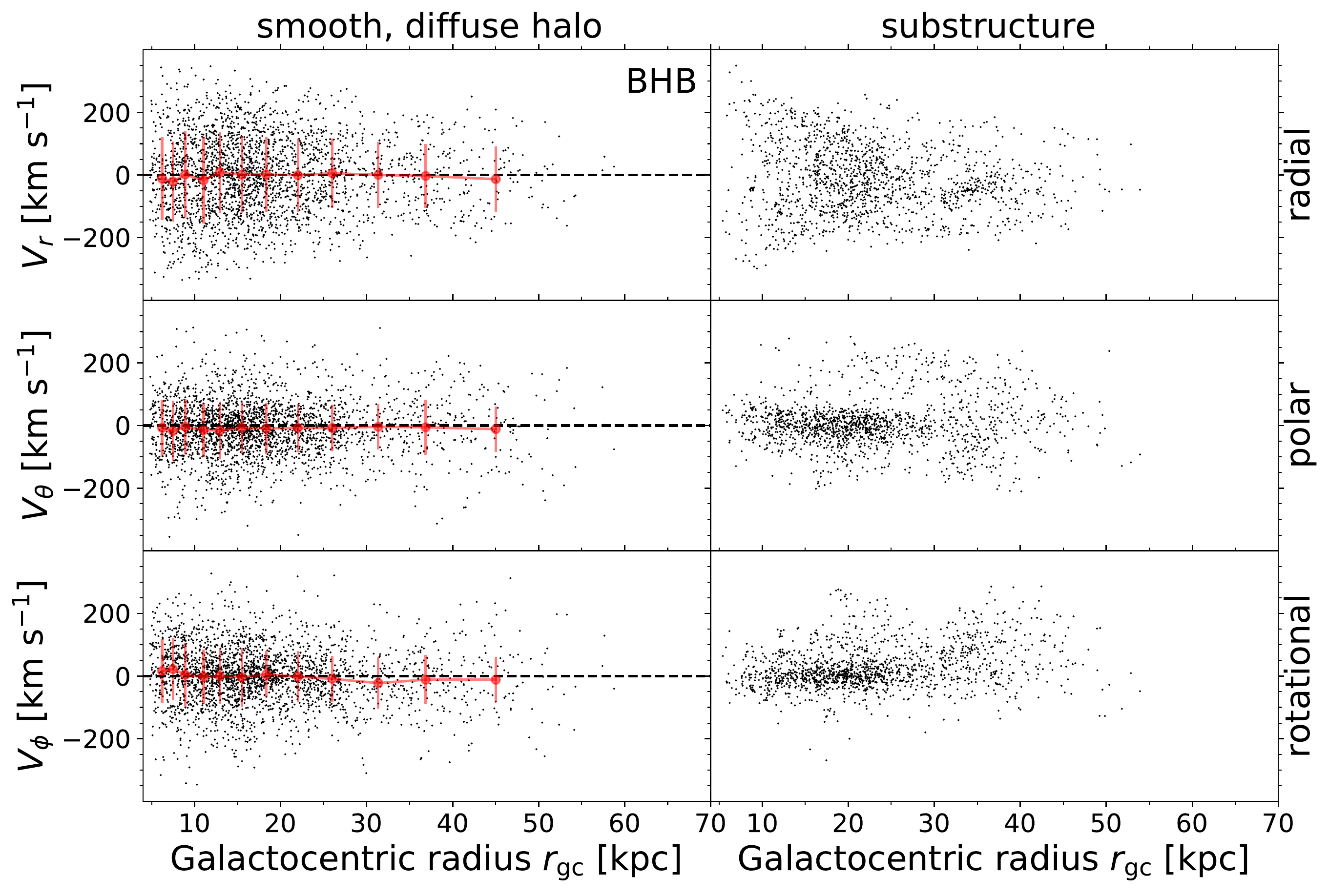}
\includegraphics[width=1.8\columnwidth]{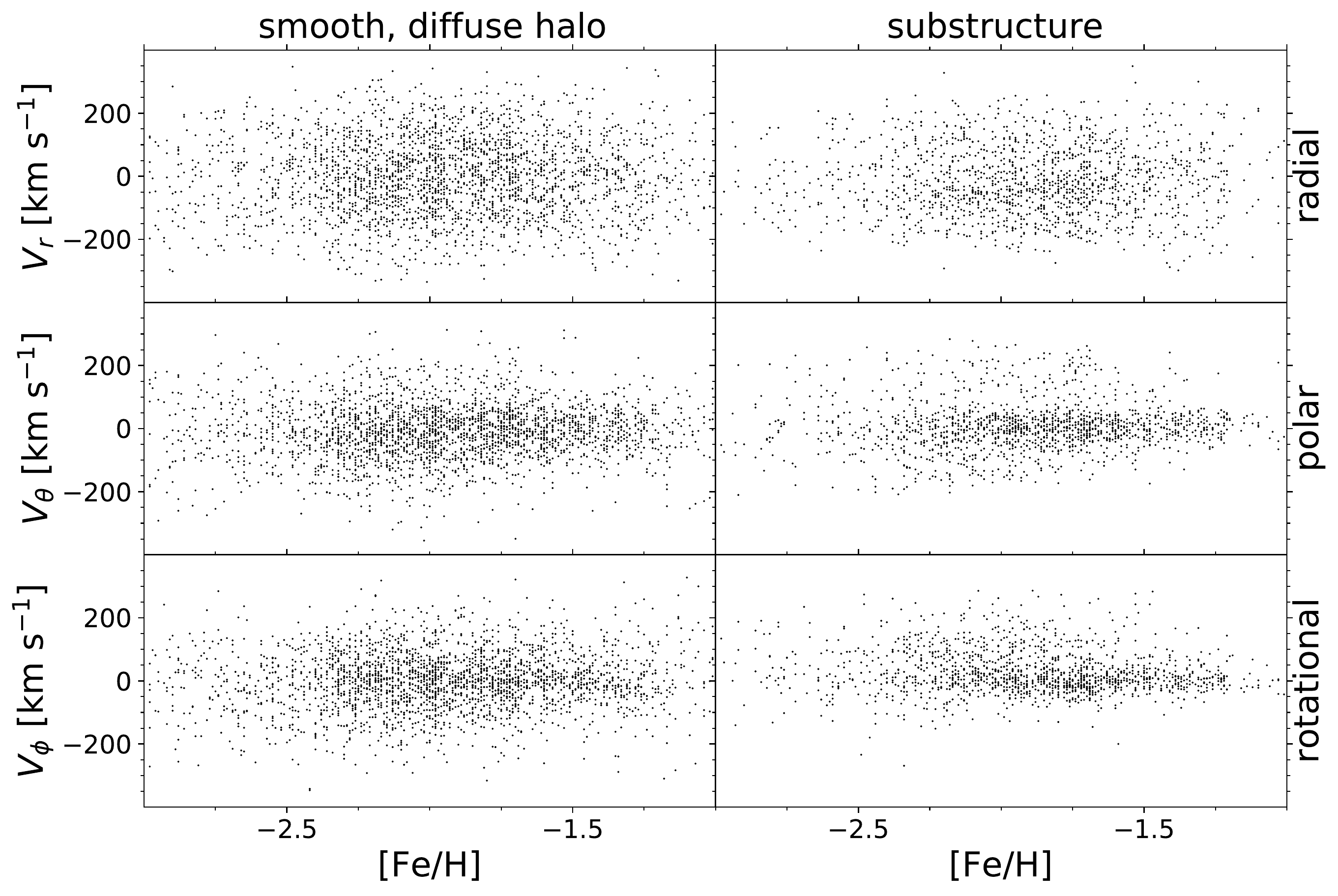}
  \caption{Space velocities in spherical coordinates
    $(r,\theta,\phi)$ for our sample of 3982 halo 
    BHB stars for which we have SDSS line-of-sight velocities and
    metallicities, and {\it Gaia} proper motions. 
The velocities are shown vs. metallicity [Fe/H] in the lower six panels, and vs. Galactocentric distance $r_\mathrm{gc}$ in the upper six panels.
Left column panels are the smooth, diffuse halo stars and the right are the removed substructure (halo stars are selected as described in Section \ref{sec:combined} and substructure is identified using the method of \citet[in preparation]{Xue2019}, as described in Section \ref{sec:sub}). Red bars on the markers indicate the respective velocity dispersion calculated in bins of $r_\mathrm{gc}$ and red markers indicate the mean velocity. The dispersion profile is further elaborated upon in Figure \ref{fig:profs} and within the text.
} 
  \label{fig:fehvel}
\end{figure*}

We next examine the extent to which substructure affects the stellar halo kinematics. 
Structures seen in
configuration and velocity space are considered likely ``un-relaxed''
and ``non-virial'' components within the Galactic halo. \citet{Bird2015mwhalo} showed that such features in velocity space can be in a transitional phase leading to relaxation. \citet{Deason2013beta} and \citet{Cunningham2016} explain that dips found in velocity space can be signatures of shells left from past mergers. As has been
shown in detail in \citet{Loebman2018}, the anisotropy parameter is
sensitive to both the presence of satellites in the halo, and to the
passage of satellites through the underlying, smooth and kinematically
relaxed stellar halo. \citet{Deason2018.862} and \citet{Simion2019} showed that anisotropy can be used to find apocentric pile-ups of halo stars left over from large satellite mergers.

In Figure \ref{fig:fehvel-kg} and \ref{fig:fehvel} 
we show the 3D velocities $V_r,V_\theta, V_\phi$ for our K giant and BHB halo samples after substructure removal and the 3D velocities of the substructure alone. The velocities are shown as functions of metallicity
[Fe/H] (lower panels) and Galactocentric radius $r_\mathrm{gc}$ (upper
panels). 
The substructure is highly clumped in velocity along $r_\mathrm{gc}$; after removal, the remaining halo sample is diffuse and smoothly distributed. No clumps remain discernible by eye. We thus refer to the remaining halo sample as the ``smooth, diffuse stellar halo.''
Along all metallicities, both the smooth, diffuse halo and the substructure have a wide spread in radial velocities. 
We use the {\tt extreme-deconvolution} algorithm \citep{Bovy2011} to fit the data in each bin with a 3D Gaussian including the errors and their covariances and plot the resulting means and dispersions with red markers and bars.

\begin{figure*}[htb]
\includegraphics[width=2\columnwidth]{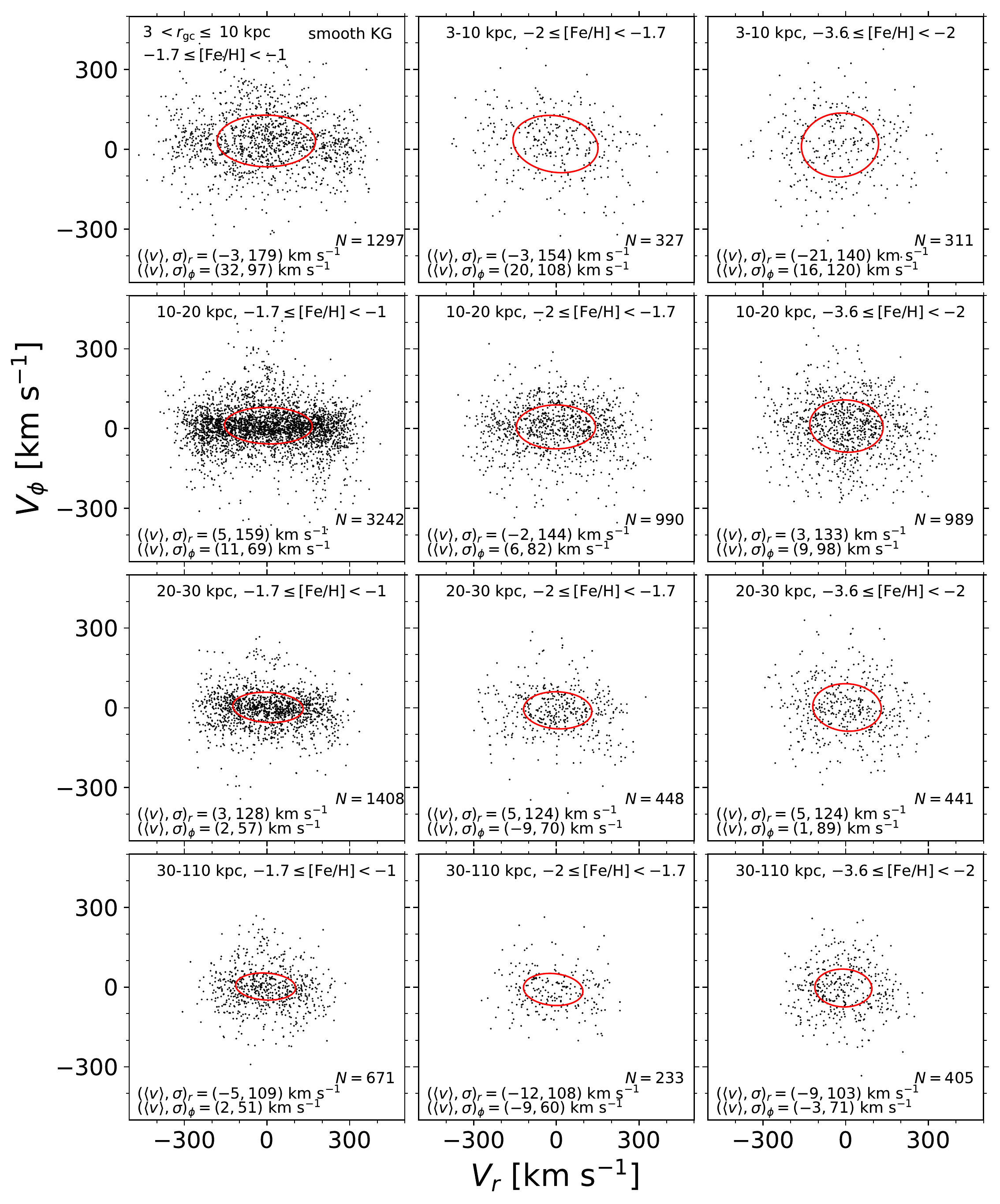}
\caption{Comparison of radial and rotational space velocities $V_r$ and $V_\phi$ (spherical coordinates)
for our combined sample of 10762 
LAMOST and SDSS halo K giants after stream removal (common stars are removed from LAMOST). 
The upper legend of each panel indicates the range of Galacatocentric distance and metallicity.
The red velocity ellipsoids mark the Gaussian component fits to the data within the Galactocentric distance and metallicity ranges displayed. The mean and dispersion, $\langle V_r \rangle$ and $\sigma_r$, are written in the lower left of each panel and below this are the mean and dispersion, $\langle V_\phi\rangle$ and $\sigma_\phi$. The number of stars is displayed in the lower right of each corresponding panel.
Fig. \ref{fig:Vr-Vphi-lamost} and \ref{fig:Vr-Vphi-bhb} 
are similar, but compare the space velocities separately for LAMOST/SDSS K giants and SDSS BHB stars, respectively.
}
\label{fig:Vr-Vphi-lamost}
\end{figure*}

\begin{figure*}[htb]
\includegraphics[width=2\columnwidth]{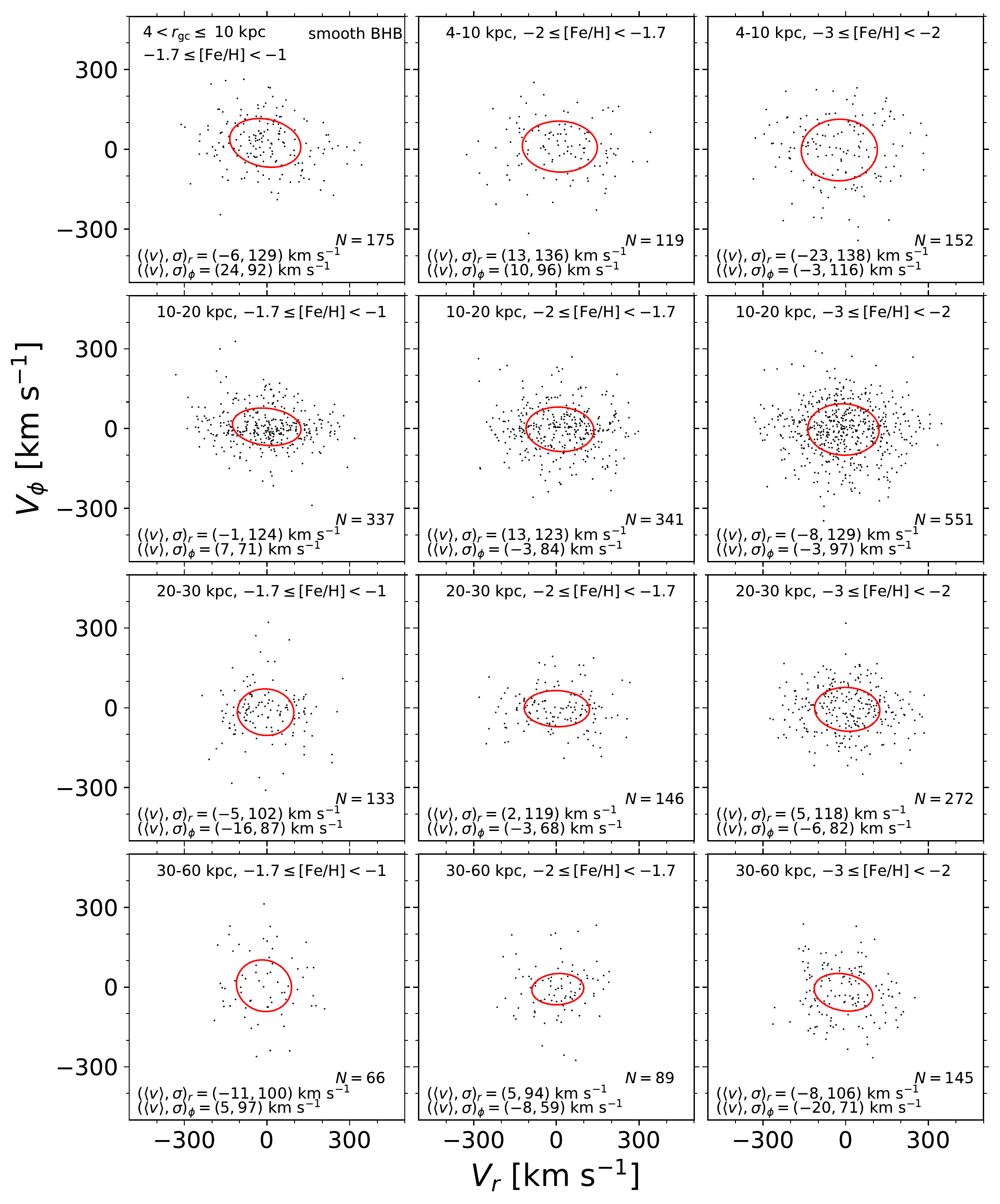}
\caption{Comparison of radial and rotational space velocities $V_r$ and $V_\phi$ (spherical coordinates)
for our sample of 2526 SDSS halo BHB stars after stream removal. 
The upper legend of each panel indicates the range of Galacatocentric distance and metallicity.
The red velocity ellipsoids mark the Gaussian component fits to the data within the Galactocentric distance and metallicity ranges displayed. The mean and dispersion, $\langle V_r \rangle$ and $\sigma_r$, are written in the lower left of each panel and below this are the mean and dispersion, $\langle V_\phi\rangle$ and $\sigma_\phi$. The number of stars is displayed in the lower right of each corresponding panel.
Fig. \ref{fig:Vr-Vphi-lamost} and \ref{fig:Vr-Vphi-bhb} 
are similar, but compare the space velocities separately for LAMOST/SDSS K giants and SDSS BHB stars, respectively.
}
\label{fig:Vr-Vphi-bhb}
\end{figure*}

After stream removal, we compare the radial and azimuthal velocities for our LAMOST and SDSS smooth, diffuse halo star samples in Figures \ref{fig:Vr-Vphi-lamost}$-$\ref{fig:Vr-Vphi-bhb}. The samples are divided into bins based on Galactocentric distance $r_\mathrm{gc}$ and metallicity [Fe/H].
We select metallicity bins motivated by the metallicity distribution of our samples (see Figure \ref{lamostsegue} upper middle panel) in order to ensure a sufficient number of stars in each bin, as well as on recent works finding evidence for a dependency between stellar halo kinematics and metallicity \citep{Myeong2018action,Deason2018,Belokurov2018,Lancaster2019} to facilitate comparison with those studies.
We use the {\tt extreme-deconvolution} algorithm \citep{Bovy2011} to fit the data in each bin with a 3D Gaussian including the error covariances and plot the resulting (tilted) ellipsoids in red.
We find that the halo's velocity dispersion is more radial closer toward the Galactic Center (upper left panels) and more nearly isotropic at further distances (lower right panels). Similarly, the halo's velocity dispersion is more radial with increasing metallicity (right panels) and more nearly isotropic with decreasing metallicity (left panels). These plots are in agreement with trends seen in the same velocity space by \citet{Belokurov2018} for a nearer sample of main sequence stars.

\begin{figure*}[htb]
\begin{tabular}{ccc}
\includegraphics[width=.65\columnwidth]{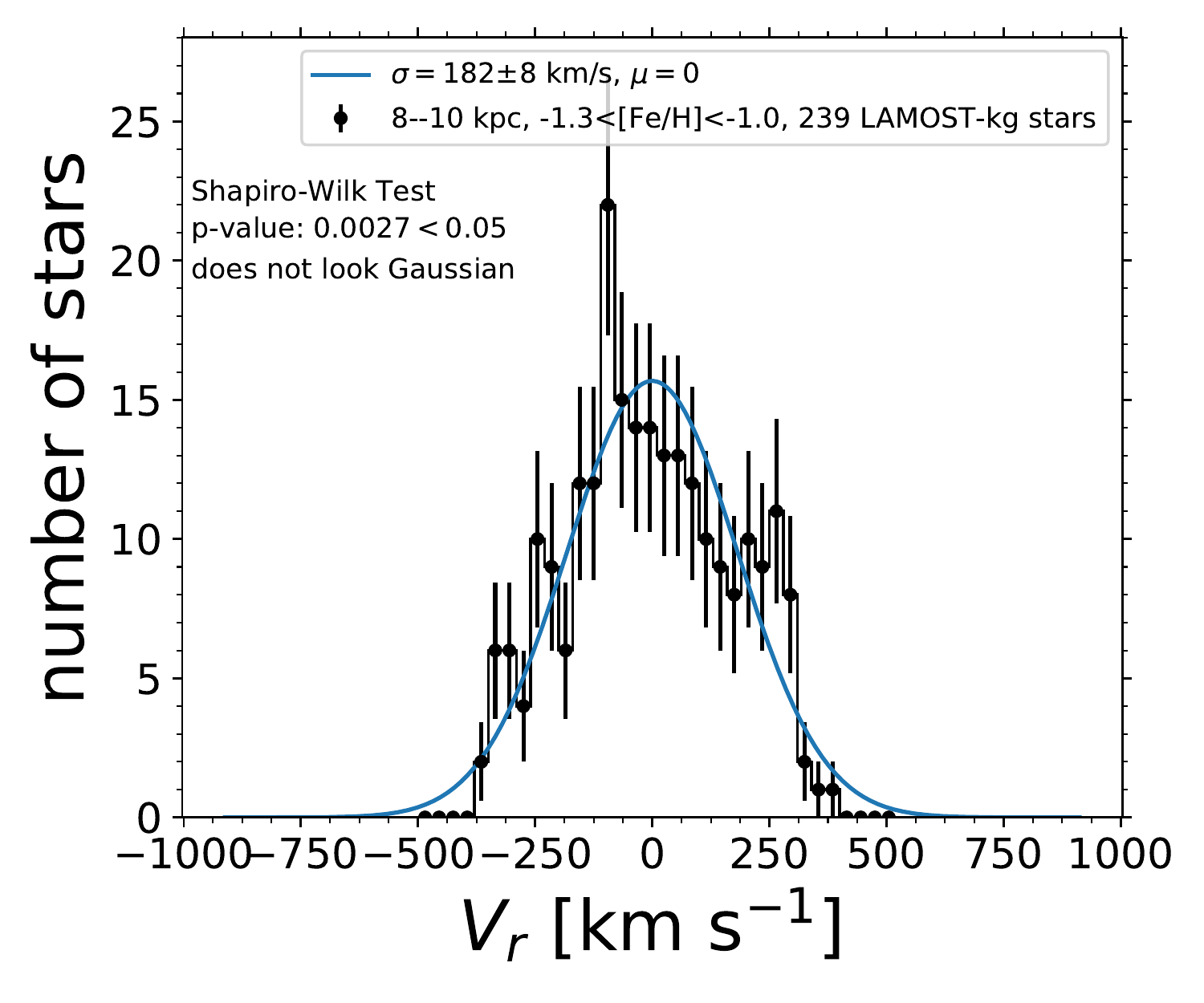}&
\includegraphics[width=.65\columnwidth]{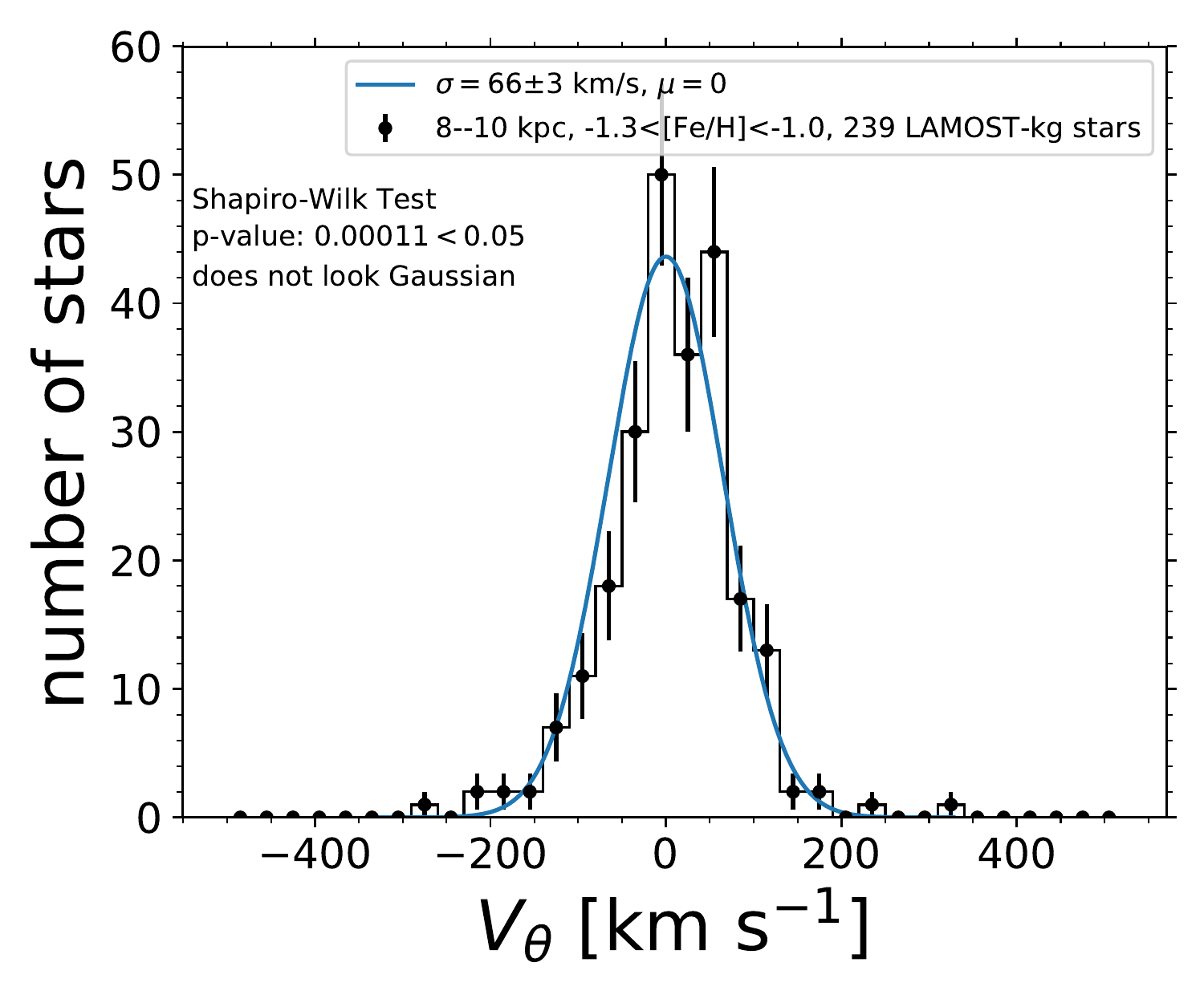}&
\includegraphics[width=.65\columnwidth]{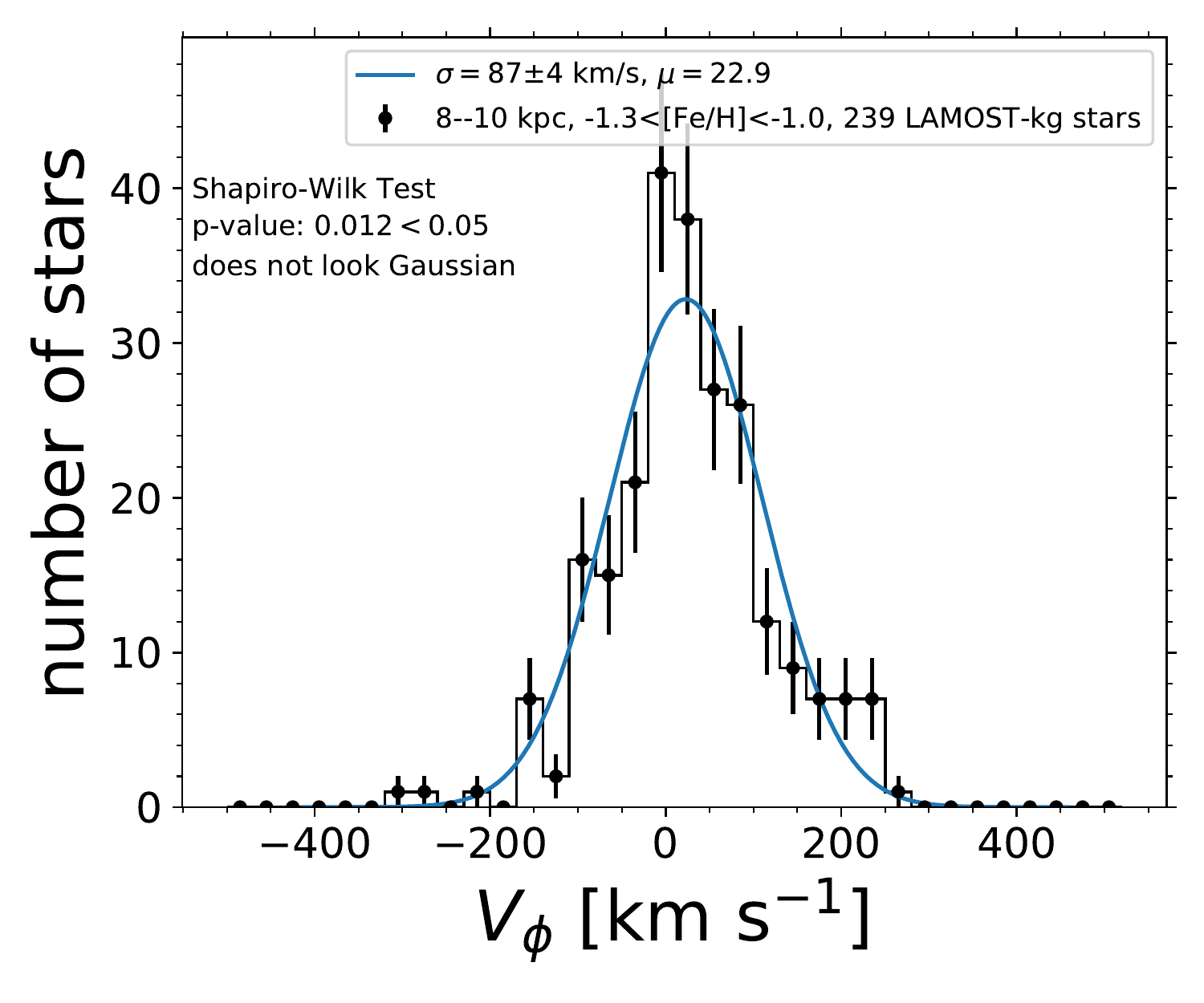}\\
\includegraphics[width=.65\columnwidth]{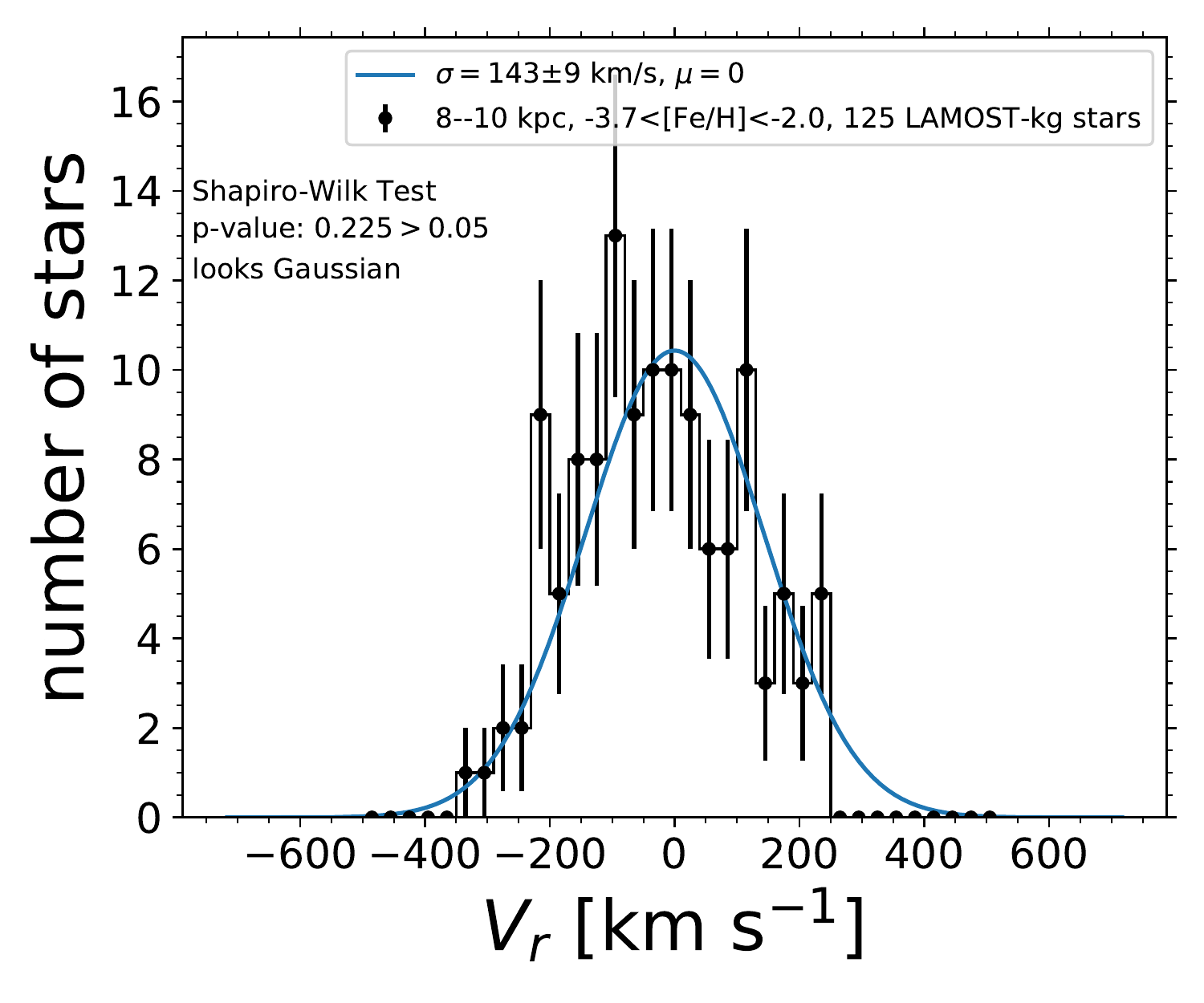}&
\includegraphics[width=.65\columnwidth]{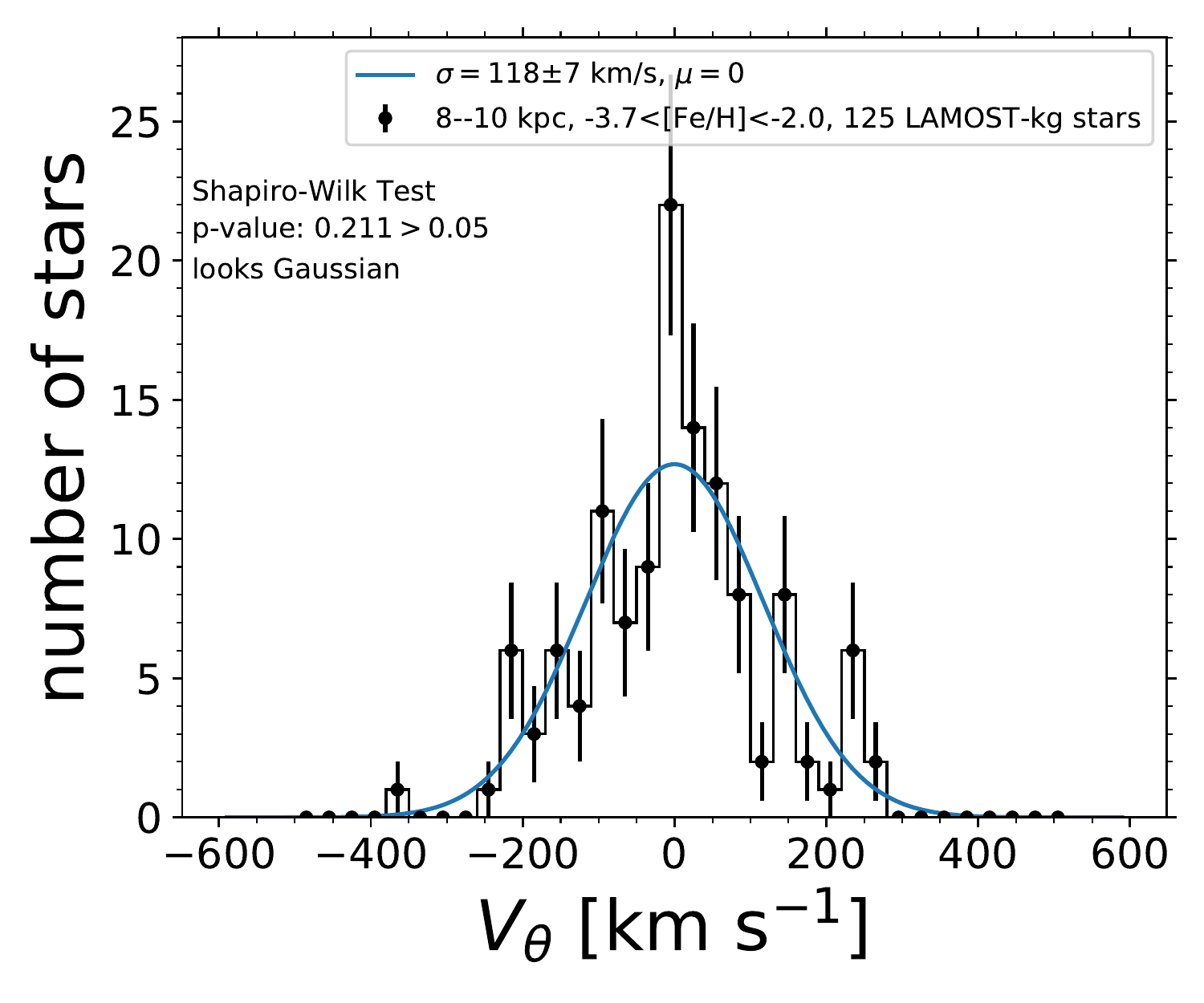}&
\includegraphics[width=.65\columnwidth]{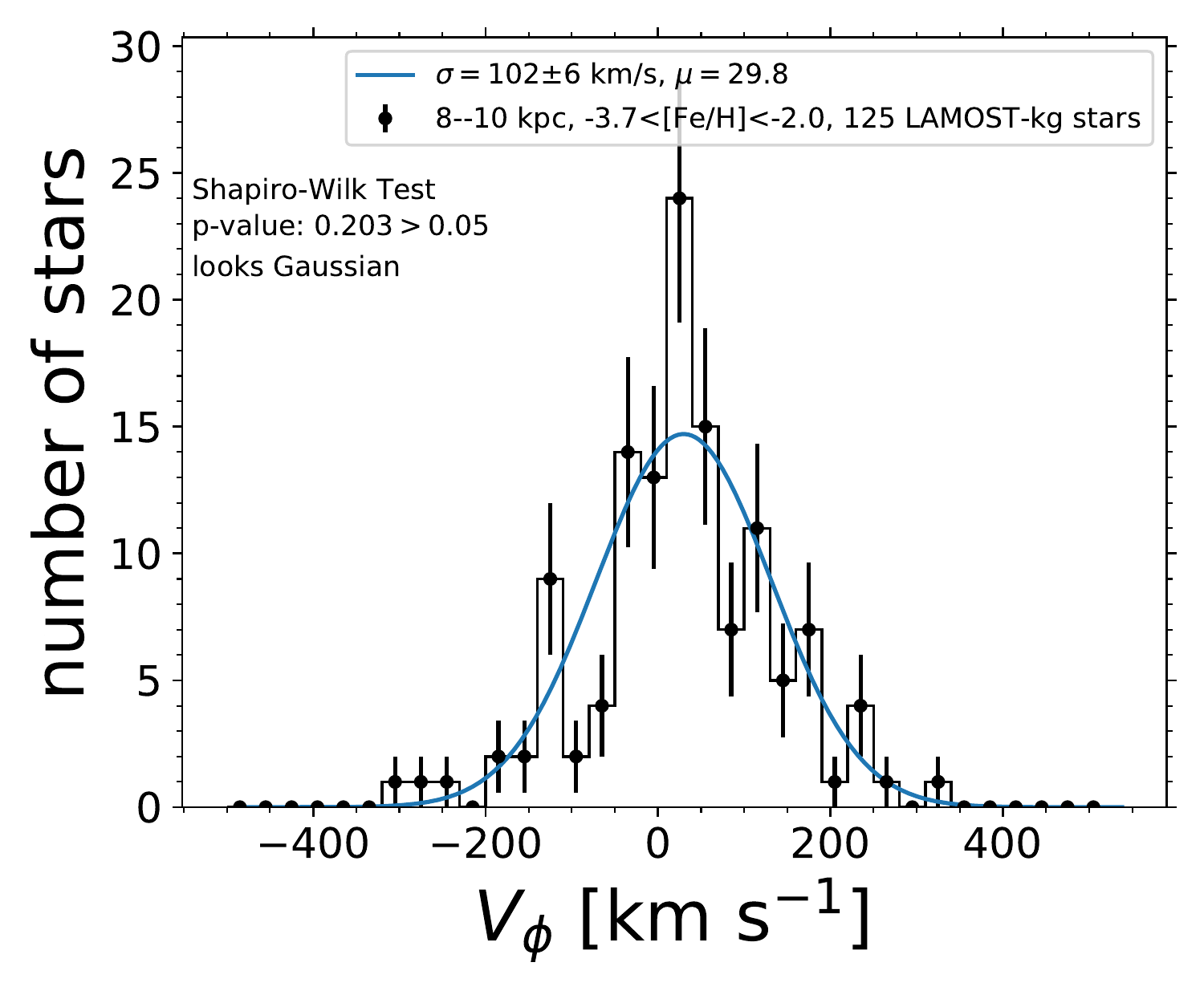}
\end{tabular}
\caption{Shapiro-Wilk test for goodness-of-Gaussian-fit. Three representative 3D velocity histograms (black histogram and error bars) and corresponding Gaussian fits (blue line). For these examples we show $(V_r,V_\theta,V_\phi)$ histograms for LAMOST K giants within the same radial bin ($8<r_\mathrm{gc}<10$ kpc) within two metallicity bins ($-1.3<$ [Fe/H] $<-1$ and  $-3.7<$ [Fe/H] $<-2$ in the upper and lower panels, respectively). According to the Shapiro-Wilk test, the data histograms in the upper panels show evidence for non-Gaussianity (p-value $<0.05$) and the data histograms in the lower panels are consistent with being Gaussian (p-value $>0.05$).
}
\label{fig:Shapiro-Wilk}
\end{figure*}

We have performed the Shapiro-Wilk test for Gaussianity on the velocity histograms in our radial and metallicity bins. We make use of the function as provided in the statistics module of {\tt SciPy} \citep{2020Virtanen}. Just over half pass this test with $p$-value$>0.05$; in many of the bins the star numbers are low and this gives aid to histograms to look more Gaussian. Six examples of the velocity distributions are shown in Figure \ref{fig:Shapiro-Wilk}. For about 40 percent of the bins there is evidence for non-Gaussianity. This could, for example, indicate that residual structure remains in the sample, or it could also be that the underlying velocity distribution is intrinsically non-Gaussian. Non-Gaussianity has also been found by, e.g., \cite{Smith2009}, \citet{King2015}, and \citet{Lancaster2019}, using the Kolmogorov-Smirnov test, Anderson–Darling test, kurtosis test, and Bayesian analysis on the velocity distributions in order to estimate the stellar halo anisotropy $\beta$. The Gaussian fits in Figure \ref{fig:Shapiro-Wilk} (blue lines) are performed using the same procedure as \citet{Bird2019beta}, namely dispersion with median uncertainty subtracted in quadrature (In the Appendix we compare the method of \citet{Bird2019beta} with the method used in this paper.).

We measure the mean velocity and the velocity dispersion (full covariance matrix) profiles using the algorithm {\tt extreme-deconvolution} \citep{Bovy2011}. This algorithm allows us to measure the 3D velocity dispersions within each radial bin, estimate the covariance between the three velocity dispersion components, and take into account corrections needed due to the known uncertainty and covariance of each star within the bin.
We measure the anisotropy in the same manner as \citet{Bird2019beta}.
Unique to the current work is our choice of bin sizes in $r_\mathrm{gc}$, selected in order to reduce covariance between bins. We select bin widths which are nearly twice the relative distance uncertainty, rather than, e.g., fixed bin width or fixed bins based on the number of stars per bin.
For the purpose of selecting the bin sizes we use 15 percent distance uncertainties for K giants and 10 percent for BHB stars. We slightly broaden the bin size for our BHB sample, which have typical relative distance uncertainty $\sim5$ percent, in order to ensure $\sim100$ or $>100$ stars per bin.
Subtle features in the velocity dispersion or anisotropy profile much smaller than these bin widths for our samples are likely unresolved due to uncertainties in distance measurements. The first bin contains stars within $r_\mathrm{gc}=2-7$ kpc. 
The remaining bins for K giants end 
at $r_\mathrm{gc}= 8,$ 10, 13, 17, 22, 29, 38, 50, 110
kpc and for BHB stars at $r_\mathrm{gc}= 8,$ 10, 12, 14, 17, 20, 24, 29, 34, 41, 80
kpc. For plotting we mark the median distance of stars within the selected radial bins.

Error bars for velocity mean, dispersion, and anisotropy are estimated from the Poissonian sampling in each bin after propagating all the sources of error. 
As analyzed by, e.g., \citet[][as discussed in their Appendix B]{Morrison1990}, we estimate the error bars using the Poissonian uncertainties defined as $\sigma n^{-0.5}$ and $\sigma (2n)^{-0.5}$ for the velocity mean and dispersion, respectively, where $\sigma$ is the velocity dispersion and $n$ is the number of stars in the bin. For each bin we calculate the uncertainty in anisotropy with 1000 Monte Carlo simulations propagating all sources of error. 

Other methods for determining the uncertainty in a measurement include, e.g., the cumulative distribution function as recently demonstrated by \citet{Cunningham2019a}, 
Bayesian techniques which incorporate treatment of contamination such as methods by \citet{Deason2013beta}, \citet{Cunningham2016}, \citet{Lancaster2019}, \citet{Cunningham2019a}, \citet{Cunningham2019b}, or that presented in the appendix of \citet{Morrison1990}.
We also must keep in mind a Gaussian may not be the best fit to the stellar halo velocity distributions.

The algorithm {\tt extreme-deconvolution} takes into account the individual velocity uncertainties for each star; we compare our results to the method of standard deviation with sigma clipping and subtracting in quadrature the median of the velocity uncertainties. We find for most bins the difference between the velocity dispersion measurements is $1-4$ percent and for large distances distances further than 40 kpc, the difference is closer to double this. Such uncertainty is comparable to that due to random error. Differences between methods of dispersion calculation have also been discussed by \citet{Lancaster2019} who find a smaller anisotropy value ($\Delta\beta\sim0.2$) for their outermost radial bin ($30.0-67.93$ kpc) when using standard deviation with sigma clipping and median velocity uncertainties subtracted in quadrature as compared to their single 3D multi-variate Gaussian method. From our tests we conclude that standard deviation with sigma clipping and subtracting in quadrature the median of the velocity uncertainties is sufficient to describe the dispersion profile for the underlying smooth, diffuse halo of the Milky Way (further discussion is included in the Appendix).

\begin{figure}[htb]
\includegraphics[width=\columnwidth]{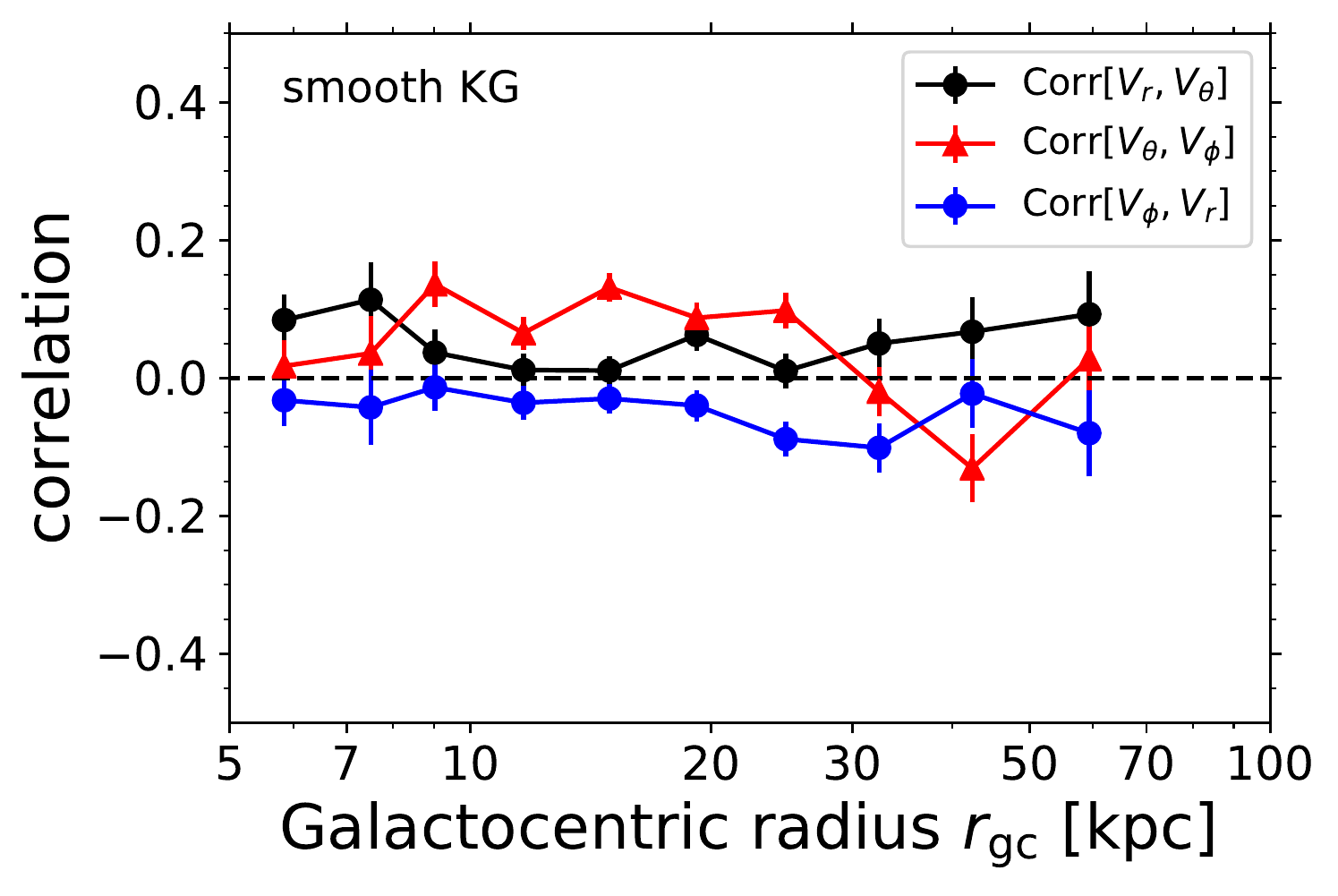}\\
\includegraphics[width=\columnwidth]{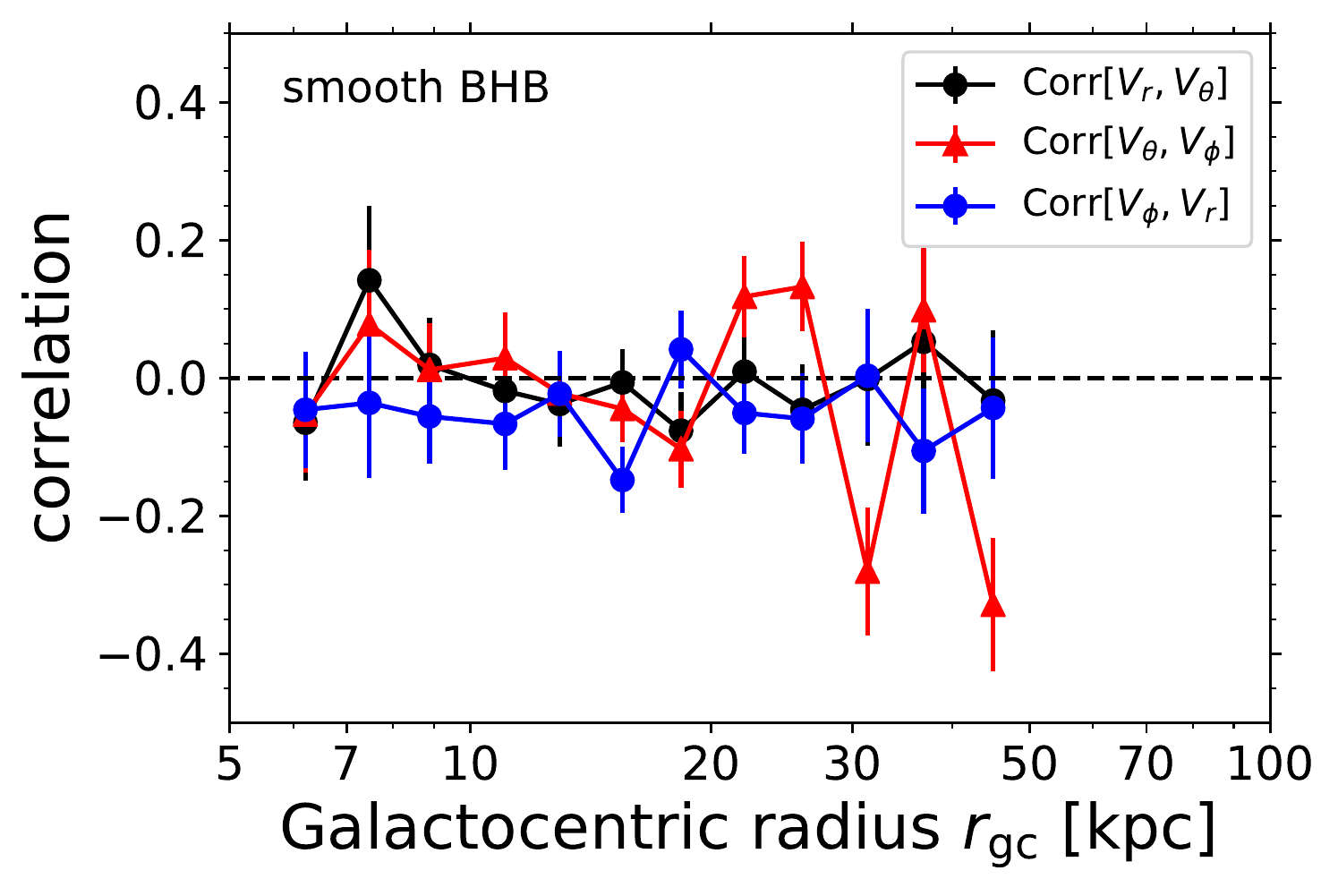}
    \caption{Correlation coefficient in bins of Galactocentric radius $r_\mathrm{gc}$, where $\mathrm{Corr}[V_i,V_j]=\frac{\sigma_{ij}^2}{\sigma_{ii}\sigma_{jj}}$.
}
  \label{fig:corr}
\end{figure}

We plot the correlation coefficient $\mathrm{Corr}[V_i,V_j]=\frac{\sigma_{ij}^2}{\sigma_{ii}\sigma_{jj}}$, where the $\sigma$ terms we attain from \\ {\tt extreme-deconvolution},
and its uncertainty \\ $\delta\mathrm{Corr}[V_i,V_j]=\sqrt{\frac{1-\mathrm{Corr}[V_i,V_j]^2}{n-2}}$ in Figure \ref{fig:corr}. The correlations are small ($<0.1$ for most bins), comparable to those measured by previous works \citep[e.g.][]{Smith2009b,Lancaster2019}.

\begin{figure*}[htb]
\includegraphics[width=\columnwidth]{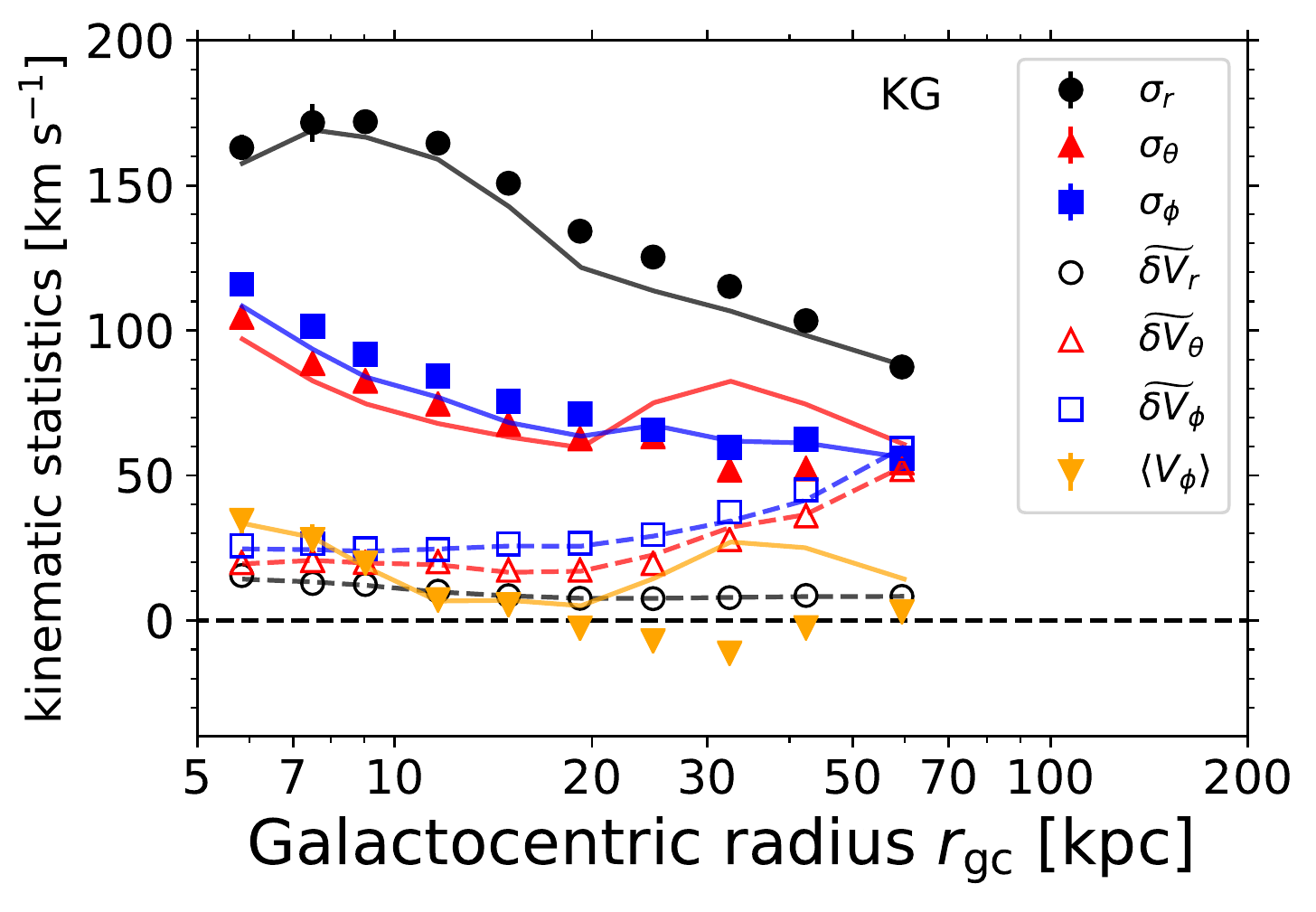}
\includegraphics[width=\columnwidth]{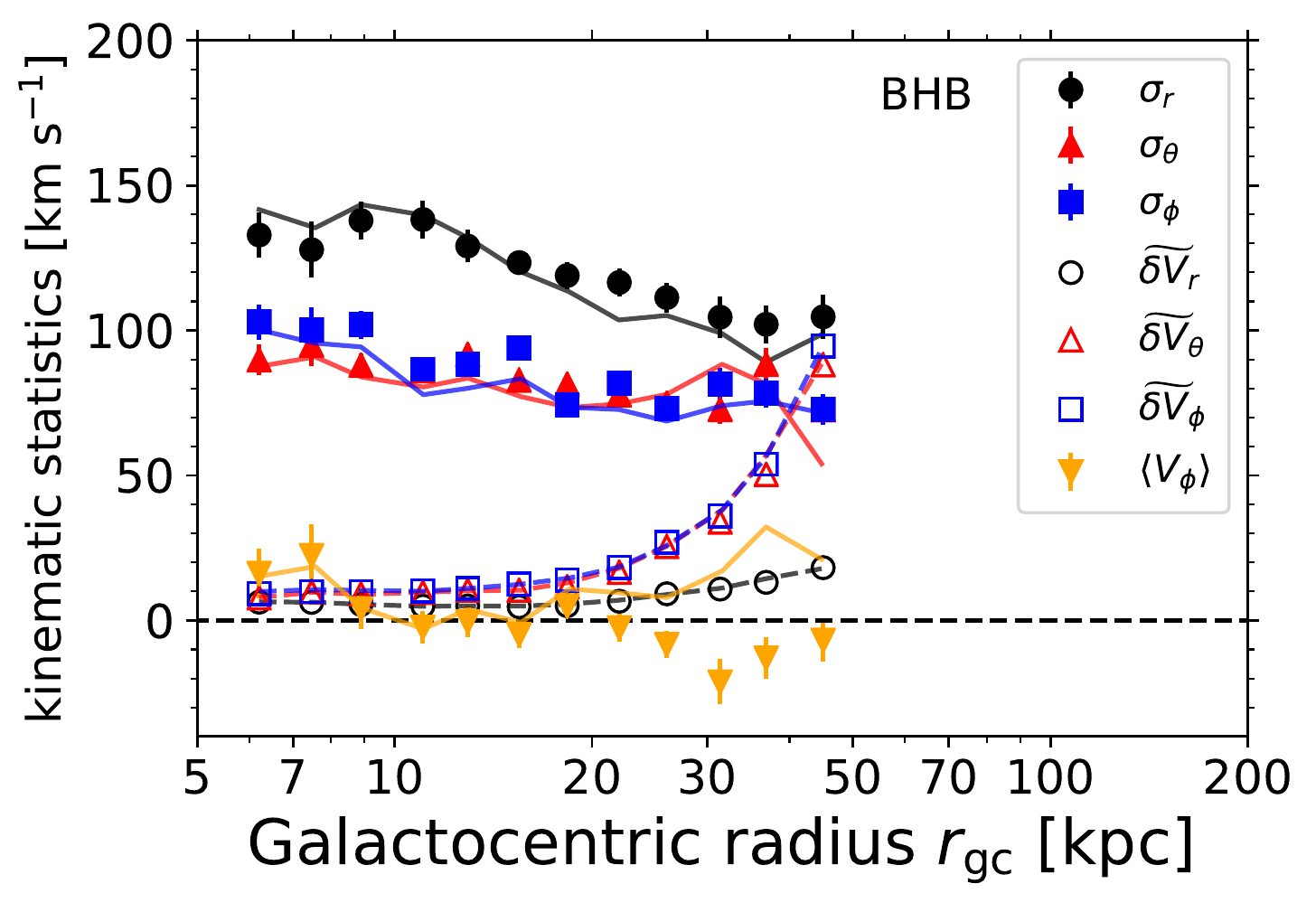}
    \caption{Total LAMOST/SDSS halo K-giant sample before substructure removal (lines, left panel) and the smooth, diffuse K-giant halo stars (symbols, left panel) after applying the method of \citet[in preparation]{Xue2019} for substructure removal, binned in Galactocentric radial bins. The same is shown for our SDSS halo BHB samples (right panel).
Filled black, red, and blue markers are the three velocity dispersion components ($\sigma_r, \sigma_\theta, \sigma_\phi$) and filled orange is the mean rotational velocity $\langle V_\phi\rangle$. 
Open black, red, and blue markers are the median errors 
($\widetilde{\delta V_r},\widetilde{\delta V_\theta}, \widetilde{\delta V_\phi}$)
on the velocities.
Each symbol represents the median radius of the stars within our selected radial bins. 
The error estimates for each star are propagated from the
distance errors, line-of-sight velocity errors, two {\it Gaia} DR2
proper motion error estimates, and {\it Gaia} DR2 proper motion covariance. 
Error bars on the dispersions and $\langle V_\phi\rangle$ are estimated from the Poissonian sampling in each bin; in a large majority of bins the error bars smaller than the symbol size.
The first bin contains stars within $r_\mathrm{gc}=2-7$ kpc. 
The remaining bins for K giants end 
at $r_\mathrm{gc}= 8,$ 10, 13, 17, 22, 29, 38, 50, 110
kpc and for BHB stars at $r_\mathrm{gc}= 8,$ 10, 12, 14, 17, 20, 24, 29, 34, 41, 80
kpc.
}
  \label{fig:profs}
\end{figure*}

We bin the stars in Galactocentric radial bins for our total halo sample (lines) and our smooth, diffuse halo sample (markers) in Figure \ref{fig:profs} for LAMOST/SDSS K giants
(left panel) and for SDSS BHB stars (right panel).
Although the bulk of the stars are within 30 kpc of
the Galactic Center, we have a sufficient number of stars to probe the kinematics 
past 100 kpc. 

The solid lines and filled symbols in Figure \ref{fig:profs} show the
velocity dispersions $(\sigma_r, \sigma_\theta, \sigma_\phi)$ and mean rotational velocity $\langle V_\phi\rangle$ as
functions of Galactocentric radius $r_\mathrm{gc}$. The most obvious influence of substructure is due to Sagittarius at distances $>20$ kpc. The polar dispersion component $\sigma_\theta$ decreases and mean rotational velocity component $\langle V_\phi\rangle$ decreases.
Both before and after removing substructure, both tangential dispersions 
are substantially less than the radial velocity dispersion $\sigma_r$
at all radii out to almost 100 kpc. The orbits are clearly 
radial throughout the entire halo. We do not discuss in detail the $\langle V_\phi\rangle$ profile, but we note that $\langle V_\phi\rangle$ is small along all Galactocentric radii and is of comparable amplitude to previous studies 
using various halo samples with various distances and methods
\citep[e.g.,][]{Morrison1990,Carney1996,Chiba2000,Smith2009,Hattori2013,Deason2017spin,Kafle2017googly,Belokurov2018,Cunningham2019b,Lancaster2019,Tian2019,Tian2020}

\begin{figure*}
  \begin{tabular}{cc}
    \includegraphics[width=\columnwidth]{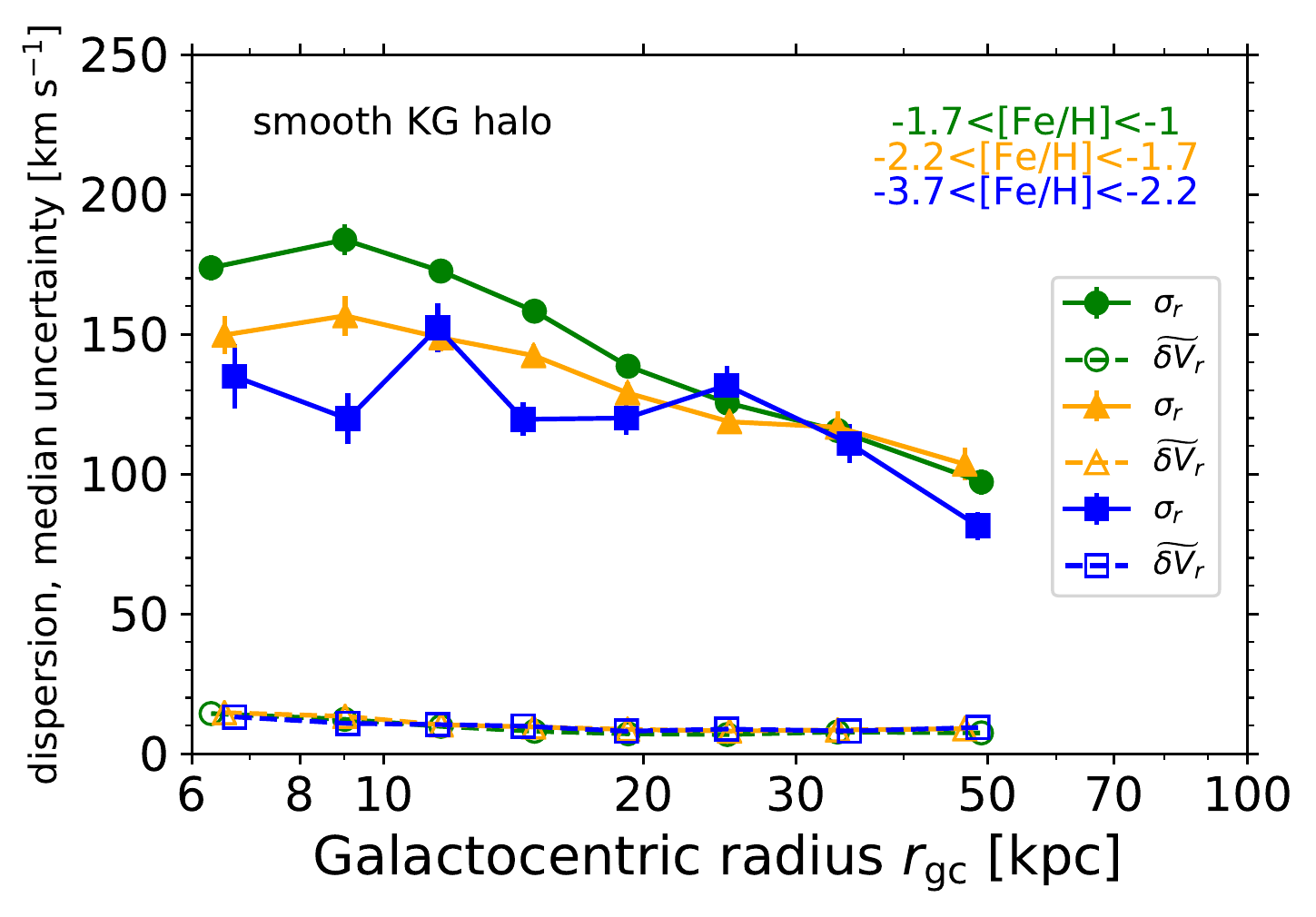}&
    \includegraphics[width=\columnwidth]{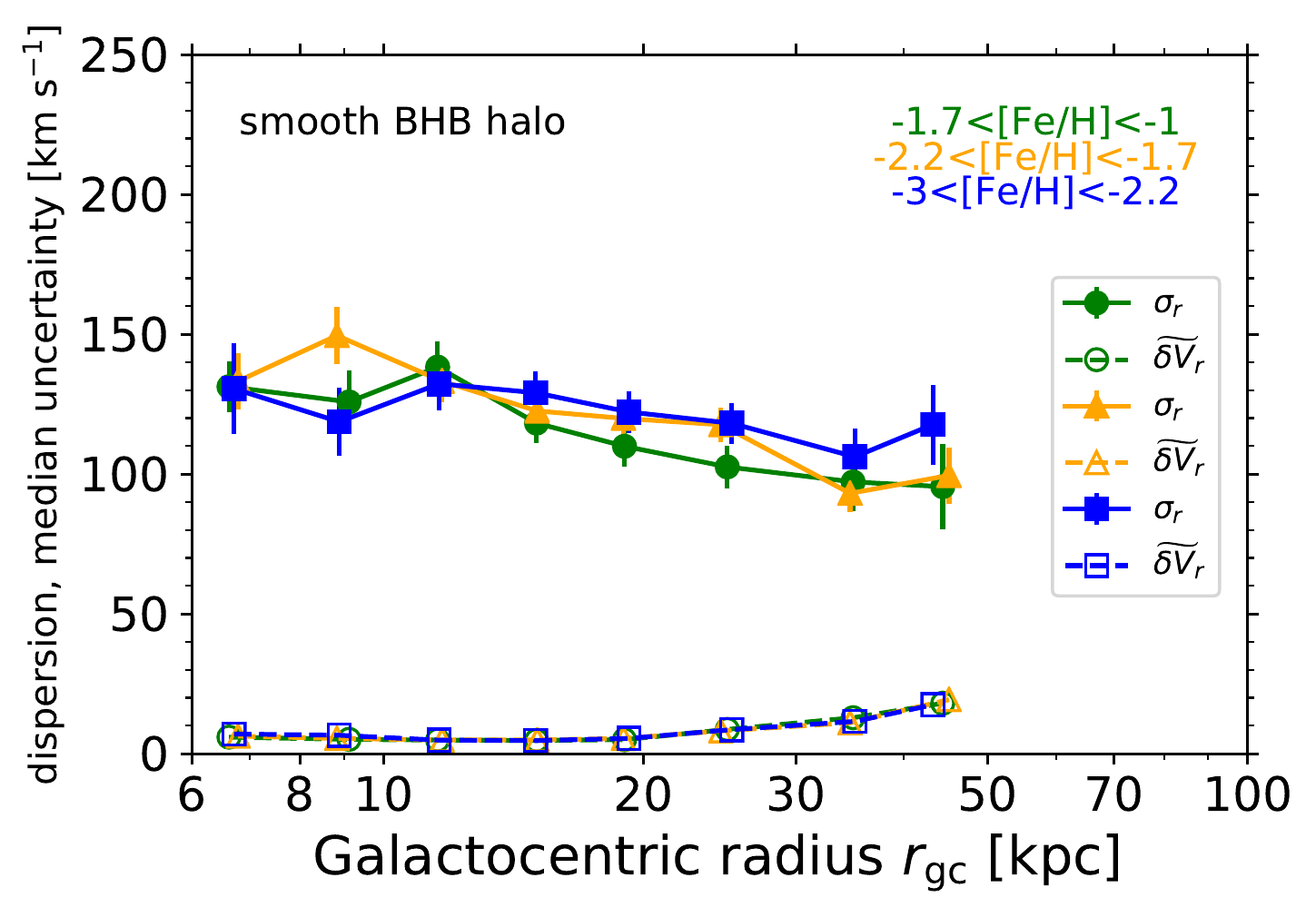}\\
    \includegraphics[width=\columnwidth]{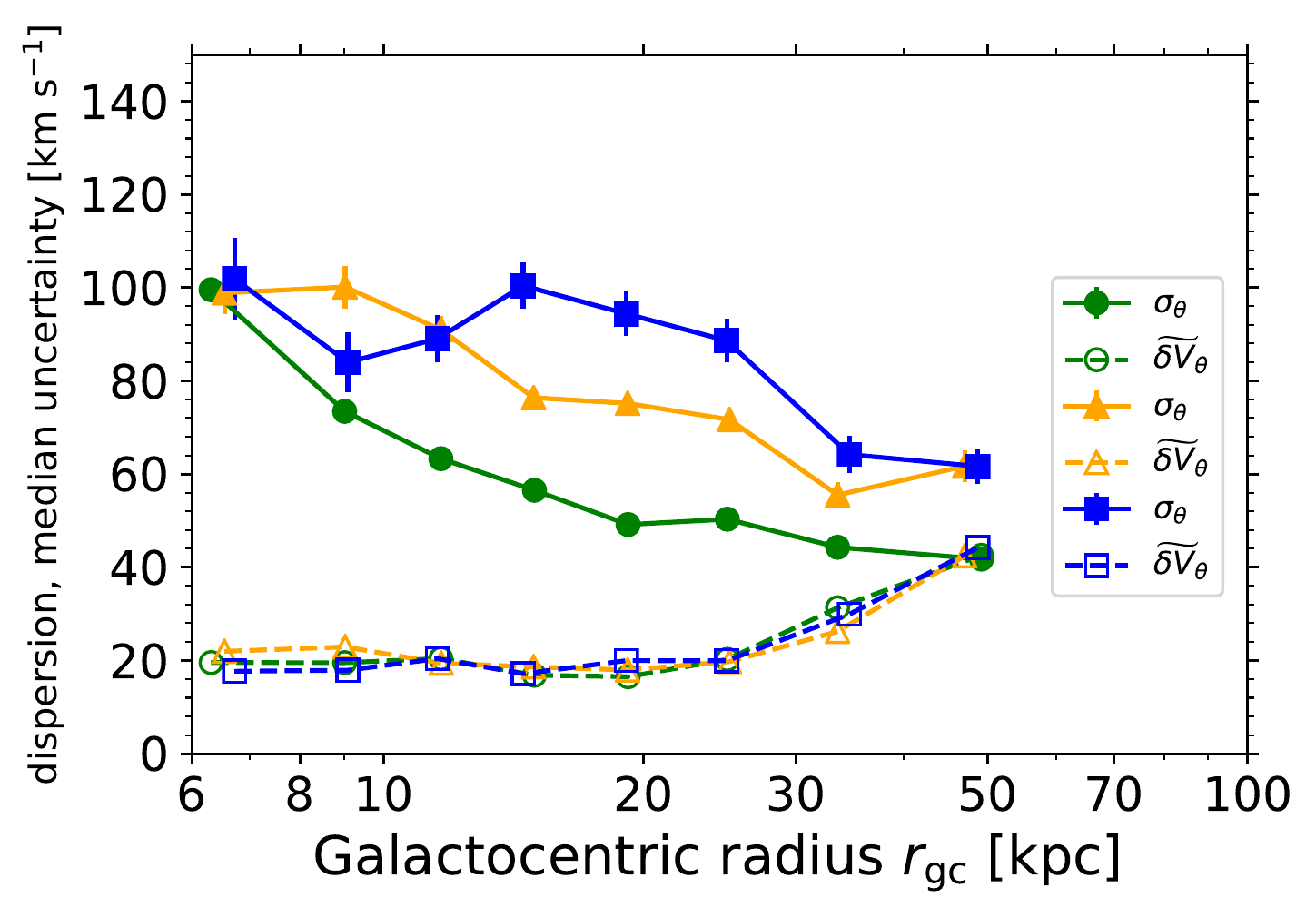}&
    \includegraphics[width=\columnwidth]{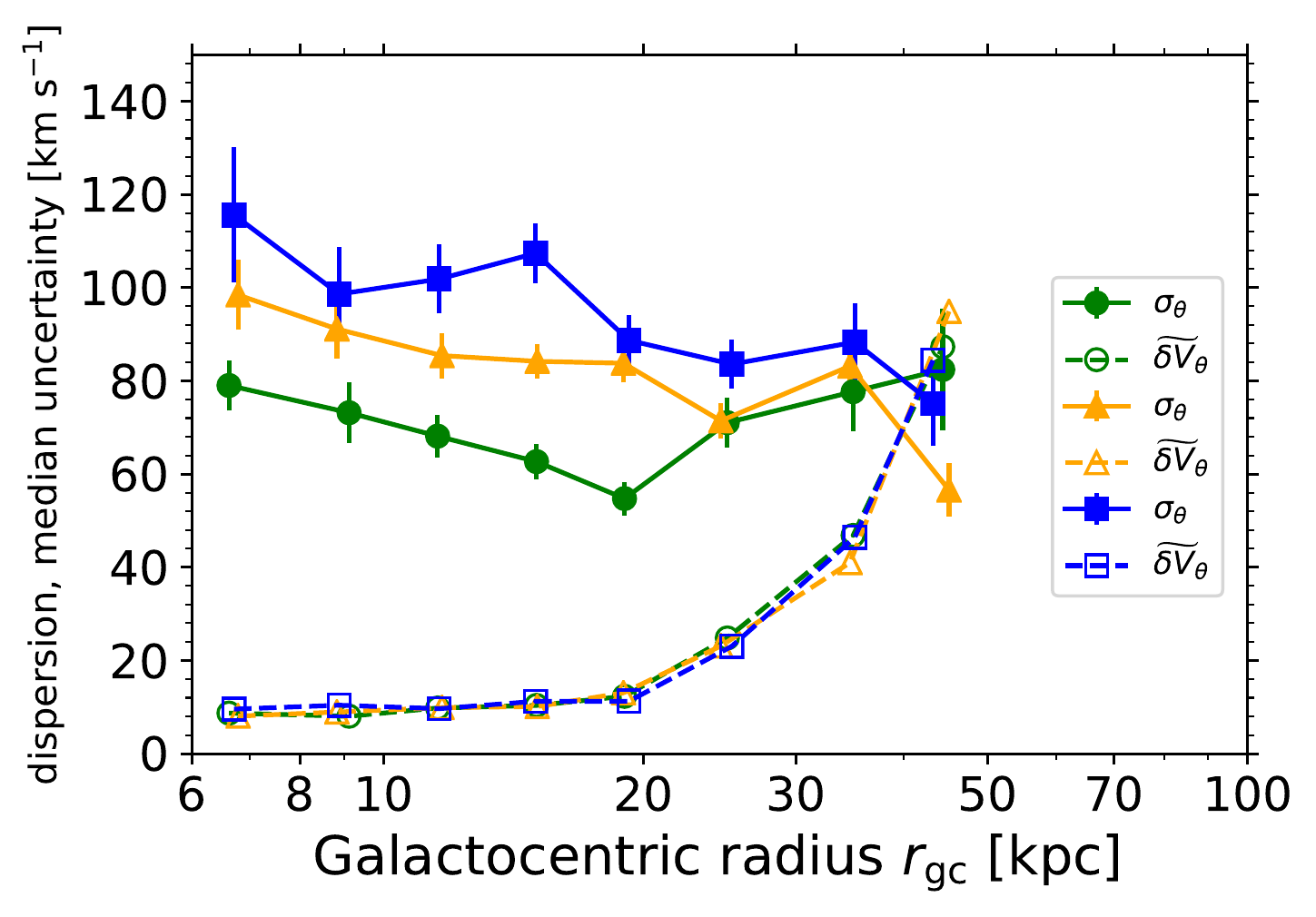}\\
    \includegraphics[width=\columnwidth]{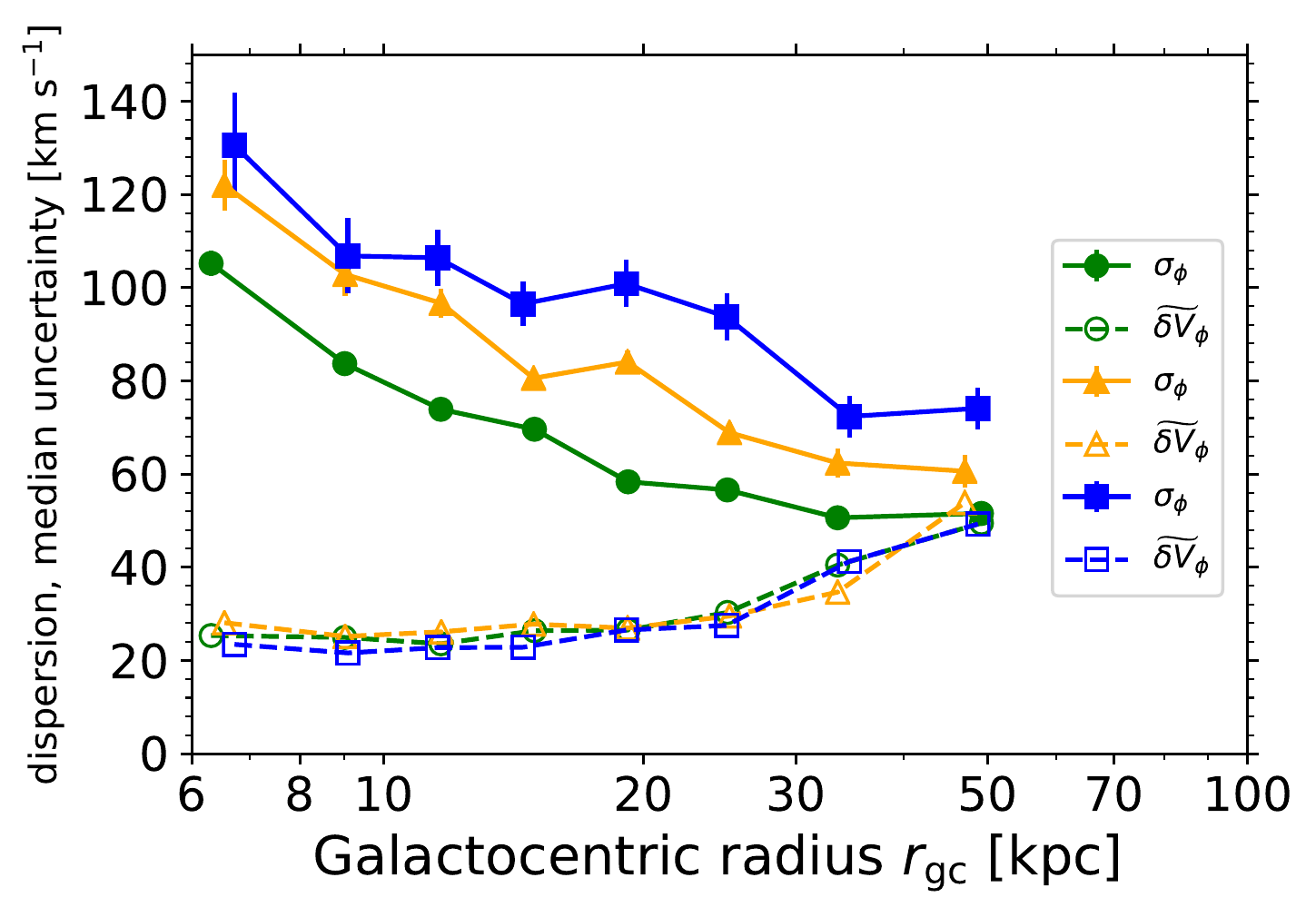}&
    \includegraphics[width=\columnwidth]{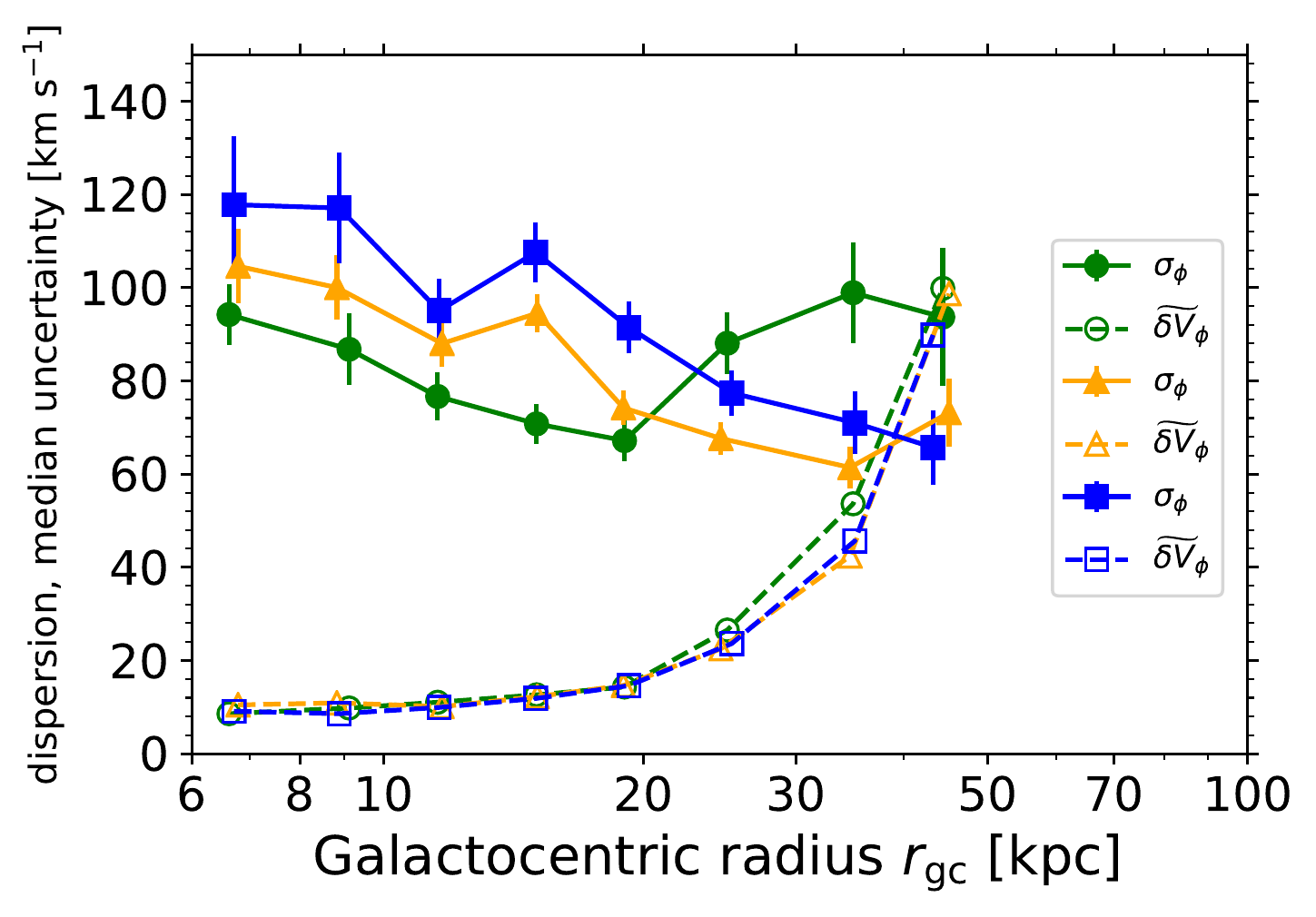}
\end{tabular}
  \caption{Metallicity dependence of velocity dispersion (filled markers connected with solid lines) for the smooth, diffuse halo LAMOST/SDSS K giant and SDSS BHB stars (left and right columns, respectively). Colors denote the metallicity range as labeled in the upper right corner of the top left and right panels.
The corresponding median velocity uncertainties within metallicity bins for the smooth, diffuse halo LAMOST/SDSS K giant and SDSS BHB stars are shown with open markers connected with dotted lines.
Upper, middle, and lower rows correspond to the dispersion and median uncertainties for $V_r,V_\theta,$ and $V_\phi$, respectively.
The same binning scheme and metallicity range is used as well in Figure \ref{betafeh} for the anisotropy $\beta$ profiles.
}
  \label{fig:sigfeh}
\end{figure*}

In Figure \ref{fig:sigfeh}, we explore the kinematic statistics and uncertainties in different metallicity bins for our smooth, diffuse halo samples. We have already seen trends between kinematics and metallicity in Figures \ref{fig:fehvel-kg}$-$\ref{fig:Vr-Vphi-bhb} and expect to also find trends in the kinematic statistics. For each metallicity range and star sample, we use the larger binning scheme in $r_\mathrm{gc}$ as used for the K giants previously mentioned. This ensures each bin has over several tens to hundreds of stars per bin.
The most apparent trend, as seen in Figure \ref{fig:sigfeh}, is the smaller spread in tangential velocities (lower two rows, filled markers and solid lines) of more metal rich halo stars compared to the larger spread in tangential velocities for more metal poor stars, the difference can be tens of km s$^{-1}$.
Looking at the profiles along $r_\mathrm{gc}$ in Figure \ref{fig:sigfeh}, we find that the trend extends along distance from the Galactic Center and survives out to great distance. The trend is largest in the tangential components ($\sigma_\theta,\sigma_\phi$) but still is seen in $\sigma_r$. In all metallicity bins the dispersions tend to decrease with increasing $r_\mathrm{gc}$.
The velocity uncertainties divided within different metallicity bins remain comparable (Figure \ref{fig:sigfeh}, open markers and dashed lines). The fact that the velocity uncertainties change little with metallicity and that the systematic trend of velocity with metallicity is seen in both K-giant and BHB samples gives confidence that dependency of the velocity dispersion with metallicity is a real feature.

\section{Results}
\label{sec:results}

\subsection{Anisotropy profile of the Smooth, Diffuse Milky Way Halo}

\begin{figure}[htb]
\includegraphics[width=\columnwidth]{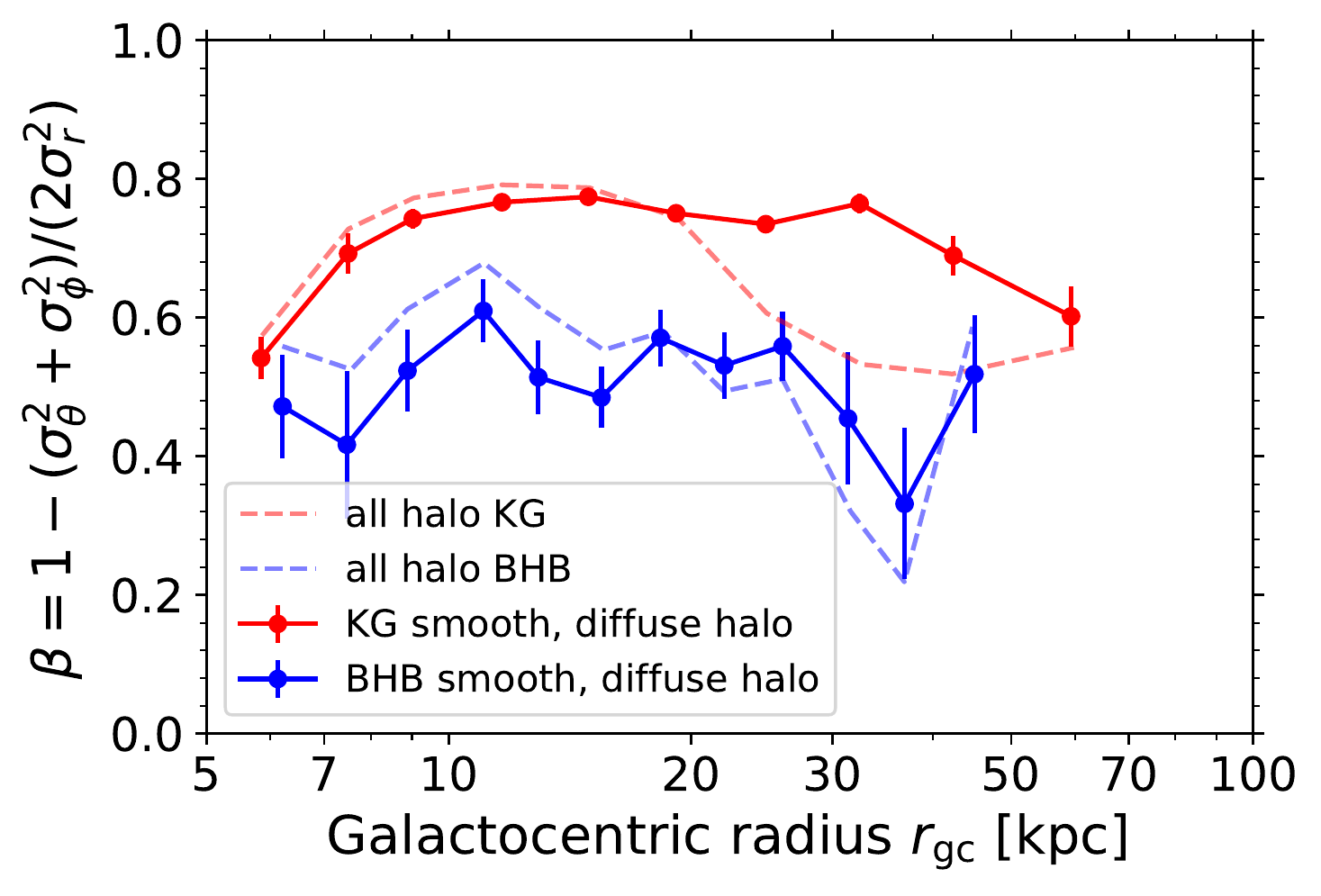}
\caption{
Velocity anisotropy profiles for the smooth, diffuse halo star samples of LAMOST/SDSS K giants and SDSS BHB stars (markers connected by solid lines) and for the total halo star samples (dotted lines).
The error bars on
    $\beta$ are propagated through from the errors in measuring
    $\sigma_r, \sigma_\theta$ and $\sigma_\phi$ (cf. Figure \ref{fig:profs}). The larger error bars for BHB stars reflect the smaller number of stars per bin as compared to K giants; although, we select a larger number of radial bins to reflect the higher relative distance accuracy of the BHB stars.
}
\label{fig:rgc-beta-multi}
\end{figure}

Our main results of the anisotropy profiles of the underlying smooth, diffuse halo star samples of LAMOST/SDSS K giants and SDSS BHB stars are shown in Figure \ref{fig:rgc-beta-multi} using the red and blue markers, respectively.
Our derived anisotropy profiles highlight the highly radial orbits seen for much of the smooth, diffuse halo.
The K giant halo orbits are strongly radial, with $\beta$
lying mainly in the range 0.6 to 0.9. The BHB halo orbits are also highly radial, but slightly less (by $\Delta\beta\sim0.1-0.3$) than the K giants. We show later that this is likely due to the different metallicity distributions of the two samples (c.f. Figure \ref{lamostsegue} middle panel upper row).
We show in Section \ref{betametals} that the seemingly discrepant anisotropy is alleviated when stars of similar metallicities are compared.
It is remarkable how nearly constant
$\beta$ is with Galactocentric radius $r_\mathrm{gc}$, out to about 40
kpc, despite the considerable change in the underlying $\sigma_r$ and
the two tangential velocity dispersions as functions of
$r_\mathrm{gc}$.
The lack of dips in the smooth, diffuse anisotropy profile reflects the characteristic of a well-mixed, virialized stellar halo.
Simulated stellar halos typically show a steady or very slowly rising $\beta$ profile for $r_\mathrm{gc}>5$ kpc 
\citep[e.g.,][]{Diemand2005,Abadi2006,Sales2007.379.1464,Kafle2012,Rashkov2013}, which is consistent with what we see for the smooth, diffuse halo in Figure \ref{fig:rgc-beta-multi} (red and blue markers) within 40 kpc. \cite{Loebman2018} have analyzed
the simulated stellar halos of \cite{Bullock2005},
\cite{Christensen2012}, and \cite{Stinson2013}, finding that, except for several cases involving major mergers, 
$\beta$ remains high (in the range 0.6 to 0.8) out to the
limits of their analysis (70 kpc). 

\subsection{Effect of Substructure on Anisotropy}

In Figure \ref{fig:rgc-beta-multi}, we also make a comparison of the anisotropy profile before and after removing substructure. 
The $\beta(r)$ profile with substructure still in the sample is shown
by the red and blue dotted lines, and the profile with substructure removed is
shown by the red and blue dots. 

Removing substructure has two main influences on the anisotropy profile for both K giants and BHB stars. The first is seen within 20 kpc from the Galactic Center, where removing substructure causes $\beta$ to become slightly less radial. This is likely due to removing obvious parts of the $Gaia$-Sausage ($Gaia$-Enceladus, Kraken), the ancient past merger with the Milky Way which has been made more clearly evident through analysis of the $Gaia$ data releases. 

The second is seen at $r_\mathrm{gc}>25$ kpc where removing substructure increases $\beta$. This is in line with expectation, as Sagittarius is on a high
latitude, tangential orbit, and biases $\beta$ to appear more
tangential (lower values of $\beta$). 

\subsection{Metallicity Dependence of Anisotropy}\label{betametals}

\begin{figure}[htb]
\begin{tabular}{cc}
\includegraphics[width=\columnwidth]{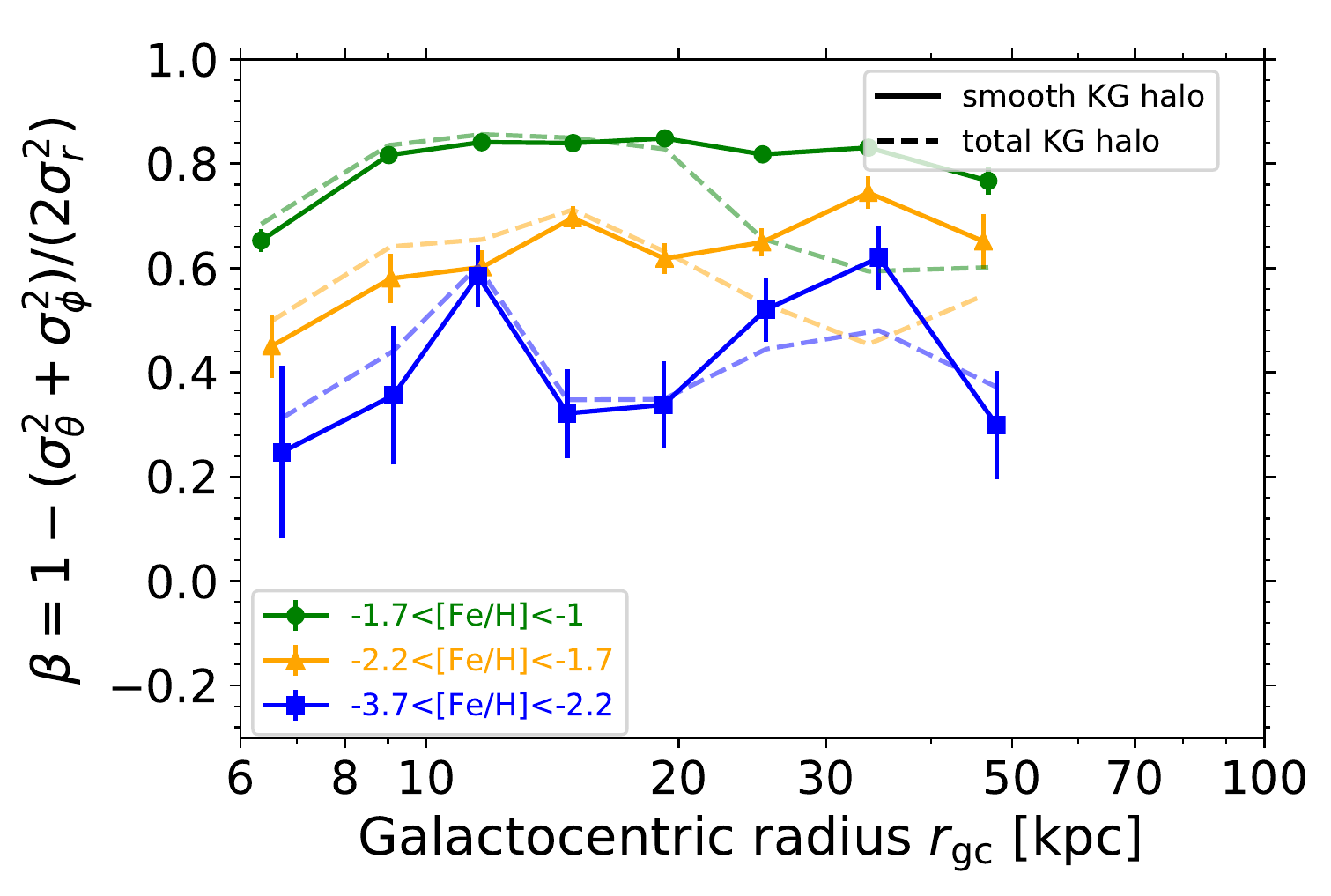}\\
\includegraphics[width=\columnwidth]{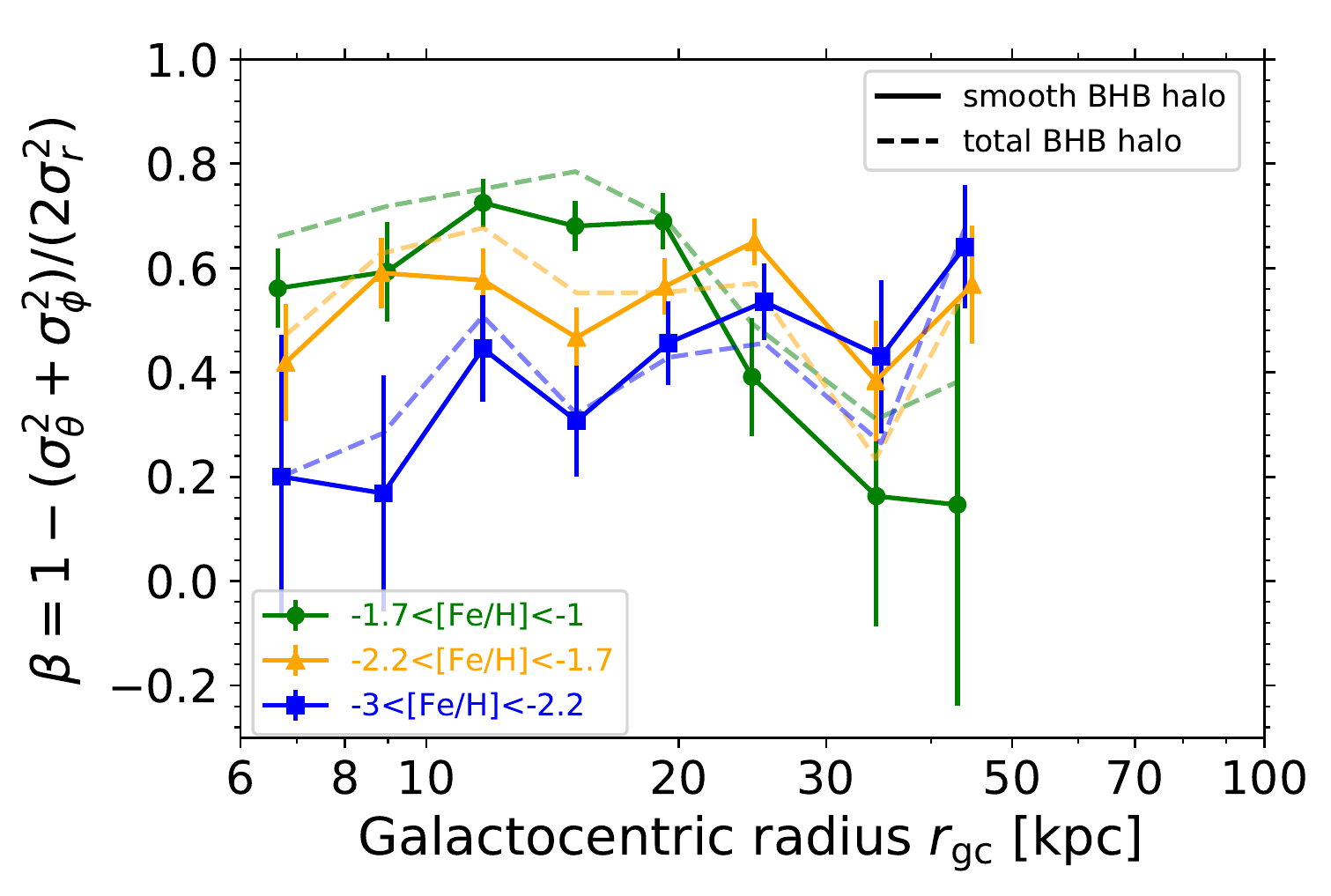}
\end{tabular}
  \caption{Metallicity dependence of anisotropy for our total sample (dotted lines) and for our smooth, diffuse halo sample (markers connected by solid lines) for LAMOST/SDSS K giants (upper panels) and SDSS BHB stars (lower panels). The metallicity dependence of anisotropy is seen regardless of stellar type or substructure removal.
}
  \label{betafeh}
\end{figure}

We now examine the behavior of $\beta(r)$ as a function of the
metallicity of the stars. 
We plot $\beta$ (Figure \ref{betafeh}) 
versus $r_\mathrm{gc}$ for three metallicity subsets. 
We investigate the metallicity dependency both in the full halo samples (dotted lines) and in the smooth, diffuse halo samples after removing substructure (filled markers and solid lines). 
After removing substructure,
selecting stars over the metallicity range $-3.6 < $ [Fe/H] $ < -1.0$,
and dividing them into three metallicity bins, we have previously shown the velocity dispersion profiles ($\sigma_r, \sigma_\theta, \sigma_\phi$) and their corresponding median velocity uncertainties (Figure \ref{fig:sigfeh}).
We note that our three halo star samples have differing limiting metallicity ranges at the metal poor end as seen in the upper row, middle panel of Figure \ref{lamostsegue}. LAMOST K giants have a hard lower limit at [Fe/H]$=-2.5$ due to the stellar parameter pipeline used \citep{Wu2011,Wu2014}. SDSS BHB stars reach to nearly [Fe/H]$=-3$ and SDSS K giants to nearly $-3.6$. 
As mentioned in Section \ref{sec:sub-rem}, we selected metallicity bins based on the metallicity distribution of our samples (see Figure \ref{lamostsegue} middle panel in the upper row) in order to ensure a sufficient number of stars, as well as on recent works finding evidence a dependency between stellar halo kinematics and metallicity \citep{Myeong2018action,Deason2018,Belokurov2018,Lancaster2019} in order to facilitate comparison which we later expound upon in Section \ref{sec:conclusion}.

The general trend, independent of substructure or stellar type, is that the relatively constant $\beta$ profile systematically decreases with metallicity. The velocity uncertainties divided within different metallicity bins remain comparable (Figure \ref{fig:sigfeh}, open markers and dashed lines). The fact that the velocity uncertainties change little with metallicity and that the systematic trend of velocity with metallicity is seen in both K-giant and BHB samples gives confidence that this is a real feature. The kinematic and chemical trend we find is in agreement with \citet{Myeong2018action,Deason2018,Belokurov2018,Lancaster2019} who also find that the more metal rich halo stars are dominated by highly radial orbits, compared to the less metal rich halo stars which are much less radially dominated to near isotropic.

Considering the smooth, diffuse halo samples (Figure \ref{betafeh}, filled markers and solid lines), the most metal poor bin ($-3.6 <$ [Fe/H]
$< -2.2$) for both K giants and BHB stars show $\beta\sim0.4$ and with a large scatter at different radius and between samples of $\pm0.2$ to $\pm0.3$. The K giants show the largest $\beta$ values of up to 
0.9 for [Fe/H]$>-1.7$. The same highest metallicity bin for BHB stars shows values of $\beta\sim0.8$,  slightly lower compared to the same metallicity range for the K giants.

Interestingly, excluding the more metal rich halo stars with [Fe/H] $>-1.7$, both the smooth, diffuse halo K giants and BHB stars (Figure \ref{betafeh}, filled markers and solid lines) have similarly valued constant anisotropy profiles to the furthest extent of our sample ($\sim100$ kpc and $\sim60$ kpc, respectively); albeit the anisotropy for the two metallicity bins which we investigate ($-2.2<$ [Fe/H] $<-1.7$ and [Fe/H] $<-2.2$) is systematically shifted by $\Delta\beta\sim0.1-0.3$. For the metallicity range $-2.2<$ [Fe/H] $<-1.7$ the samples show nearly constant $\beta$ of $0.7-0.5$; for [Fe/H] $<-2.2$ the samples show $\beta\sim0.2-0.6$, nearly constant within the uncertainties. The more metal rich samples with [Fe/H] $>-1.7$ show a declining trend with distance in anisotropy which becomes apparent $\sim20$ kpc, a slow decrease for K giants totaling $\Delta\beta\sim0.3$ at the extent of our sample although quite drastically for BHB stars (see our note concerning increasing uncertainties concluding this subsection) with a drop $\Delta\beta\sim0.8$!

The full sample of K giants and BHB stars show similar dependency on metallicity, except the most apparent difference is the influence of the substructure at distances $r_\mathrm{gc}>20$ kpc which is less radial at all metallicities compared to the smooth, diffuse halo. Removing the substructure extends the constant anisotropy profile along $r_\mathrm{gc}$. 
An exception to the general trend is the anisotropy of the last bin for the full BHB sample, which is more radial compared to the smooth halo sample for the highest and lowest metallicity bins considered.

Beyond 40 kpc, 
we note that the sample size decreases and uncertainties in the distance and velocities increase, all of which
influence $\beta$ and any trends should be treated with caution.

\subsection{Uncertainties}

Three factors play large roles in influencing the anisotropy profile presented in this work.

The first is the stream removal method. We find the $\beta$ profile is constant out to larger Galactocentric distances before gently falling compared to the profile of \citet{Bird2019beta}. The $E-L$ stream removal employed by \citet{Bird2019beta} was not intended to be a complete stream removal method, but rather a demonstration showing how removing a large portion of the Sagittarius Stream changes the anisotropy profile. The currently presented anisotropy profile uses integrals of motion to remove all obvious substructure. Binning the data in similar metallicity bins enhances the flatness of $\beta$ along $r_\mathrm{gc}$.

Even before stream removal, the anisotropy in the outer-most radial bins presented in the current work differ by $\sim0.1-0.2$ to those of \citet{Bird2019beta}. The reason for the difference is likely due to the slightly different selection criteria based on the \citet[in prep.]{Xue2019} method. Those stars with $E>0$ and semi-major axis $a>300$ km s$^{-1}$ are removed from the current data sample, whereas in \citet{Bird2019beta}, these stars remained in the sample. These stars have higher tangential velocity values and thus tend to decrease anisotropy.

Secondly, as analyzed and discussed in \citet{Hattori2017} (as well as the appendices of \citet{Loebman2018} and \citet{Cunningham2019b}), influencing the anisotropy profile is the lack of stars in distant bins. The mocks shown in the Appendix of this paper investigate the amount of uncertainty which is added to our analysis due solely to Poisson uncertainties as well as Poisson uncertainties combined with measurement uncertainties estimated from the real data. We recover the underlying profile well, but the uncertainty of $\beta$ in each bin indeed increases with decreasing stars per bin.

Thirdly, \citet{Lancaster2019} and \citet{Cunningham2019b} point out that the gentle decrease in \citet{Bird2019beta} anisotropy profile for distances past $\sim30$ kpc may be enhanced due the increasing uncertainties in distances and proper motions. The mocks in the current work agree that the increasing measurement uncertainties cause the $\beta$ profile to gently drop by $\Delta\beta<0.1$. We also find, as mentioned first among the three uncertainties here discussed, that the stream removal method also influences the amount $\beta$ decreases at large radii by $\Delta\beta\sim0.1-0.2$.  

\section{Discussion and Conclusions} 
\label{sec:conclusion}

\subsection{Comparison with \citet{Bird2019beta}}
\label{sec:beta1beta2}

\begin{figure*}[htb]
\includegraphics[width=\columnwidth]{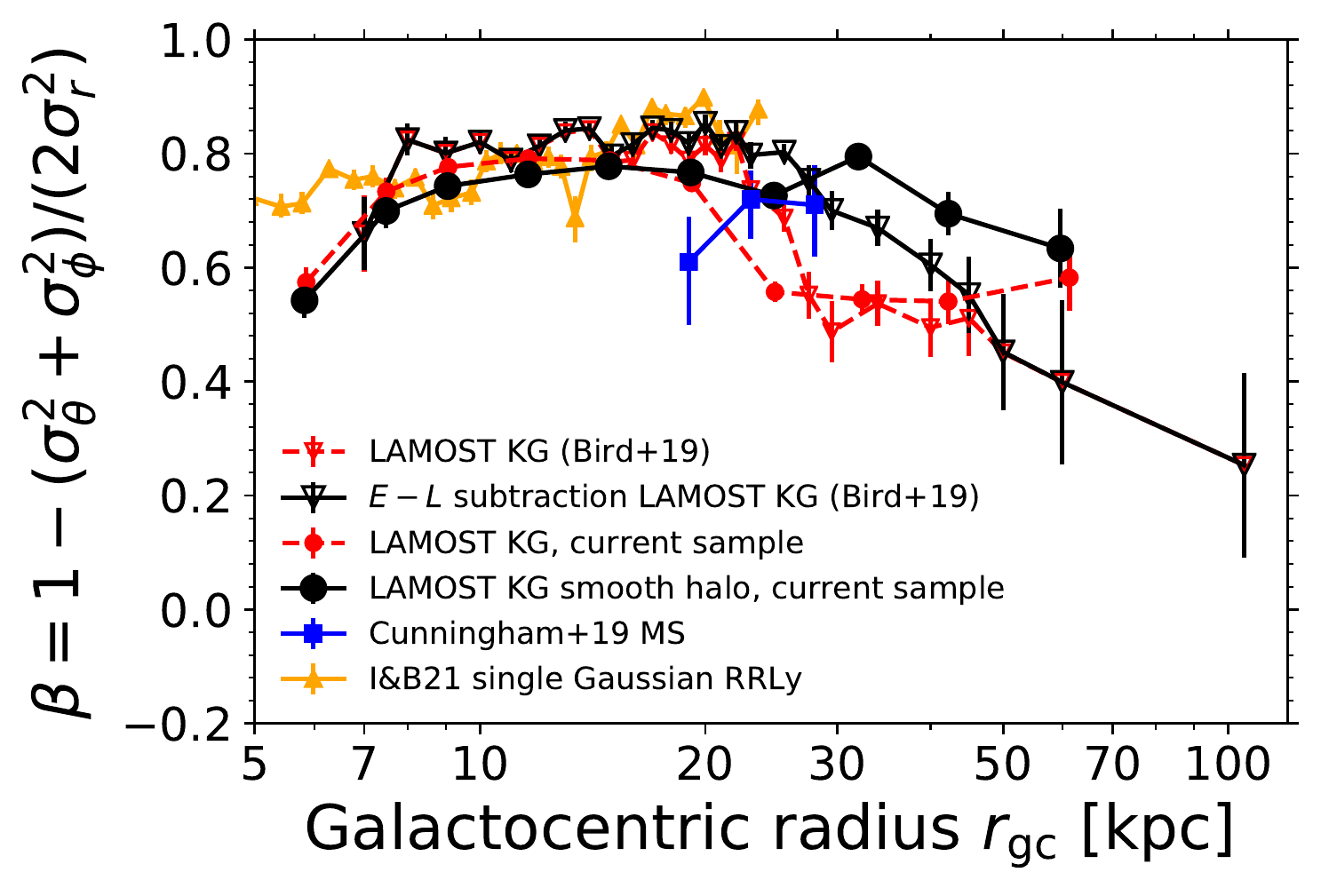}
    \includegraphics[width=\columnwidth]{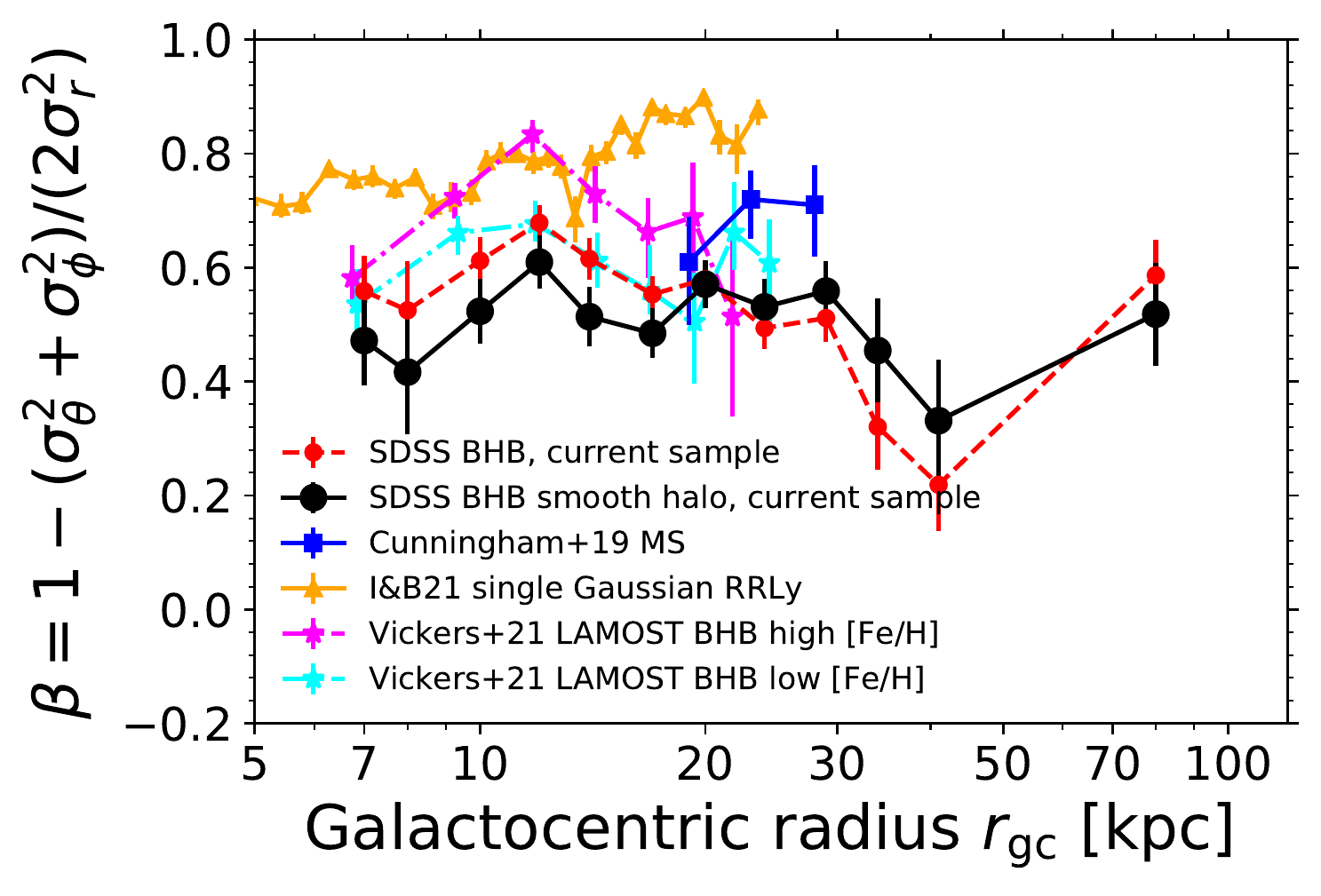}
\caption{{\bf Left panel}: LAMOST K giant velocity anisotropy profile comparison of \citet[black and red open triangles]{Bird2019beta} and this current work (black and red closed circles). We plot the total LAMOST K giant samples of both studies in red. After employing the \citet{Bird2019beta} $E-L$ Sagittarius removal method and the current complete integrals-of-motion substructure removal method \citet[in preparation]{Xue2019}, we plot the resulting anisotropy profiles in black open triangles and black closed circles, respectively. 
{\bf Right panel}:
Anisotropy parameter $\beta$ as a function of Galactocentric radius $r_\mathrm{gc}$ for our BHB sample.
 As in the left panel, the black circles are the smooth, diffuse halo star sample after substructure was flagged and removed using the method of \citet[in preparation]{Xue2019}, the red circles are the entire BHB halo star sample. Magenta and cyan star-shaped markers show the anisotropy profiles of halo BHB stars selected from LAMOST by \citet{Vickers2021} divided into more metal rich and more metal poor subsets, respectively (the poor [Fe/H] $r_\mathrm{gc}$-axis coordinates have been shifted sightly to avoid overlapping with the rich [Fe/H] sample).
{\bf Both panels}: Blue squares are the spherically averaged $\beta$ estimates of main sequence (MS) stars from \citet{Cunningham2019b}. Orange triangles are the spherically averaged $\beta$ estimates of RR Lyrae from \citet{Iorio2021} (I\&B21, RRLy) using a single Gaussian component (we do not plot the RR Lyrae profile for $r_\mathrm{gc}<5$ kpc, the $r_\mathrm{gc}$-axis limit to our plot, nor do we plot the points for $r_\mathrm{gc}>25$ kpc as these fall below the $\beta$-axis of our plot). The RR Lyrae anisotropy profile of \citet{Iorio2021} is in especially close agreement with the K giants for $r_\mathrm{gc}\sim 8-20$ kpc.
The anisotropy of the main sequence stars of \citet{Cunningham2019b} fall in between our results for K giants and BHB stars and is consistent with our results within the uncertainties. The LAMOST BHB samples of \citet{Vickers2021} are in good agreement with our full sample of SDSS BHB halo stars and show similar dependency on metallicity as we show in Figure \ref{betafeh}.
}
\label{fig:rgc-beta2020_compare}
\end{figure*}

Both \citet{Bird2019beta} and the current study use LAMOST halo K giants and find the same main results, first of a quite constant $\beta$ profile, second, which retains constant values to larger distances after substructure is removed, and third of a systematic decrease in the constant $\beta$ profile with decreasing metallicity. We compare the anisotropy profiles in the top panel of Figure \ref{fig:rgc-beta2020_compare}.

In \citet{Bird2019beta} we very conservatively cut the LAMOST K-giant sample to metallicities
[Fe/H] $< -1.3$ in order to probe the halo only.
This limit was based on our estimated metallicity errors of 0.1 dex and an
examination of a fuller data set of
LAMOST K giants (for metallicities up to Solar ([Fe/H] $ = 0$) 
the disk-to-halo transition occurs at [Fe/H] $ = -1$ for the
stars analyzed (for which $|Z|>5$ kpc).
The [Fe/H]$<-1$ K-giant velocities are
clearly those of a nearly non-rotating, pressure supported population,
with no sign of a much faster rotating thick disk (or disk).
In our current study, we aim to add to our sample more nearby halo K giants and BHB stars, $2<|Z|<5$ kpc. By selecting only stars with [Fe/H]$<-1$
we ensure that the added stars belong to the metal poor halo.

Kinematic substructures appear prominently in $E$-$L$ space, due to common
    energy and angular momentum.
A large bulk of Sagittarius stars are found at
$E = -90000$ (km s$^{-1}$)$^2$ and $L = 5000$ kpc km s$^{-1}$. 
In \citet{Bird2019beta}, we
selected a large portion of Sagittarius using the criteria $E < -80000$ (km s$^{-1}$)$^2$ and $4000 < L < 6000$ kpc km s$^{-1}$ \citep[red markers in Figure 6]{Bird2019beta}, and find that the removal of these stars increases the value of $\beta$ and past 20 kpc the $\beta(r)$ profile becomes more radial (Figure \ref{fig:rgc-beta2020_compare} top panel and \citet{Bird2019beta} Figure 7).
In this analysis, we seek to remove as much of the remaining substructure as possible. To do this we use the method of \citet[in preparation]{Xue2019} which selects stars sharing common integrals of motion. This method effectively removes all obvious substructure in $E-L$ space. This additional effort to remove all obvious substructure enhances the constant $\beta$ profile with radius, removing more radially dominated substructure within 10 kpc and more tangentially dominated substructure past 20 kpc. BHB stars noticeably have less tangentially dominated substructure to be removed.

The results of these two works are complementary and the similarities between both K giants and BHB stars strengthens confidence in the results.

\subsection{Comparison with \citet{Cunningham2019b} and \citet{Vickers2021}}

HALO7D \citep{Cunningham2019a,Cunningham2019b} is a project which allows the measurement of $\beta$ in $HST$ CANDELS fields using main-sequence turnoff stars. Line-of-sight velocities are measured with Keck II / DEIMOS spectroscopy and proper motions are measured using multi-epoch fields from $HST$. The first results show $\beta$ varies from one field to another, but the general trend (measured over similar magnitude bins equivalent to distances of $r_\mathrm{gc}\sim20-30$ kpc) show radial anisotropy increasing with distance. Their spherically averaged anisotropy estimates show $\beta\sim0.6-0.7$ which is in excellent agreement to our estimates with K giants and BHB stars. We compare these results in Figure \ref{fig:rgc-beta2020_compare}. As \citet{Cunningham2019b} do not remove substructure from their sample, we make the comparison using our samples both with and without substructure removed. We overplot the spherically averaged, binned in magnitude, main sequence star anisotropy (blue squares) of \citet{Cunningham2019b} over our anisotropy measurements from the sample K giants (upper panel) and BHB stars (lower panel), both for our samples with substructure (red circles) and with substructure removed (black circles). The main sequence star anisotropy profile lies between our K giant and BHB star anisotropy profiles, regardless of removing substructure or not. This shows the dependency of the anisotropy profile with stellar population, which may be a proxy of many different factors, e.g., average metallicity, age, or properties of the satellite in which they formed (infall time, orbital characteristics, etc.).

Considering the field-to-field anisotropy variation found by \citet{Cunningham2019b}, their most extreme value for one of their individual fields, slightly below $\beta=0.25$, is comparable to, although slightly lower than, the most metal poor halo stars in our K giant and BHB samples, [Fe/H] $<-2.2$. Their field may primarily be dominated by stars with such low metallicity and thus similarly lower $\beta$ values, or, alternatively, as they postulate, may be dominated by substructure or stars kicked-up from the disk, which would have their own unique metallicity perhaps different from the smooth, diffuse halo stars. 

A detailed comparison with this work will more easily be made once their metallicities are analyzed and substructure is removed with similar methods.
Until then, the the samples are in agreement within the scatter (due to number statistics, differing average metallicities, and substructure removal). We do not explore within these data sets localized $\beta$ structure such as the fields of HALO7D. We leave this for future work. The $\beta$ profile is used in the Jeans equation to estimate the mass of the Galaxy, although the impact on the total mass estimate due to localized changes in $\beta$ will need further exploration.

\citet{Vickers2021} use machine learning to extract from LAMOST one of the largest samples of BHB stars to date. Their final sample of 13,693 stars is expected to be 86\% pure and includes metallicity and absolute magnitude estimates for each star. 
From this sample they remove duplicate stars and those with $|Z|<3$ kpc to select halo stars and find 2,692 high confidence BHB stars with well-measured kinematics.
We plot their measured anisotropies in the right panel of Figure \ref{fig:rgc-beta2020_compare} along Galactocentric distance $r_\mathrm{gc}$, where they divide their sample into a lower and higher bin of metallicity (cyan and magenta star-shaped markers, respectively).
The anisotropy profiles of BHB stars from LAMOST \citep{Vickers2021} and from our SDSS sample (before stream removal, red circles) are in agreement within the uncertainties. \citet{Vickers2021} also find a similar chemodynamical trend as seen in Figure \ref{betafeh} such that their more metal rich sample has higher anisotropy compared to their lower metallicity sample.

\subsection{Comparison with \citet{Lancaster2019}}\label{sec:Lancaster}

\begin{figure*}
    \includegraphics[width=\columnwidth]{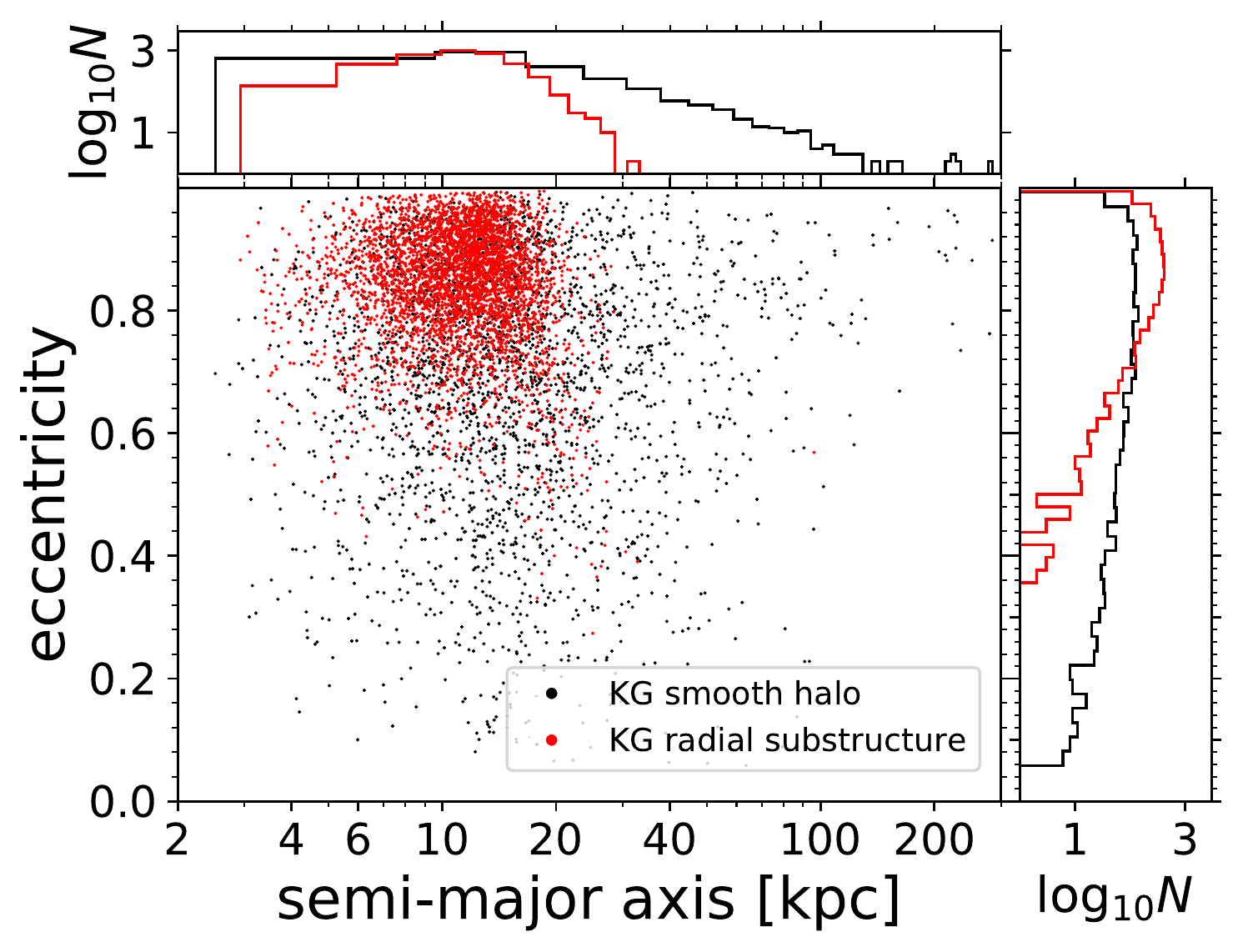}
    \includegraphics[width=\columnwidth]{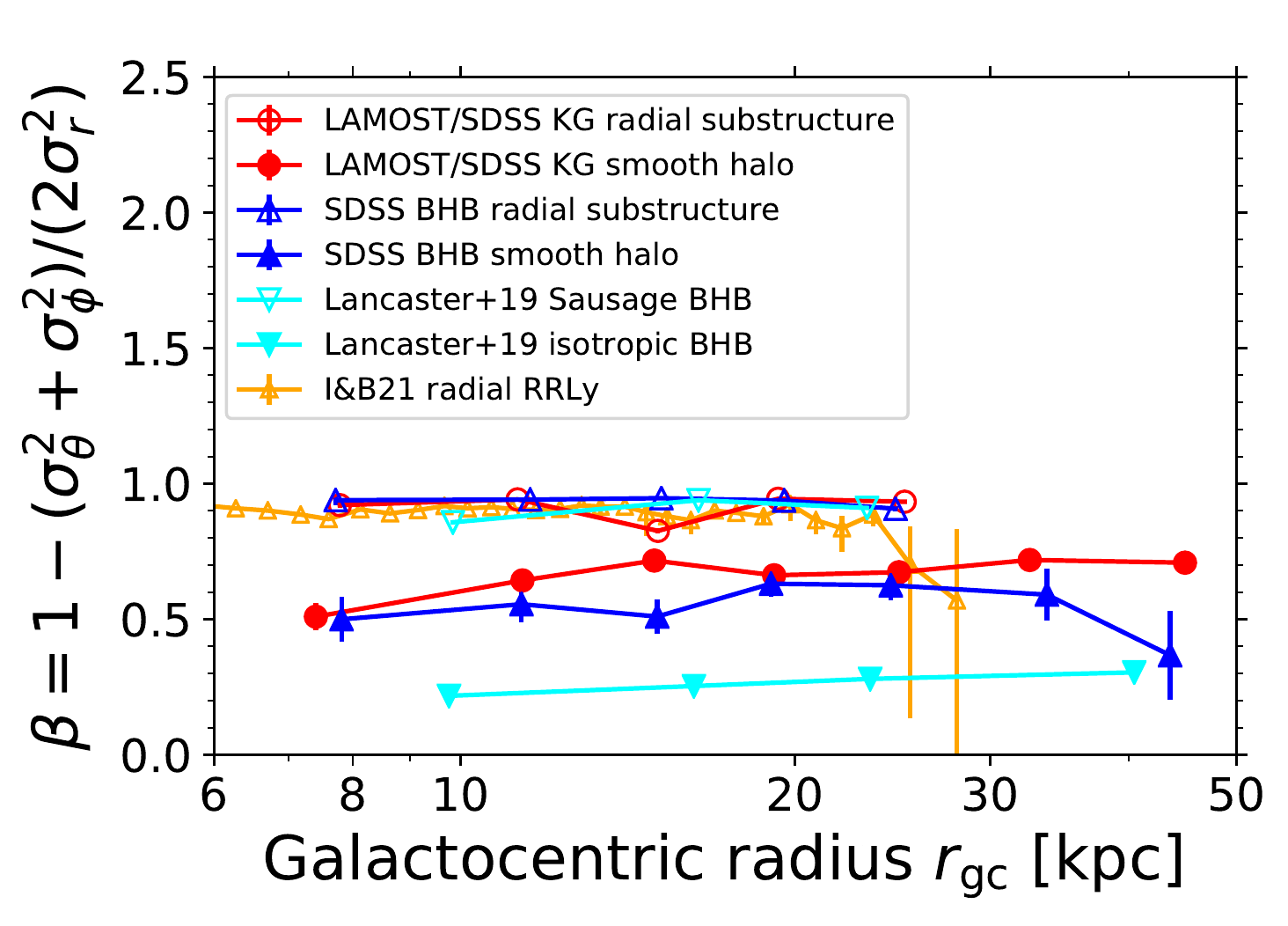}
    \caption{Comparison of the K-giant and BHB anisotropy profiles $\beta(r_\mathrm{gc})$ with the work of \citet[][BHB]{Lancaster2019} and \citet[][RR Lyrae]{Iorio2021}. {\bf Left panel}: Semi-major axis and eccentricity distribution for our smooth, diffuse halo stars (black) and for our substructure stars (red) most comparable to the Sausage (highly-radial) component of the \citet{Lancaster2019} and \citet{Iorio2021} Gaussian mixture model. K giants are shown; selected BHB stars share a similar distribution. This parameter space is used to select stars characteristic of the highly radial Sausage component. The side panels show number histograms along the axes.
{\bf Right panel}: 
Filled markers are the smooth, diffuse halo star samples selected from the current sample of LAMOST/SDSS K giants and SDSS BHB stars (filled red circles and blue upward-triangles, respectively) within a metallicity range comparable to the near-isotropic component from the Gaussian mixture model of \citet{Lancaster2019} (filled cyan downward-triangles). In the range $-2.1<$ [Fe/H] $<-1.7$, we find within our smooth, diffuse halo star sample a total of 2544 LAMOST/SDSS K giants and 893 SDSS BHB stars. 
Open markers are the LAMOST/SDSS K giant and BHB stars in kinematically highly radial substructure (open red circles, 4113 stars, and blue upward-triangles, 830 stars, respectively) as defined by the \citet[in preparation]{Xue2019} method. These substructure stars are comparable in kinematics to the highly radial Sausage component of \citet{Lancaster2019} (open magenta downward-triangles) and \citet{Iorio2021} (open orange upward-triangles; we do not plot the RR Lyrae profile for $r_\mathrm{gc}<6$, the x-axis limit to our plot, nor do we plot the point for $r_\mathrm{gc}>30$ as it fall below the y-axis of our plot). We plot our substructure stars such that $>60$\% member stars have eccentricity $>0.7$ with semi-major axis $<20$ kpc.
To attain the substructure stars, neither we nor \citet{Lancaster2019} and \citet{Iorio2021} apply metallicity restrictions except for the exclusion of spurious SDSS BHB measurements (i.e., not measured or [Fe/H] $\le-3$).
Note in both these lower panels, the BHB stars selected by \citet{Lancaster2019} are from the \citet{Xue2011} catalog, which we also use, although the selected stars from the catalog are slightly different and the analysis methods are different.
}
  \label{betafeh-compare}
\end{figure*}

\citet{Lancaster2019} use a Gaussian mixture model to fit the velocities and metallicities of halo BHB stars. In their model, the highly radial, more metal rich component characterizes stars belonging to the ancient {\it Gaia}-Sausage merger event. The second component is less radial and more metal poor; this fit represents halo stars built up over time from minor mergers.

In order to compare our analysis with the results of \citet{Lancaster2019}, we carefully select two subsamples from our halo LAMOST/SDSS K giant and BHB stars. We select these subsamples such that they most likely have the same origin as the \citet{Lancaster2019} model representative stars, namely either originating from an ancient radial merger or from the smooth, diffuse halo which is built up over time from many mergers.

We select a subsample from the substructure found in our halo LAMOST/SDSS K giant and BHB stars using the method of \citet[in preparation]{Xue2019}. 
Our selection criteria are semi-major axis $<20$ kpc and eccentricity $>0.7$. These substructure stars are likely members of large ancient radial mergers in our Galaxy, one of the most likely candidates being the {\it Gaia}-Sausage. As the \citet{Xue2019} method defines groups of stars most likely part of the same substructure based on their integrals of motion, some of the stars within the same group (i.e., part of the same substructure) may not meet our requirements for the semi-major axis and eccentricity. We keep all groups such that $>60$\% of the stars meet the requirements. In Figure \ref{betafeh-compare}, we see in the left panel that the majority of our selected {\it Gaia}-Sausage candidates (red markers and red lines) have semi-major axis $<20$ kpc and eccentricity $>0.7$, but indeed some substructure members are outside of these criteria. In the right panel of Figure \ref{betafeh-compare}, we plot the anisotropy profile of our selected LAMOST/SDSS K giant stars from the left panel, apply the same criteria to our BHB substructure stars, and make our comparison with the highly radial component of \citet{Lancaster2019} which they find using their Gaussian mixture model. The anisotropy profiles match very well and both methods select similar highly radial stars ($\beta\sim0.9$).

We next compare a subsample of our smooth, diffuse halo stars to the second component fit, which is near-isotropic, from the \citet{Lancaster2019} Gaussian mixture model; these 
represent the stars built up over time by many mergers. We plot our selected stars for this comparison in the left panel of Figure \ref{betafeh-compare} in the space of semi-major axis and eccentricity. \citet{Lancaster2019} remove the obvious substructure from Sagittarius before they apply their method; as our sample has had substructure removed by the method of \citet{Xue2019}, we simply select stars in a comparable metallicity range $-2.1<$ [Fe/H] $<-1.7$ to those from the isotropic component of \citet{Lancaster2019} (mean $\mu_\mathrm{[Fe/H]}\sim-1.9$, dispersion $\sigma_{[Fe/H]}\sim0.1$). As seen in the right panel of Figure \ref{betafeh-compare}, the anisotropy profiles from the two methods differ very substantially, by $\Delta\beta\sim0.3-0.4$.

Differences between the anisotropy profile for the smooth, diffuse halo stars in Figure \ref{betafeh-compare} are most likely due to the different methods used to model the velocity dispersions, which lead to different criteria for assigning  a star to one model fit or another. 
The Gaussian mixture model strives to fit all highly radial stars with one component and a second component is used to fit those stars remaining. The \citet{Xue2019} method groups stars which share similar integrals of motion; thus stars with highly radial orbits remain in our smooth halo sample if they are not obvious members of an integral-of-motion group. These results imply that many stars with highly radial orbits no longer are obviously identifiable as members of the same substructure in integrals of motion.

If members of the ancient $Gaia$-Sausage are no longer obviously selected as substructure, they can be categorized as members of the smooth, diffuse halo component of the Milky Way. This is an important determination as one would like to use the virialized halo stars for mass estimation.

\subsection{Comparison with \citet{Iorio2021}}

\citet{Iorio2021} use the full sample of RR Lyrae stars found in {\it Gaia} DR2; these include distances and proper motions and \citet{Iorio2021} estimate the 3D velocities through modeling techniques. They use an initial Gaussian mixture model to assign stars to a disk and halo component. Further from the halo component, they measure the velocity anisotropy $\beta$ profile along Galactocentric radial bins in two ways, using a single Gaussian and using the Gaussian mixture model developed by \citet{Lancaster2019} and \citet{Necib2019_874}. In Figure \ref{fig:rgc-beta2020_compare} (as shown by \citet{Iorio2021} in their Figure 9, middle panel), we plot the RR Lyrae anisotropy using the single Gaussian model in Galactocentric radial bins with orange triangles. The profile compares well with our LAMOST/SDSS K-giant sample within $r_\mathrm{gc}\sim 8-20$ kpc, although profiles from the full samples of K giants and RR Lyrae are more radially biased compared to the BHB stars.
In Figure \ref{betafeh-compare} (as shown by \citet{Iorio2021} in their Figure 7, middle left panel), we plot the radial RR Lyrae halo component in orange triangles. The radial component measured by the Gaussian mixture model for RR Lyrae is in good agreement with the highly radial substructure selected (see Section \ref{sec:Lancaster} for details) by the integral-of-motion method of \citet{Xue2019} for both our samples of K giants and BHB stars.

\subsection{Clues to Milky Way's Dynamics}\label{betametal-explain}

\begin{figure*}[htb]
\begin{tabular}{cc}
\includegraphics[width=\columnwidth]{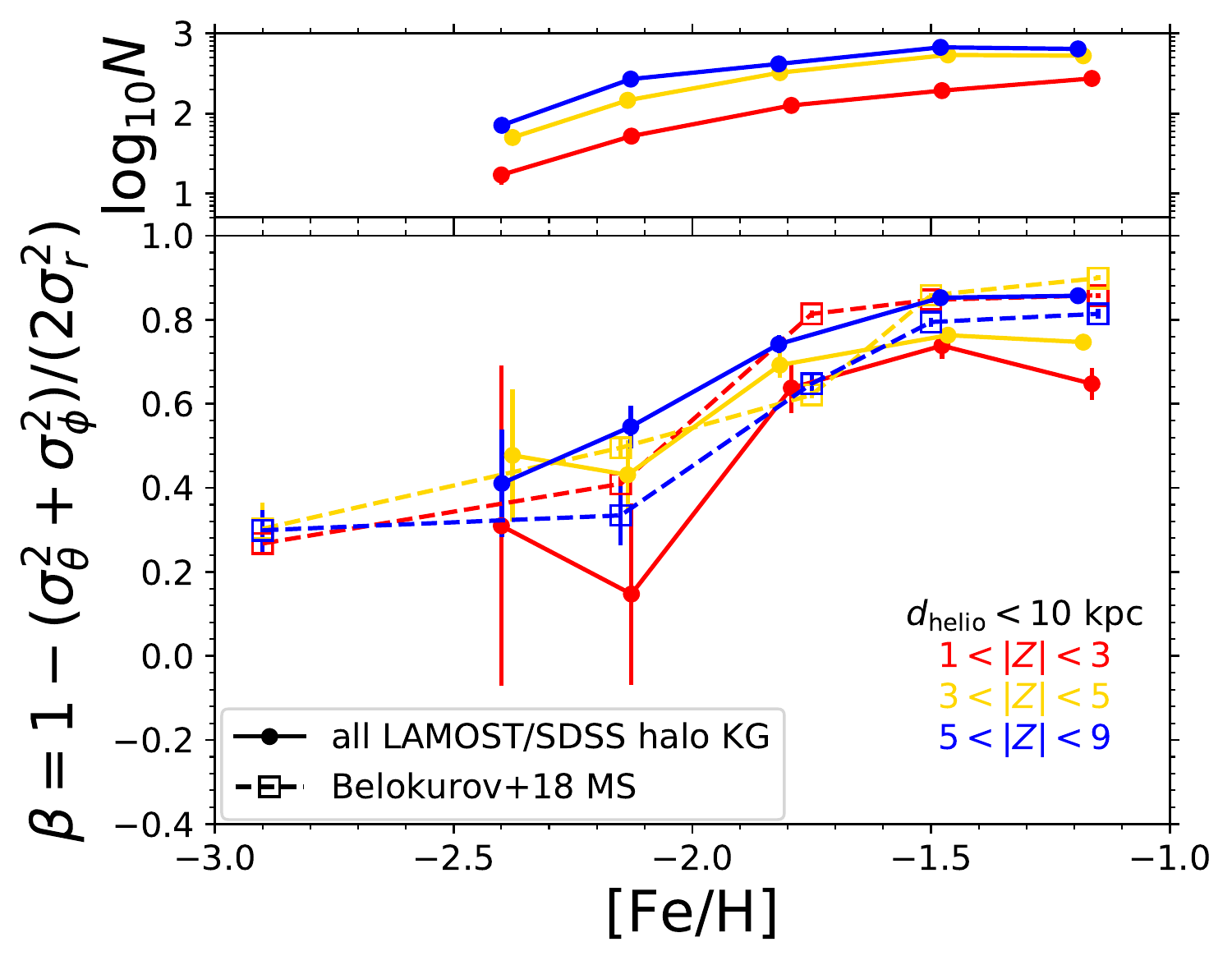}&
\includegraphics[width=\columnwidth]{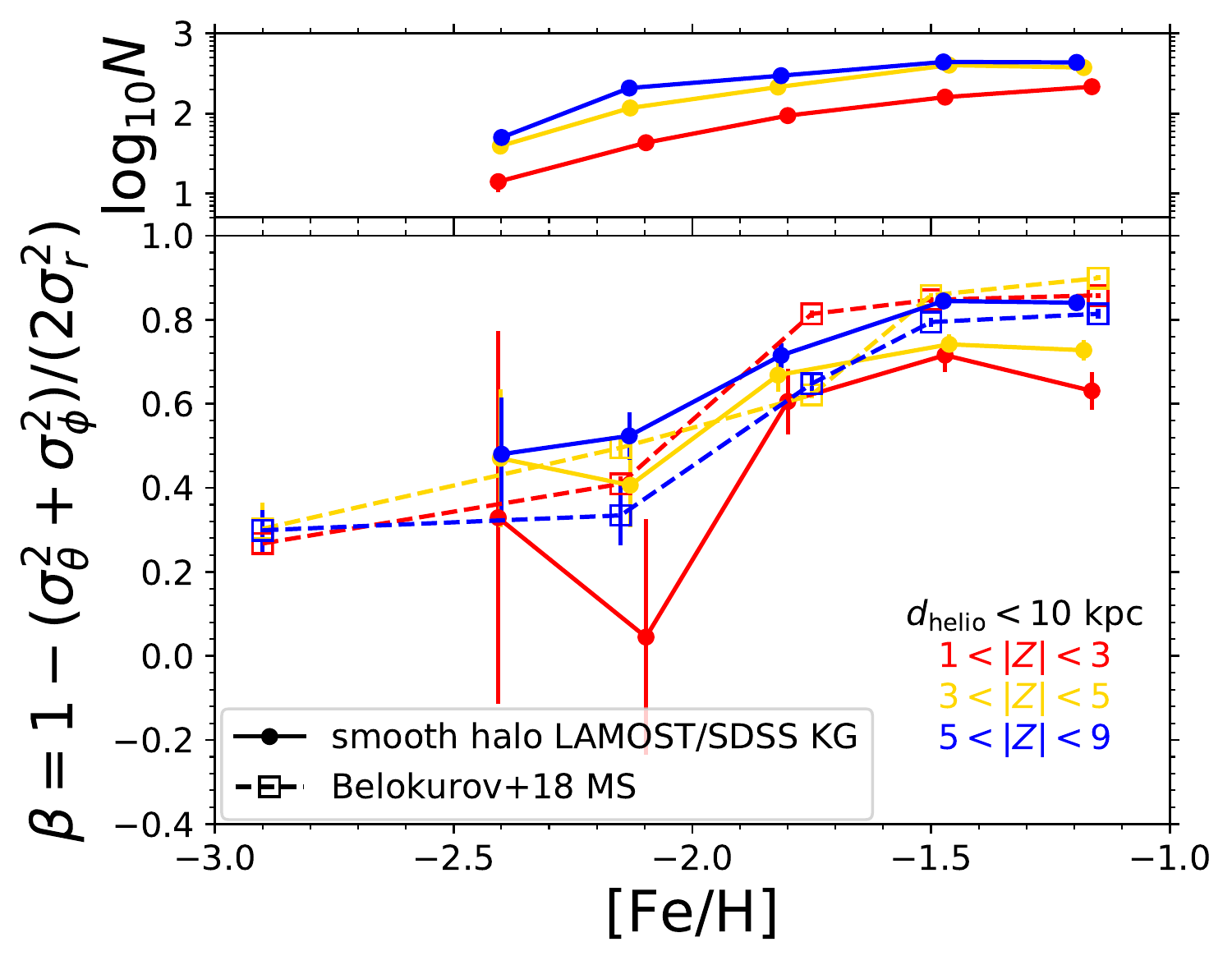}\\
\includegraphics[width=\columnwidth]{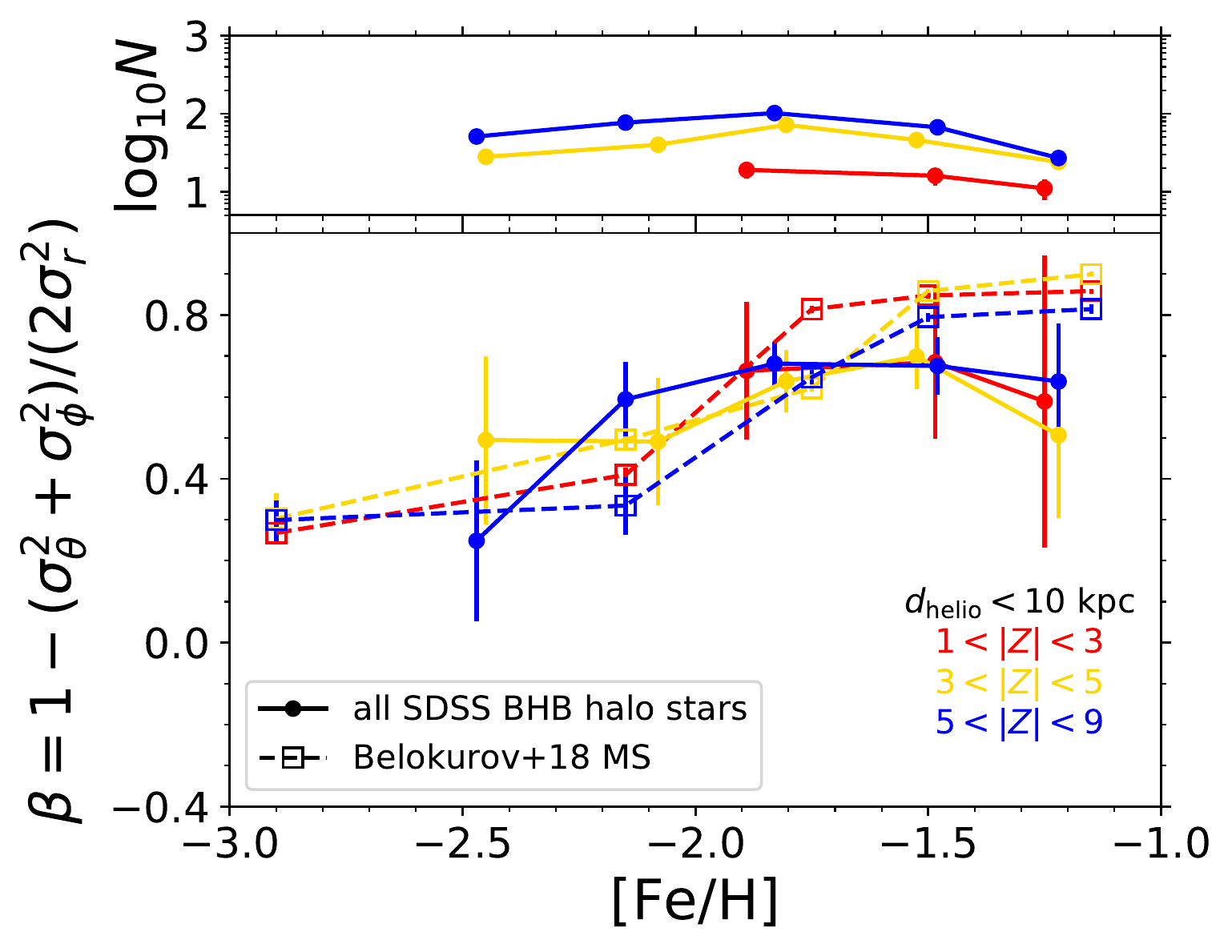}&
\includegraphics[width=\columnwidth]{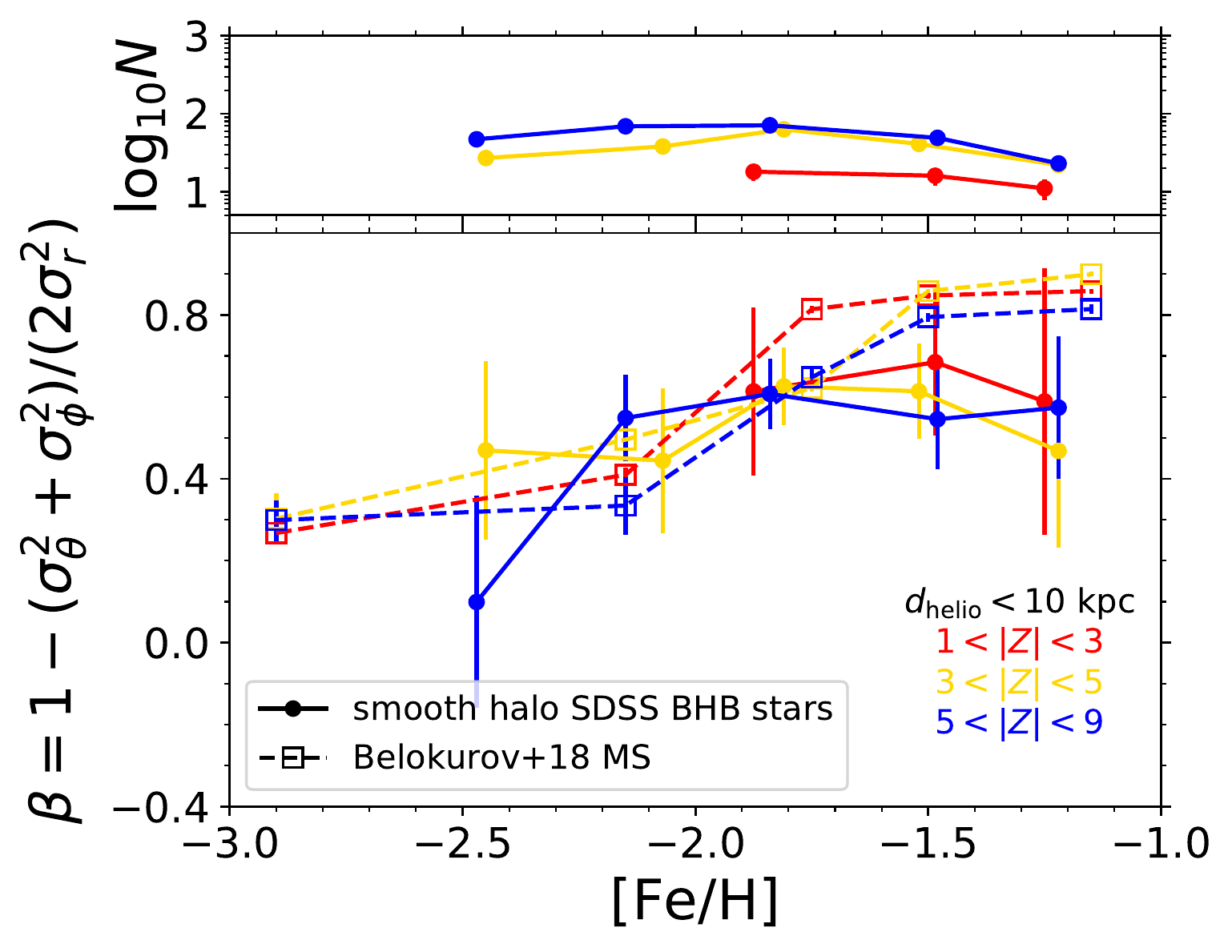}
\end{tabular}
  \caption{Anisotropy comparison of the main sequence stars (MS) of \citet{Belokurov2018} with our total halo sample (left column) and smooth, diffuse halo sample (right column) for LAMOST/SDSS K giants (upper panels) and SDSS BHB stars (lower panels). We find broad agreement between the two studies for the behavior of $\beta$ with metallicity. The dependence of $\beta$ on metallicity is independent of substructure and stellar type. The number of stars per bin for the K giant and BHB samples of the current work are shown in the upper part for each of the four panels. Note, for the samples of this work, we plot neither stars with $|Z|<2$ kpc nor bins with a number of stars $N<10$. We plot the median metallicity of our sample stars within each bin.
}
  \label{fig:Belokurov}
\end{figure*}

Independent of star type, metallicity, or substructure, we find the spherically averaged stellar halo anisotropy profile to be radially biased. Such a radially biased anisotropy is in agreement with simulations, although in this study we have not yet detected a definite rise in the profile, as is commonly found in simulations \citep{Diemand2005,Abadi2006,Sales2007.379.1464,Rashkov2013}. Instead,
for our smooth, diffuse K giant and BHB halo samples, we find a steady decline of $\beta$ with increasing $r_\mathrm{gc}$, and for different metallicity bins, a fairly constant anisotropy profile, within the uncertainties, of $\beta\sim0.6$ and $0.2<\beta<0.6$ for $-2.2<$ [Fe/H] $<-1.7$ and $-3<$ [Fe/H] $<-2.2$, respectively.
A shallow rise in the profile may exist and remain hidden within the uncertainties of our sample.

Our results are in good agreement with 
\citet{Belokurov2018}, who have analyzed the anisotropy
of a sample of main sequence halo stars, for which, as
here, full kinematical and chemical data are available. Their
sample has been selected from SDSS and {\it Gaia} DR1, and are within
$\approx 10$ kpc of the Sun. We select a sample of our stars with $d_\mathrm{helio}<10$ kpc and consider the same metallicity bins, [Fe/H] within the range $(-3,-2.3,-2,-1.66,-1.33,-1)$, and height from the Galactic midplane as in \citet{Belokurov2018}. We compare the results in Figure \ref{fig:Belokurov}.

The orbits of the
of the nearby halo main sequence stars of \citet{Belokurov2018} with [Fe/H] $>-1.66$ show highly radial $\beta \approx 0.9$,
with $\beta$ declining to more mildly radial orbits (0.2 $< \beta <$
0.4) down to the lowest metallicities probed ([Fe/H]$ \approx
-3$). Regardless of removing substructure, our halo K giants and BHB stars show similar anisotropy trends with metallicity. 

This decline in $\beta$ with metallicity for the nearby
stars is comparable to what we have found in the distant halo for $r_\mathrm{gc}>20$ kpc. 
After removing substructure,  
we find the anisotropy profiles
for our two lower metallicity bins (Figure \ref{betafeh}, $-2.2<$ [Fe/H] $<-1.7$ and $-3<$ [Fe/H] $<-2.2$)
 remain constant ($\beta\sim0.6$ and $0.2<\beta<0.6$, respectively) to the extent of the samples, the last bins reaching $\sim100$ kpc for K giants and $\sim60$ kpc for BHB stars).

\citet{Lancaster2019} find that the fraction of stars contributing to the more radial anisotropic component of BHB stars, those which they characterize as {\it Gaia}-Sausage members, drops sharply beyond $r_\mathrm{gc}\sim30$ kpc and point out that this corresponds to the same distance as the break radius in the BHB stellar density profile found by \citet{Deason2011.416} and the 
apogalacticon of an ancient merger found by \citet{Deason2018}.
We see a similar feature such that the number of stars in our most metal rich sample of BHB stars ($-1.7<$ [Fe/H] $<-1$) drops at beyond $r_\mathrm{gc}\sim30$ kpc. As seen in Figure \ref{betafeh} (lower panel), we measure a significantly lower value of anisotropy for these stars $>30$ kpc, $\beta\sim0.0-0.2$ as compared to $\beta\sim0.6-0.8$ within 30 kpc.
The large drop in anisotropy for these more metal rich stars may be enhanced by low number statistics as there are only a few tens of BHB stars at this distance. A larger sample of BHB stars will help clarify the picture. On the other hand, we measure $\beta\sim0.7-0.9$ for the more metal rich smooth, diffuse LAMOST/SDSS K giant stellar halo ($-1.7<$ [Fe/H] $<-1$) reaching to the end of our sample beyond $r_\mathrm{gc}=50$ kpc with several hundred stars in the last radial bin (anisotropy profile in Figure \ref{betafeh} upper panel, green filled markers and solid line). 

Our finding of radial anisotropy past the break radius matches the recent predictions of \citet{Elias2020}. They analyze a suite of $\sim150$ Milky Way analogs from Illustris. The characteristic highly radial orbits, high metallicity for halo stars, and compact apocentric radius $r_\mathrm{gc}<30$ kpc is uncommonly found in the simulations. They find that the one simulated radial merger, which produces the most similar stellar halo compared to the recent results found in our Galaxy, deposits stars past the bulk concentration of the merger remnants ($r_\mathrm{gc}>30$ kpc). 
The distant ($r_\mathrm{gc}>30$), more metal rich ($-1.7<$ [Fe/H] $<-1$), kinematically radial ($\beta>0.7$) halo stars which we find in our smooth, diffuse LAMOST/SDSS K giant samples are intriguing. 
A detailed analysis of their chemical properties, such as can be provided by long spectroscopic exposures will help clarify the origin of these stars.

Despite the differences (samples, methods) between the current work and those of \citet{Belokurov2018} and \citet{Lancaster2019}, the general trend is in agreement such that the more metal rich halo samples are on highly radial orbits and the more metal poor are still on slightly radial orbits, but to a much less degree, a difference in anisotropy of $\Delta\beta\sim0.3-0.6$.

In earlier studies of the Milky Way, e.g., \citet{Carollo2007} and \citet{Deason2011.411},
a picture has emerged of a two-component halo which
differs in spatial distributions of metallicity, age, and kinematics.
The same formation and evolution processes responsible for the 
dichotomy found by these previous studies help to form the velocity anisotropy profile dependency on metallicity which we find in this study. 
Finding similar trends, \citet{Hattori2013} and \citet{Kafle2013}, in studies of halo BHB stars within a few tens of kpc, have noted
that anisotropy is a function of metallicity. In both studies, the
halo stars are subdivided into two metallicity bins at [Fe/H] $=-2$, and the difference in
the kinematics (i.e. anisotropy and the bulk rotation) of the two
sub-samples is argued to support the two-component picture for the
halo. They find that the lower metallicity stars have rounder orbits
--- although the anisotropy obtained is significantly lower than what
we find for our metal poor stars (we find $\beta \approx 0.6$ whereas
they find much rounder orbits, with $\beta \approx 0$ to $-1$). We find a strong dependency of the anisotropy profile with metallicity; such a trend could be caused by the overlapping of two or more components to the stellar halo, although we here do not further investigate the best fitting number of components needed to model the K giants and BHB stars. We leave this for future work. 

The cause of the anisotropy dependency with metallicity is likely a complex correlation with the hierarchical merging history of the Galaxy. 
Previous studies 
\citep[e.g.,][]{Fattahi2019,Amorisco2017atlas,Mackereth2019,Kruijssen2019} 
analyze the dependency of the chemical composition of the merging satellites, their mass and concentration, as well as their infall time in relation to the chemodynamical properties of the stellar halo. The last significant merger dominates the stellar halo chemodynamical properties. As described by the relation between galaxy metallicity and stellar mass for
dwarf galaxies, larger satellites tend to have higher metallicities \citep{Kirby2013}. These larger satellites are typically more concentrated and tend to plunge deep into the more central regions of the Galaxy, dispersing their member stars on radial orbits throughout the halo \citep[e.g.,][]{Amorisco2017atlas,Deason2013}. 

Because of the various unique characteristics of different Milky Way-type stellar halos, both seen in observations \citep[e.g.,][]{Mouhcine2005b,Mouhcine2005c,Monachesi2016ghosts,Merritt2016,Harmsen2017} and simulations \citep[e.g.,][]{Renda2005,Cooper2010,D-Souza2018,Elias2018}, we need observations of our Galaxy in order to constrain and disentangle the processes and events which have shaped our stellar halo to such as we now observe it. For example, in a recent analysis of stellar anisotropy measurements, \citet{Cunningham2019b} detail the stellar halo anisotropy from two Milky Way-type galaxies in FIRE in order to compare to their observed radially dominated main sequence halo stars; they find one simulated stellar halo is radially dominated and the other is tangentially dominated, thus showing the need for observational constraints to decipher from simulations which formation histories lead to the observed data. 
These observational constraints allow us to place the Milky Way on the cosmological scale in order to compare to other similar galaxies, giving us clues as to which events lead to which stellar halo characteristics.

\citet{Robertson2005}, \citet{Cooper2010}, \citet{Deason2016}, \citet{Amorisco2017atlas}, and \citet{D-Souza2018} find that the last (or last few) significant merger(s) dominate the features of Milky Way-type galaxies such as the mass, metallicity, and dynamics of the stellar halo.
\citet{Deason2013} and \citet{Belokurov2018.477} put forward the case that Milky Way experienced an ancient satellite merger $\sim10$ Gyr ago which reveals itself today in left-over shells (detected in density) and through the mixture of RR Lyrae type ab with $r_\mathrm{gc}$. \citet{Deason2018.862} and \citet{Simion2019} show that in the Milky Way the apocentric pile-ups of the ancient merger can be detected through features found in anisotropy. 
\citet{Belokurov2018} analyze simulations and further develop the picture of an early merger, finding that the highly radial halo component they detect with nearby halo main sequence stars results in the simulations when large satellites merge early on and the highly radial anisotropy is enhanced when the merger occurs during the disk's formation around $z=2$. 
Our results of the smooth, diffuse halo anisotropy profile dependency on metallicity is also in line with this early merger picture. 
A large majority of stars from this early, massive, radial merger with our Galaxy can be expected to have virialized and contributed to the smooth, diffuse halo component (perhaps the vast majority that of), e.g., such as the mixed components of halo stars seen in simulations by \citet{Johnston2008}. 
The fact that we see similar anisotropy-metallicity dependency for both our total halo samples (before removing substructure) and smooth, diffuse halo samples (as seen in Figures \ref{betafeh} and \ref{fig:Belokurov}) likely results from the same dominating contributor to the stellar halo, the last (or last few) significant merger(s).
Further analysis of stellar halo anisotropy in simulations will be useful to develop  a clearer understanding of events leading to anisotropy-metallicity dependence of the smooth halo component.  

As a clue to the dynamical mass modeling of the Milky Way, since the anisotropy depends on metallicity, we can assume they trace the same Milky Way potential and that the density profile of these Milky Way stellar samples also depends on metallicity. Simultaneously using different stellar populations with different anisotropy and density profiles helps leverage against the degeneracies in galaxy mass estimation \citep[e.g.,][]{Battaglia2008,Walker2011,Agnello2012,Amorisco2012a,Amorisco2013,Agnello2014b,Napolitano2014,Pota2015mass,Zhu2016}.
The aforementioned works sought to alleviate such degeneracies as between the
concentration--virial mass \citep{Humphrey2006,Kafle2014},
dark halo mass--velocity anisotropy \citep{Merrifield1990,Dekel2005,Deason2012,Kafle2012,Agnello2014b,Kafle2014},
luminous mass--dark halo mass \citep{Agnello2014b,Napolitano2014,Pota2015mass},
dark halo density--dark halo scale radius \citep{Napolitano2011},
dark halo mass slope--virial mass \citep{Walker2011,Agnello2012},
and the debate over cored vs. cusped central regions of dark halos \citep{Kleyna2002,Koch2007,Walker2009,Walker2011,Agnello2012,Amorisco2012a,Breddels2013.433,Breddels2013.558}.
The main degeneracies which influence the estimate of the Milky Way dark halo mass are those of the
stellar density slope--velocity anisotropy
(now largely alleviated through measuring $\beta$ in this work),
dark halo density--dark halo scale radius,
and dark halo mass slope--virial mass.
Currently the strength of the aid from different tracers with different anisotropy is unclear and future work is planned to fully utilize the data towards mass estimation.

\subsection{Conclusions}

We summarize in the following four results and then make our conclusions.

First, we use the largest sample of halo stars (21958 stars) to measure the radially dominated anisotropy profile out to the furthest reached distances ($r_\mathrm{gc}\sim100$ kpc) to date; we find a nearly constant radially dominated anisotropy profile for K giants and BHB stars within a Galactocentric radius of 20 kpc. We note that the K giants and BHB stars peak at different metallicities, which explains the two seemingly discrepant anisotropy profiles when considering the total samples over all metallicities ($0.6<\beta<0.9$ for K giants and a downward shift of $\Delta\beta\sim0.1-0.3$ for BHB stars).

Second, we find the stellar halo anisotropy profile is dependent on metallicity. More relatively metal rich halo stars in our sample are on more radially dominated orbits compared to the less metal rich halo stars. This results is in agreement with previous observational and theoretical works. 

Third, different stellar types within similar metallicity ranges share similar anisotropy profiles ($\Delta\beta\sim0.1-0.2$) regardless of substructure removal. We find this for our halo K giants and BHB stars, as well as in a comparison with the main sequence stars analyzed by \citet{Belokurov2018}.

Fourth, the 3D velocity dispersion profiles decline in such a manner that the anisotropy profile remains constant. We use a more robust substructure removal method compared to \citet{Bird2019beta} and show that resulting anisotropy profile after substructure removal remains constant out to a further distance ($\sim20$ kpc) than compared to before removing substructure. Our smooth, diffuse K giant and BHB samples share constant anisotropy to the extent of the profiles for [Fe/H] $<-1.7$, with $0.2<\beta<0.7$ depending on the metallicity range probed. The constant anisotropy profiles for the smooth, diffuse halo peak at $\beta\sim0.9$ for K giants and $\beta\sim0.8$ for BHB stars within 20 kpc for the more metal rich samples [Fe/H] $>-1.7$; at larger distances the anisotropy profile declines to $\beta\sim0.7$ and $\beta\sim0.0-0.2$ for K giants and BHB stars, respectively.

The data and our analysis lead to two main constraints on the assembly history of the Milky Way stellar halo.

First constraint, an underlying, radial $\beta$, virialized halo exists. Lots of substructure needs removing. We can hopefully use the smooth, diffuse halo star sample to measure the mass profile $M(r)$. The remarkable agreement of the observed radially-dominated anisotropy profile and those of simulations gives support for the picture where the properties of Milky Way-type stellar halos are strongly dependent on the last (few) significant merger(s).

Second constraint, the dependency of metallicity at all radii probed (the furthest stars reaching 100 kpc) is complementary and may be a (or one) natural outcome of the picture where the last significant mergers dominate the bulk of the stellar halo mass and kinematical and chemical properties for Milky Way-type galaxies. Further analysis of metallicity dependency of anisotropy in simulations will help clarify the stochastic processes involved which form galaxies most similar to our own Milky Way. Because of the diversity seen in Milky Way-type stellar halos, the Galaxy's metallicity-dependent velocity anisotropy profile gives a key constraint for simulations to explain through the formation history.
With the large variety seen in Milky Way-type galaxy stellar halos, both in observations and simulations, the dependency of the anisotropy profile with metallicity is an important constraint for theories of Galactic formation and evolution.

Future prospects such as efforts by LAMOST, SDSS-V \citep{Kollmeier2017}, DESI \citep{DESICollaborationAghamousa2016}, and Subaru PFS \citep{Takada2014,Tamura2016}, will add new halo star line-of-sight velocities and efforts such as from updated $Gaia$-releases, Subaru HSC \citep{Qiu2021}, and Rubin-LSST \citep{LSSTScienceCollaborationAbell2009,Juric2017,Ivezic2019} will provide tangential velocities with smaller uncertainties, especially regarding the more distant halo stars. We will use these anisotropy profile results to estimate the mass of the Galaxy using the 3D Jeans equation.

\acknowledgments

We thank Warren Brown, Monica Valluri, Emily Cunningham, Huang Yang, Lachlan Lancaster, Timothy Beers, Wenbo Wu, Kai Zhu, and Ling Zhu for useful discussions, and John Vickers additionally for sharing results used for plotting and comparison. We thank the referee for a very thorough reading of the paper and a number of very helpful suggestions which greatly improved the paper.
We thank P. L. Lim for the 11/2009 {\tt robust\_sigma} port from {\tt IDL} to {\tt Python}.
This work is supported by the National Key R\&D Program of China under grant no. 2019YFA0405500 and 2018YFA0404501;
by the National Natural Science Foundation of China under grant no. 11988101, 11873052, 11890694, 11835057, 11773052, 11761131016, and 12025302;
by the Chinese Space Station Telescope project;
and by the ``111'' Project of the Ministry of Education under grant no. B20019.
This work made use of the Gravity Supercomputer at the Department of Astronomy, Shanghai Jiao Tong University, and the facilities of the Center for High Performance Computing at Shanghai Astronomical Observatory, and
 was developed in part at the 2019 $Gaia$-LAMOST Sprint workshop, supported by the NSFC under grants 11873034, U1731108, U1731124, and Hubei Provincial Outstanding Youth Fund (2019CFA087), and at the 2018 Heidelberg Summer School for $Gaia$ Data and Science.
S.A.B. acknowledges support from the Aliyun Fellowship, Chinese Academy of Sciences President's
International Fellowship Initiative Grant (no. 2016PE010 and 2021PM0055), and Postdoctoral Scholar's Fellowship
of LAMOST.  
C.F. acknowledges financial
support by the Beckwith Trust. 
Guoshoujing Telescope (the Large Sky
Area Multi-Object Fiber Spectroscopic Telescope LAMOST) is a National
Major Scientific Project built by the Chinese Academy of
Sciences. Funding for the project has been provided by the National
Development and Reform Commission. LAMOST is operated and managed by
the National Astronomical Observatories, Chinese Academy of
Sciences. This work has made use of data from the European Space
Agency (ESA) mission {\it Gaia}, processed by the {\it Gaia} Data Processing and
Analysis Consortium (DPAC). Funding for the DPAC has been provided by
national institutions, in particular the institutions participating in
the {\it Gaia} Multilateral Agreement. 
Funding for SDSS-III has been provided by the
Alfred P. Sloan Foundation, the Participating Institutions, the
National Science Foundation, and the U.S. Department of
Energy Office of Science. The SDSS-III Web site is \url{http://www.sdss3.org/}. SDSS-III is managed by the Astrophysical
Research Consortium for the Participating Institutions of the
SDSS-III Collaboration, including the University of Arizona,
the Brazilian Participation Group, Brookhaven National
Laboratory, University of Cambridge, Carnegie Mellon University, University of Florida, the French Participation Group,
the German Participation Group, Harvard University, the
Instituto de Astrofisica de Canarias, the Michigan State/Notre
Dame/JINA Participation Group, Johns Hopkins University,
Lawrence Berkeley National Laboratory, Max Planck Institute
for Astrophysics, Max Planck Institute for Extraterrestrial
Physics, New Mexico State University, New York University,
Ohio State University, Pennsylvania State University, University of Portsmouth, Princeton University, the Spanish
Participation Group, University of Tokyo, University of Utah,
Vanderbilt University, University of Virginia, University of
Washington, and Yale University.
This research has made use of NASA's
Astrophysics Data System Bibliographic Services. 

\software{{\tt Astropy} \citep[v2.0.2;][]{astropy2013,astropy2018},
{\tt galpy} \citep[1.3.0;][]{Bovy2015}, {\tt matplotlib.pyplot} \citep[2.1.0;][]{Hunter2007}, {\tt NumPy} \citep[v1.13.3;][]{Oliphant2006,Walt2011,Oliphant2015}, {\tt ROBUST\_SIGMA} \citep{Freudenreich1990}, {\tt SciPy} \citep[v1.2.2;][]{2020Virtanen}, {\tt TOPCAT} \citep[v4.3-3;][]{Taylor2005}
}

\bibliography{}

\appendix

\begin{figure}[htb]
\begin{tabular}{cc}
\includegraphics[width=.45\columnwidth]{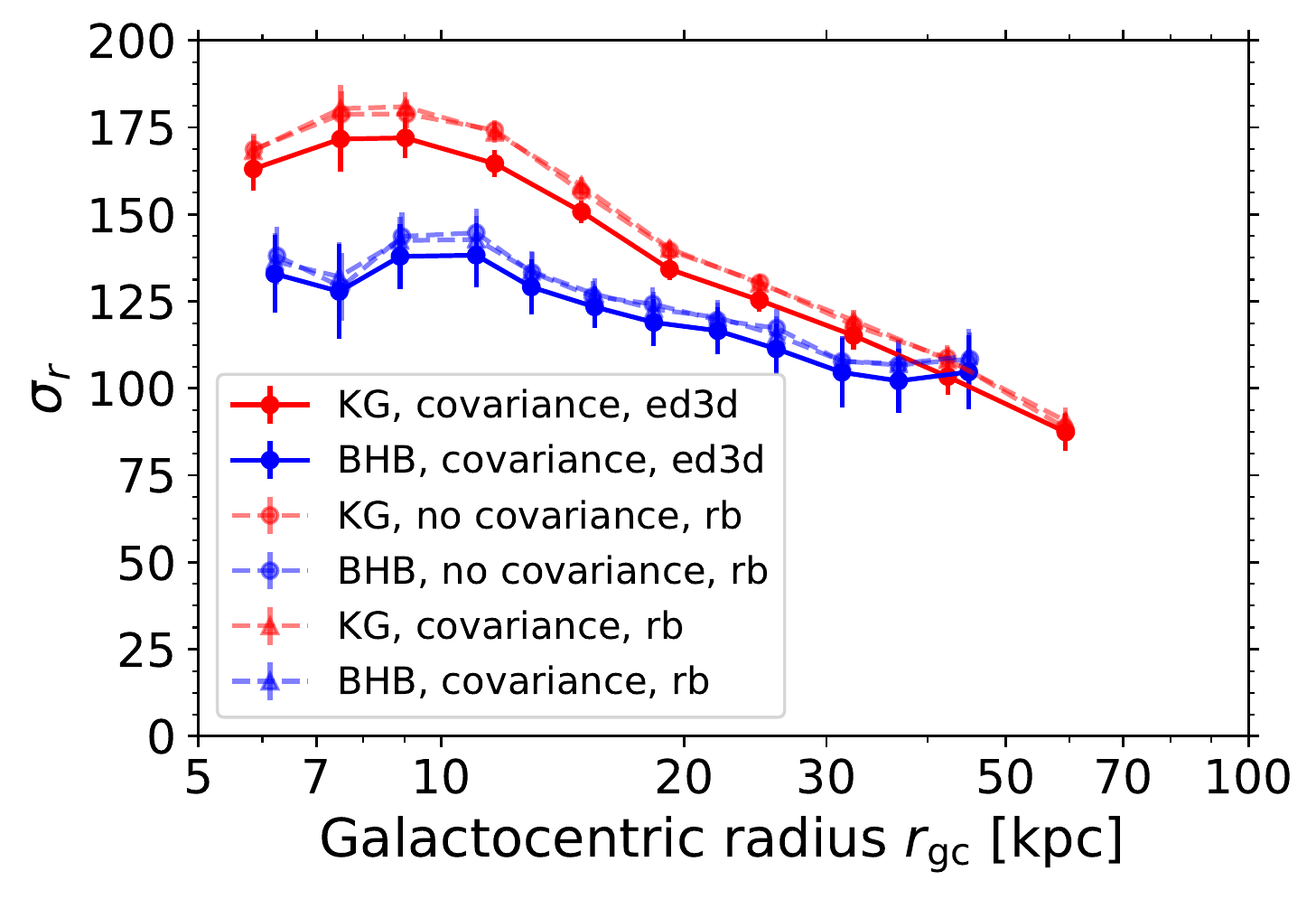}&
\includegraphics[width=.45\columnwidth]{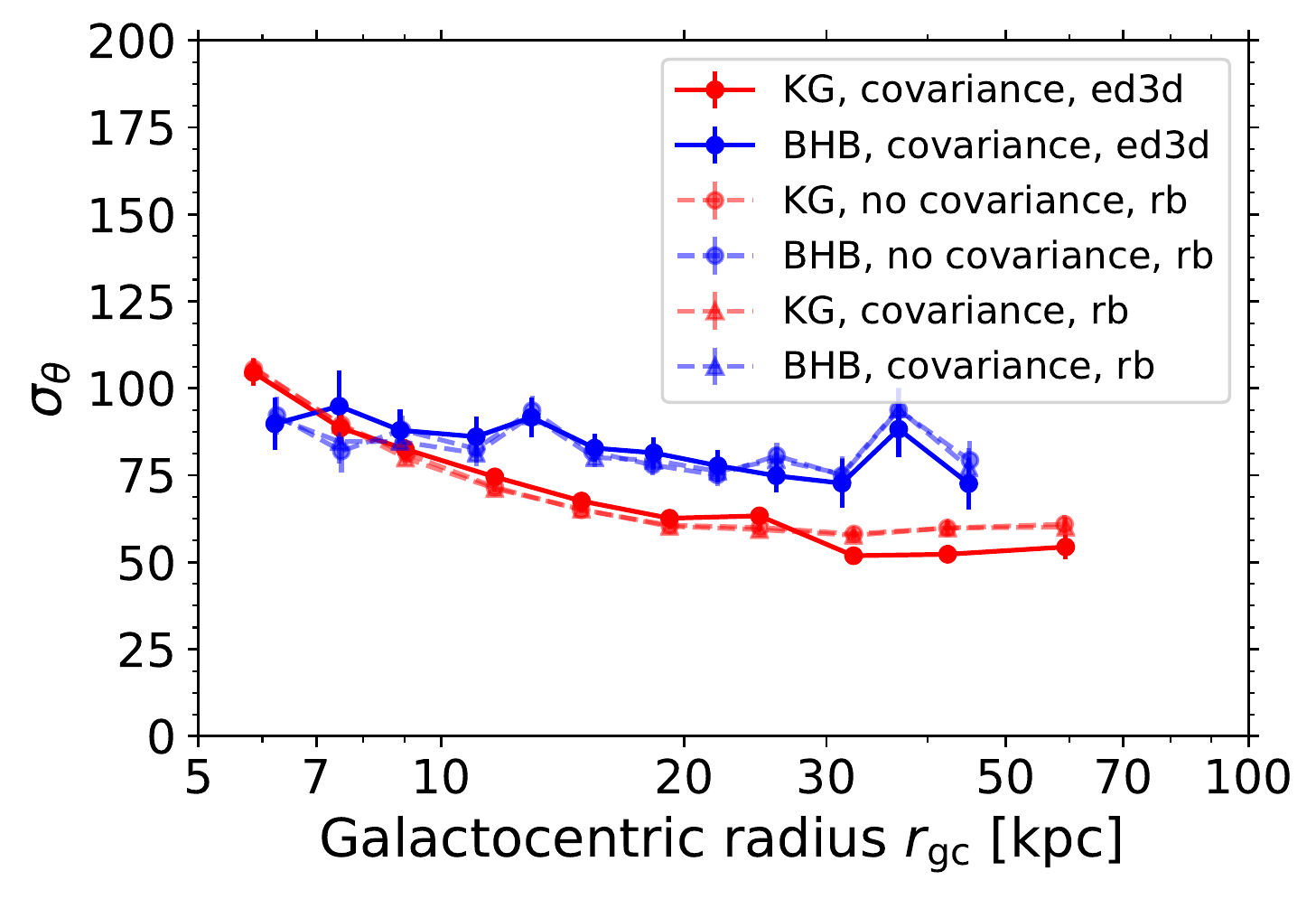}\\
\includegraphics[width=.45\columnwidth]{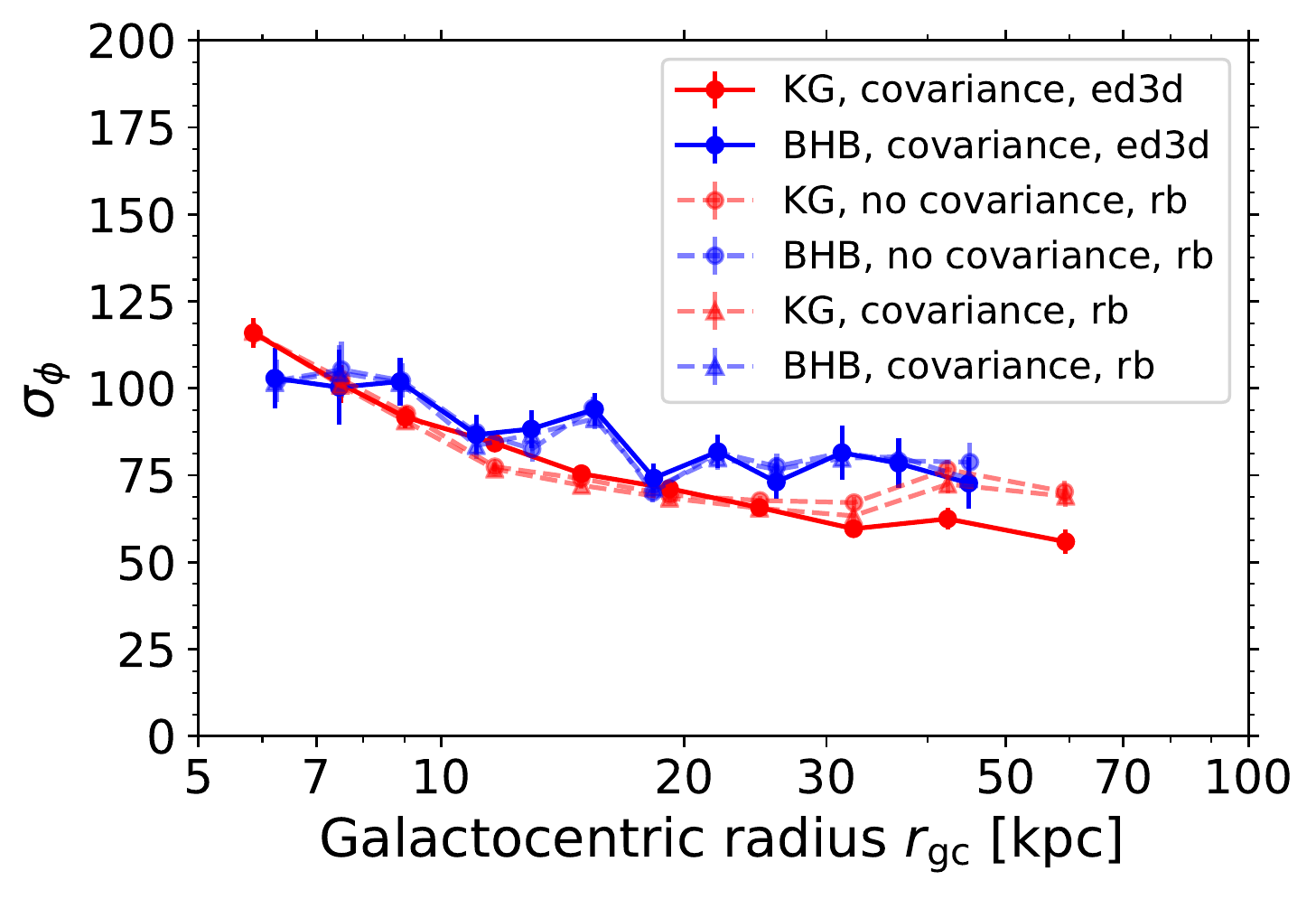}&
\includegraphics[width=.45\columnwidth]{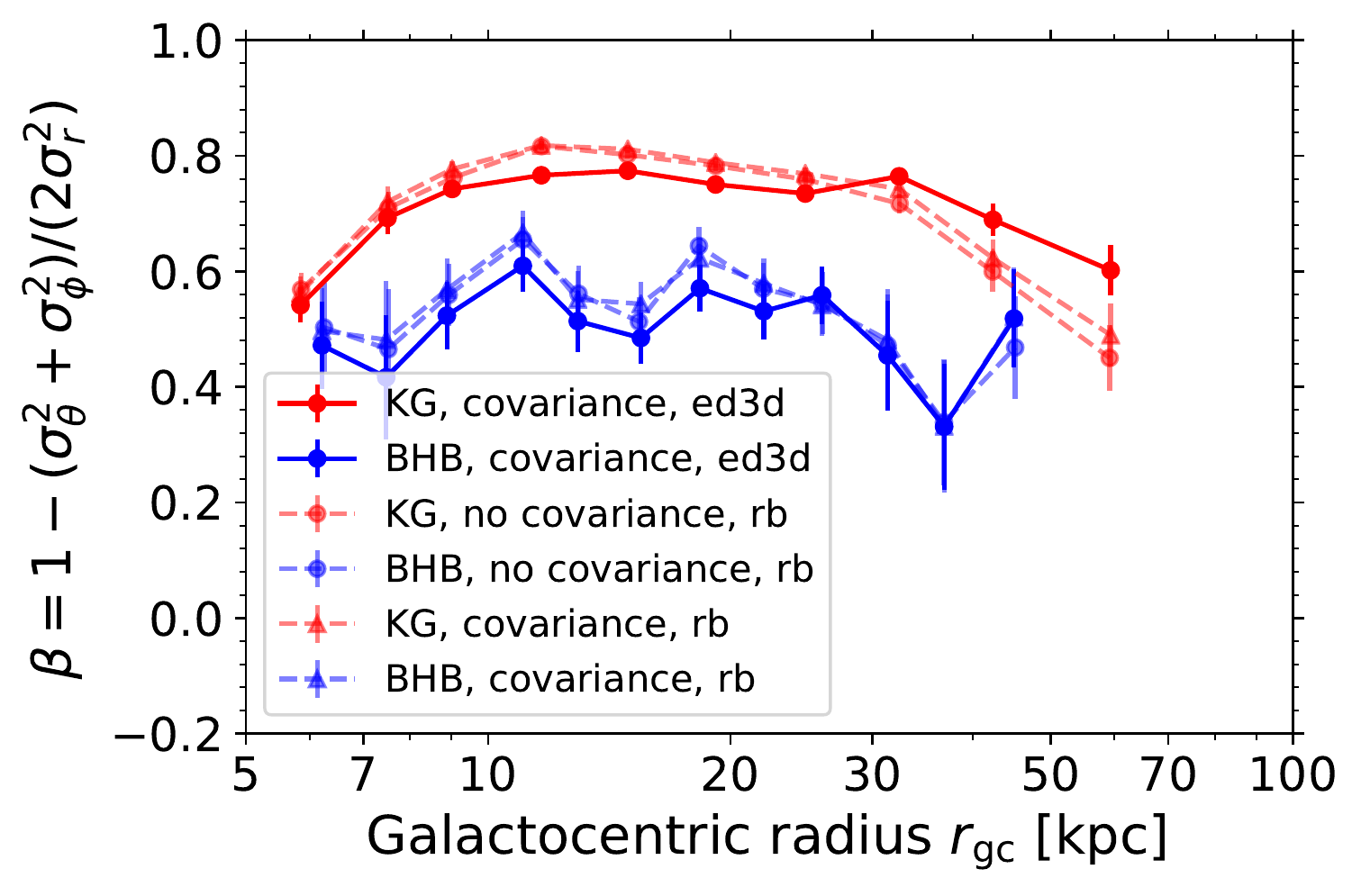}
\end{tabular}
\caption{Three profiles for each of the KG (red) and BHB (blue) $\sigma_r$, $\sigma_\theta$, $\sigma_\phi$, and $\beta$ profiles in the upper left, upper right, lower left, and lower right panels, respectively. The solid line profiles are the results from this work, propagating the $Gaia$ proper motion covariance to the uncertainties in the 3D Galactocentric spherical velocities $(V_r,V_\theta,V_\phi)$, and using the 3D {\tt extreme-deconvolution} (ed3d) algorithm. The remaining two profiles, dashed lines with dot or triangle markers, similarly use {\tt ROBUST\_SIGMA} (rb) with median velocity uncertainty subtracted in quadrature but contrastly ignore or propagate the {\it Gaia} proper motion covariance, respectively. We find that ignoring or propagating the proper motion covariance causes negligible difference in the profiles (the red dashed line profiles share negligible differences and the blue dashed line profiles share negligible differences). The two methods using {\tt ROBUST\_SIGMA} with median velocity uncertainty subtracted in quadrature and 3D {\tt extreme-deconvolution} (ed3d) produce similar profiles with differences along $r_\mathrm{gc}$ between the two methods all within the estimated uncertainties. In conclusion, for the purpose of Galactocentric radially binned profiles, either method is acceptable.
}
\label{fig:compare_covar}
\end{figure}

We use our current smooth, diffuse halo KG and BHB samples and apply the method of \citet{Bird2019beta} for determining the velocity anisotropy $\beta$ as a
function of Galactocentric radius. \citet{Bird2019beta} measure velocity dispersions with the median velocity uncertainty subtracted in quadrature. They measure dispersion using {\tt ROBUST\_SIGMA} \citep{Freudenreich1990} which is a routine from {\tt IDL
ASTROLIB} \citep{Landsman1993} and a port to {\tt Python} found at 
\url{https://home.fnal.gov/~stoughto/build/ARCONS/html/_modules/util/robust_sigma.html}.
In Figure \ref{fig:compare_covar}, we compare the smooth, diffuse halo KG and BHB results using the method of \citet{Bird2019beta} with the results of this paper which, besides propagating the uncertainties in distance, proper motion, and line-of-sight velocity, additionally propagates the uncertainties due to the covariance between the {\it Gaia} proper motions to the uncertainty estimates for the 3D Galactocentric spherical velocities $(V_r,V_\theta,V_\phi)$ as well as uses {\tt extreme-deconvolution} to measure $\sigma_r$, $\sigma_\theta$, $\sigma_\phi$, and $\beta$. Figure \ref{fig:compare_covar} includes three profiles for each of the KG and BHB $\sigma_r$, $\sigma_\theta$, $\sigma_\phi$, and $\beta$ profiles. The solid line profiles are the results from this work, propagating the $Gaia$ proper motion covariance and using the 3D {\tt extreme-deconvolution} (ed3d) algorithm. The remaining two profiles (dashed lines with dot or triangle markers) similarly use {\tt ROBUST\_SIGMA} with median velocity uncertainty subtracted in quadrature but contrastly ignore or propagate the {\it Gaia} proper motion covariance, respectively. We find that ignoring or propagating the proper motion covariance causes negligible difference in the two profiles (the two red KG dashed line profiles and the two blue BHB dashed line profiles). The two methods (dashed lines compared to solid line) using {\tt ROBUST\_SIGMA} with median velocity uncertainty subtracted in quadrature and 3D {\tt extreme-deconvolution} (ed3d) produce similar profiles with differences along $r_\mathrm{gc}$ between the two methods all within the estimated uncertainties. In conclusion, for the purpose of Galactocentric radially binned profiles, either method is acceptable. As a caution, further testing is needing to investigate the extremity of differences caused, for example, when calculating individual orbits or profiles including bins in the azimuthal or polar axes. 

\begin{figure}[htb]
\begin{tabular}{cc}
\includegraphics[width=.45\columnwidth]{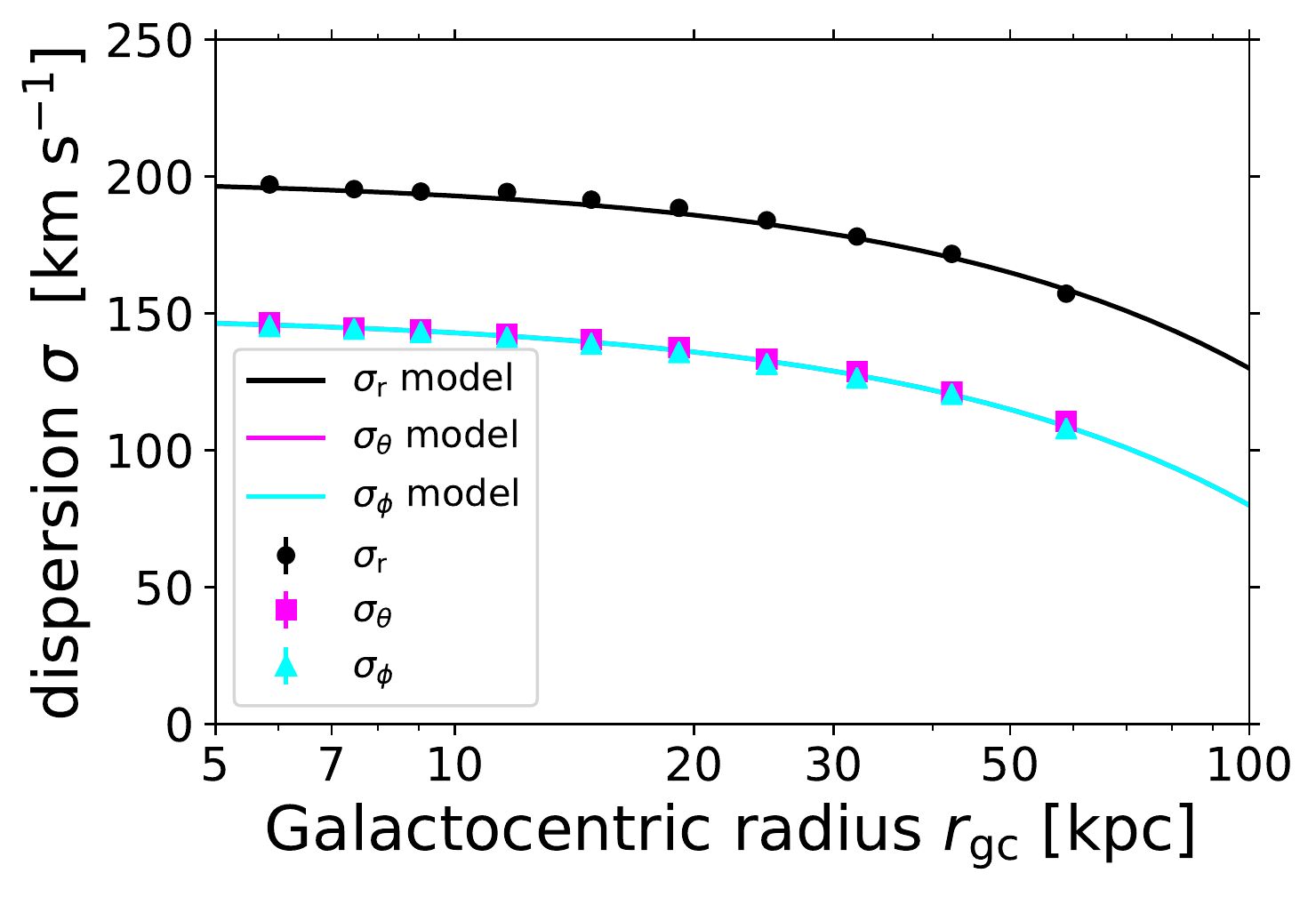}&
\includegraphics[width=.45\columnwidth]{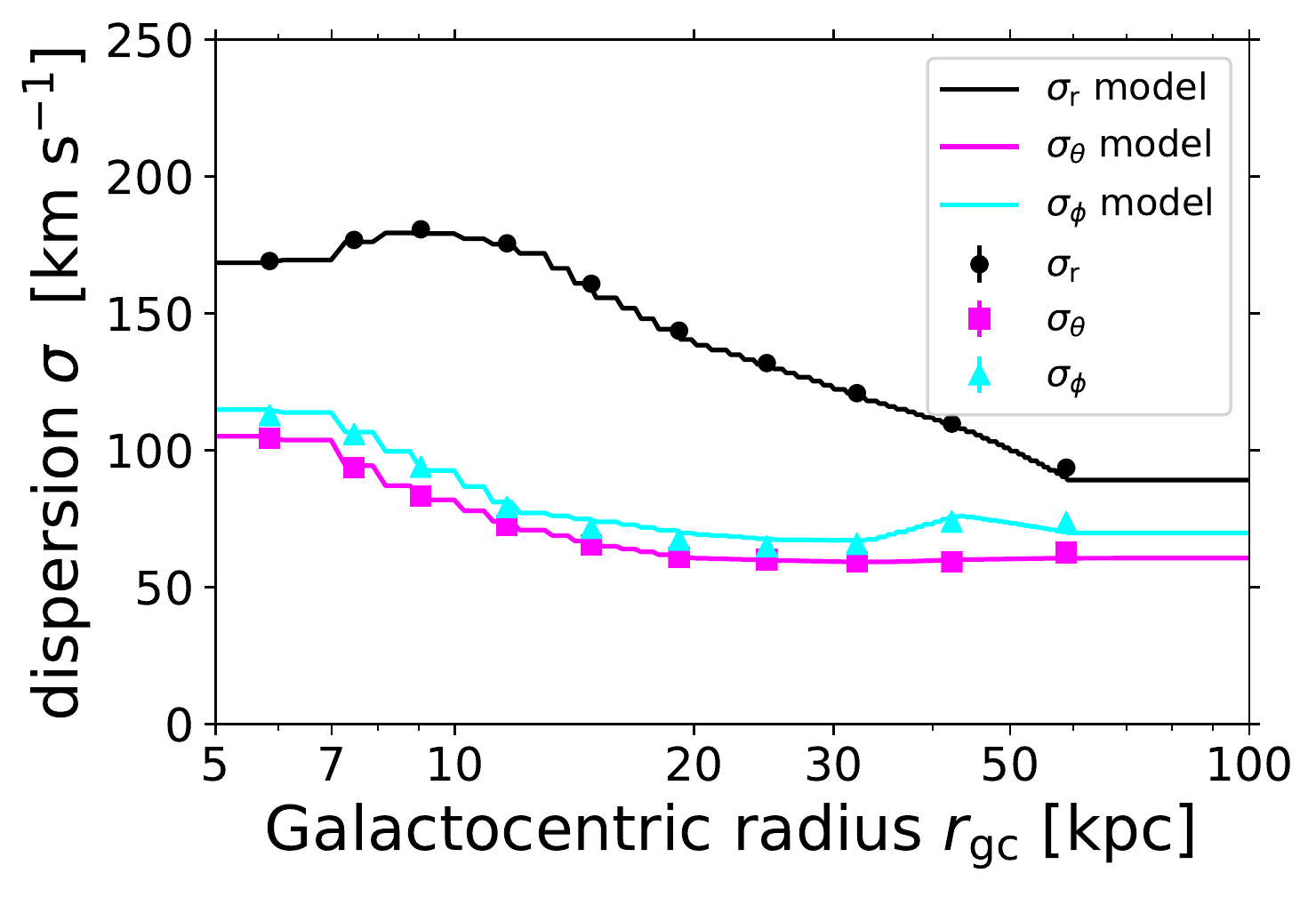}
\end{tabular}
\caption{Mock star velocity dispersion profiles: lines are the input models for radially rising anisotropy (left panel) and for LAMOST/SDSS K giant anisotropy, markers are the measured dispersions for 100 mock stars for each real K giant. Mocks have been scattered by the measurement 
uncertainties in distances, line-of-sight velocities, and proper motions.
}
\label{fig:veldisp100}
\end{figure}

\begin{figure}[htb]
\begin{tabular}{cc}
\includegraphics[width=.45\columnwidth]{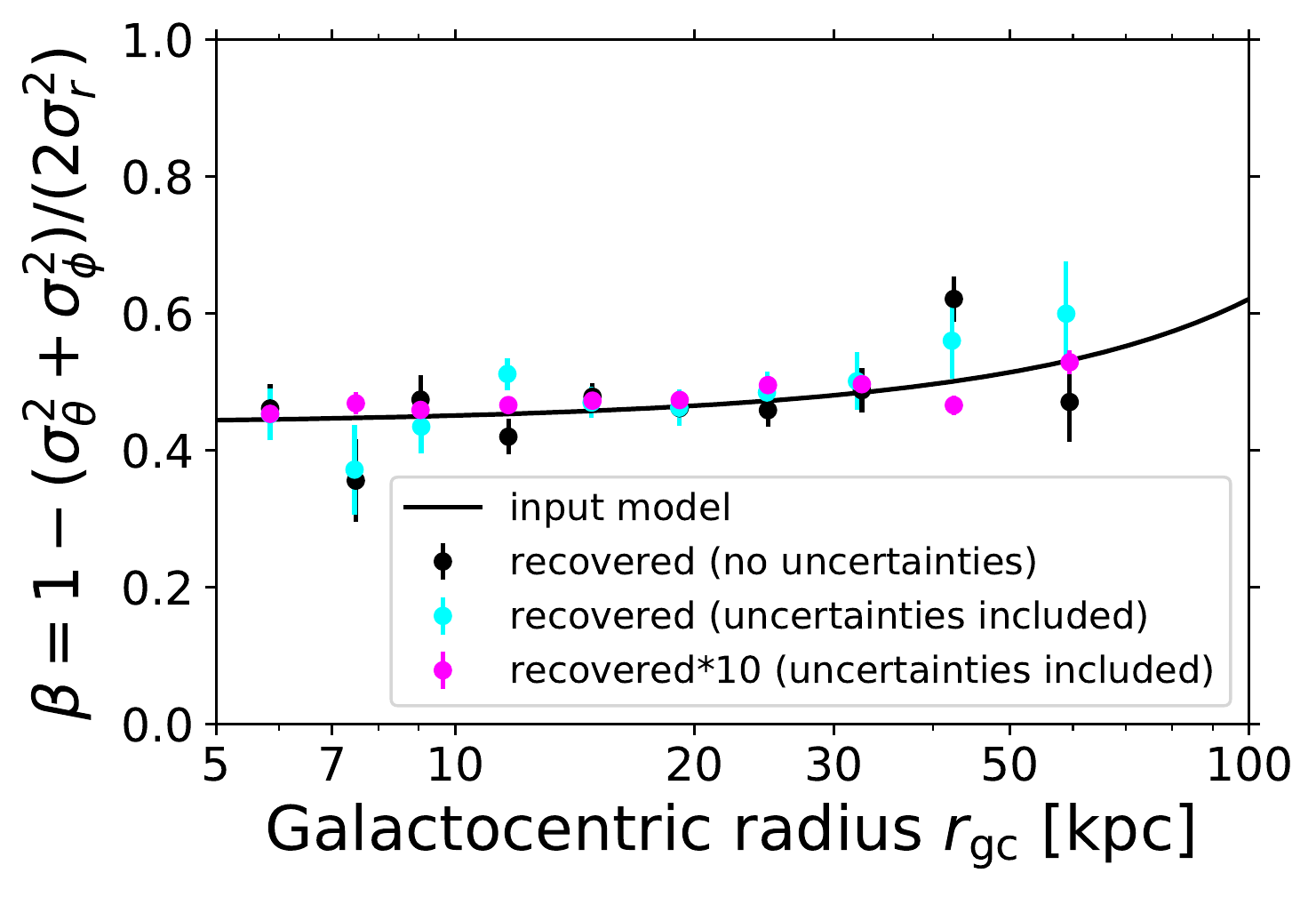}&
\includegraphics[width=.45\columnwidth]{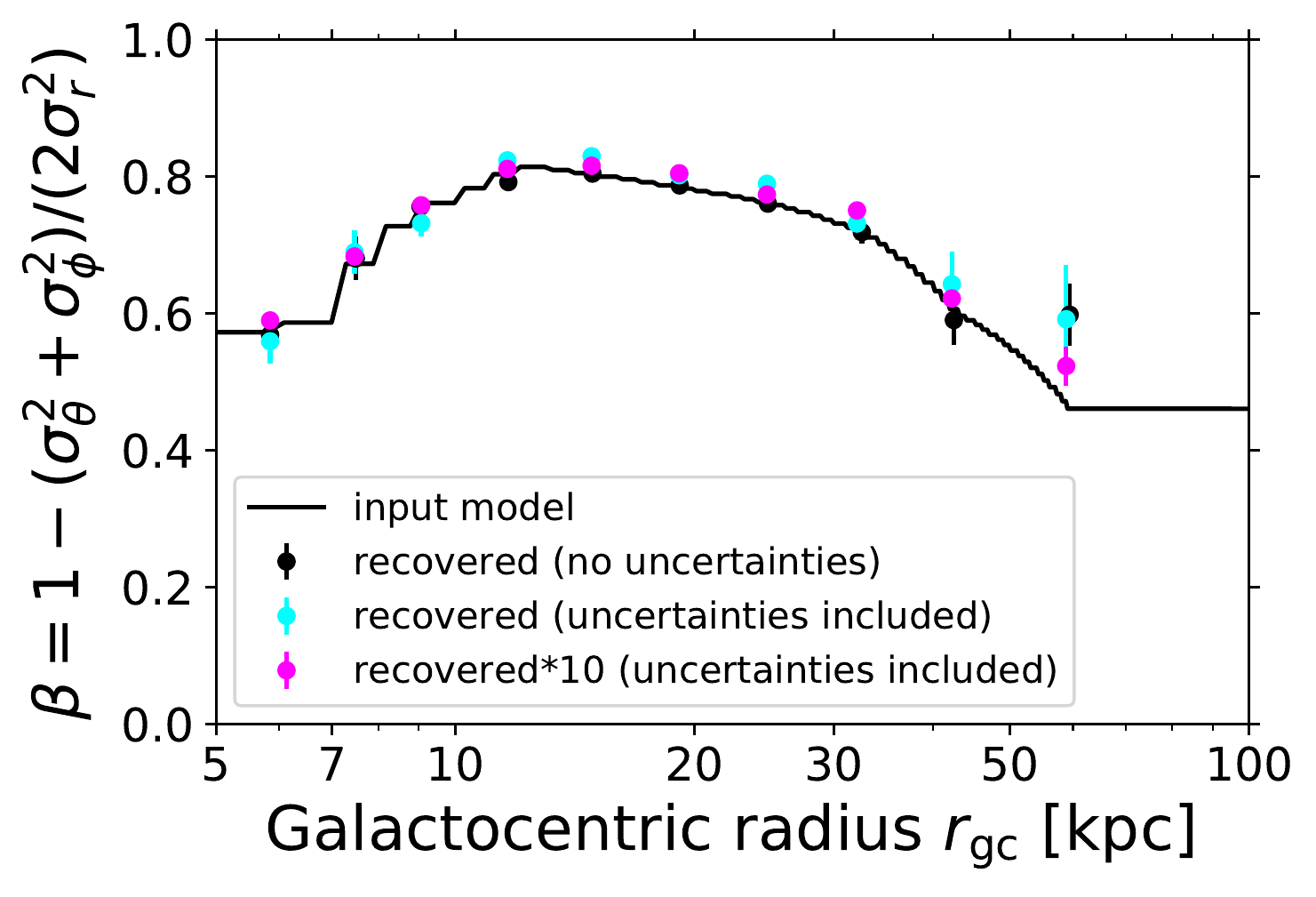}\\
\includegraphics[width=.45\columnwidth]{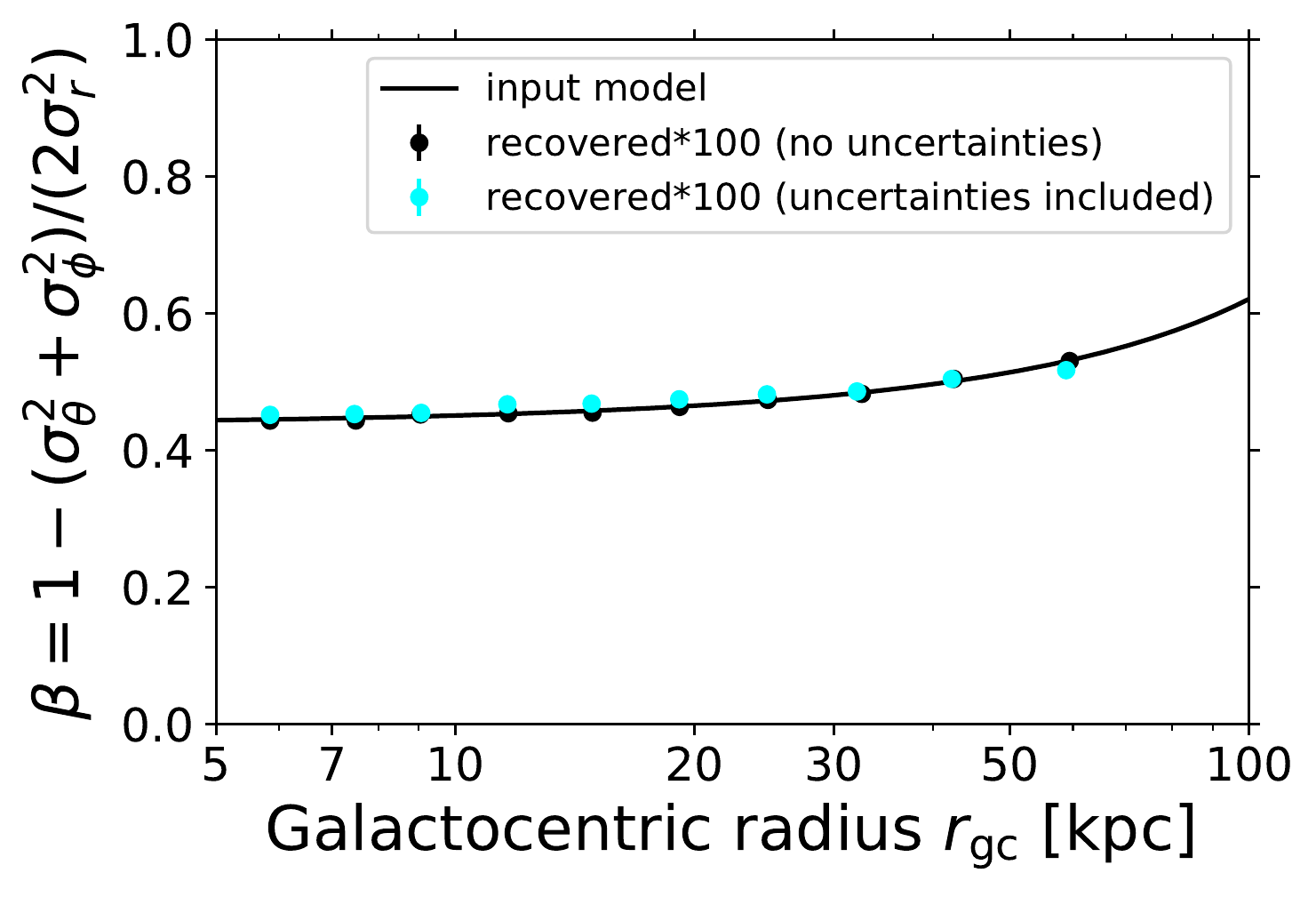}&
\includegraphics[width=.45\columnwidth]{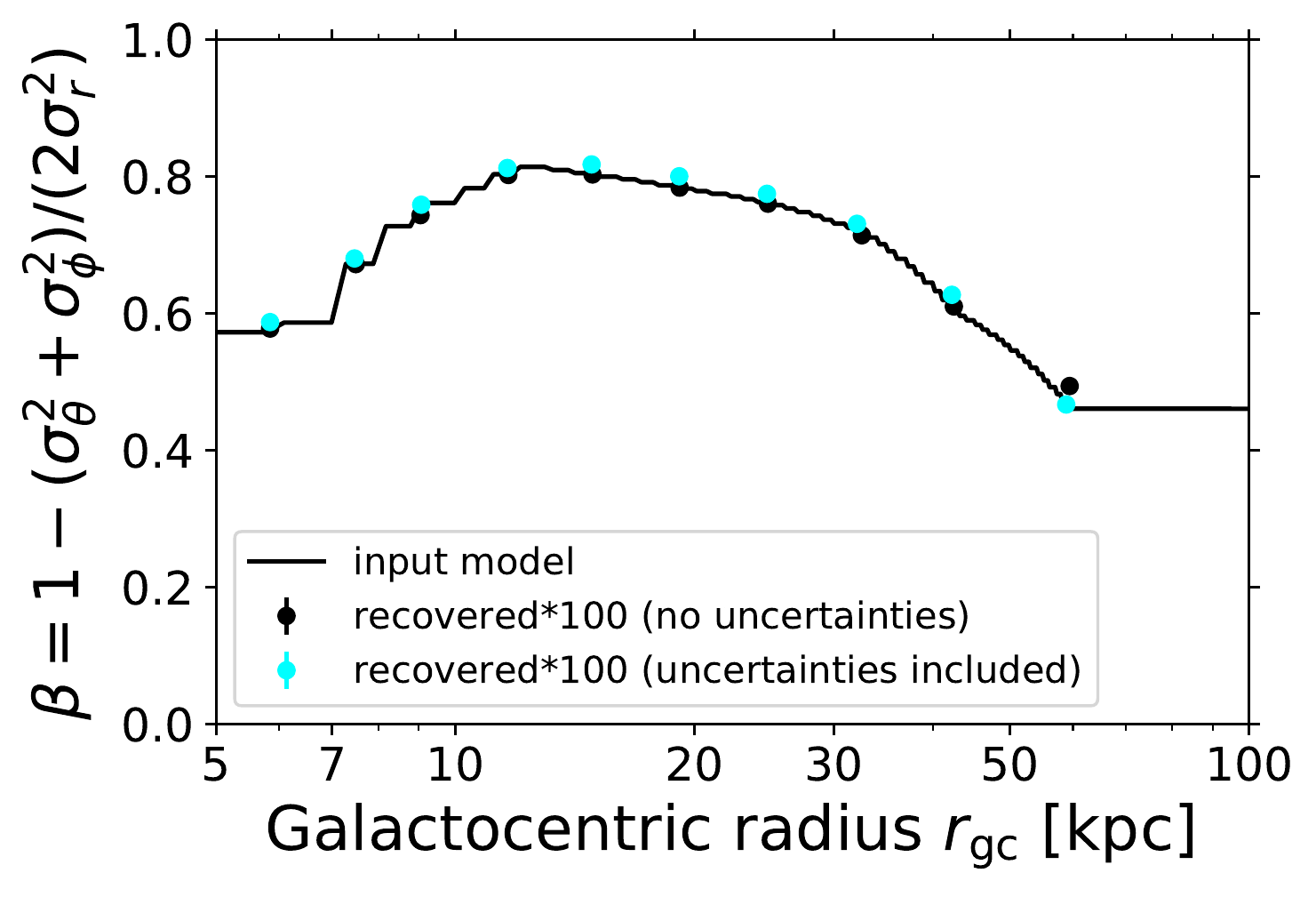}
\end{tabular}
\caption{Anisotropy profiles of the input models (black line) and of that recovered from the mocks (markers). Left panels show the radially rising model and right panels show the K giant model.
Tests presented include mocks with $1\times$ (upper panels), $10\times$ (upper panels), and $100\times$ (lower panels) the number of K giants, both with and without scattering by uncertainties (as indicated in the legends). 
We recover the underlying models well with little bias. The similarity in scatter between recovered models with and without uncertainties in distances, line-of-sight velocities, and proper motions included signifies that we are properly able to subtract the velocity uncertainties from our kinematic dispersions. 
}
\label{fig:beta_compare}
\end{figure}

\begin{figure}[htb]
\begin{tabular}{cc}
\includegraphics[width=.45\columnwidth]{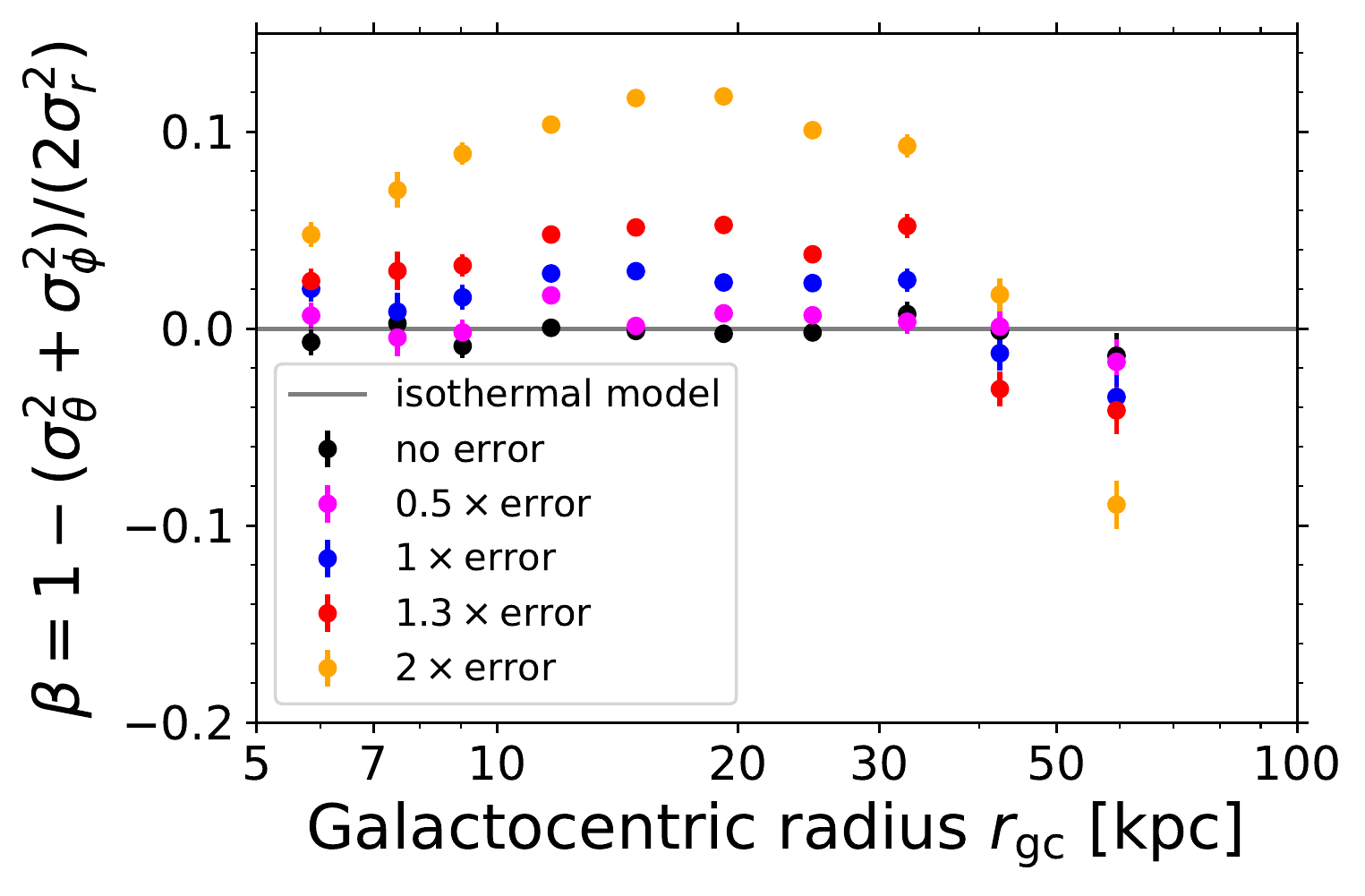}&
\includegraphics[width=.45\columnwidth]{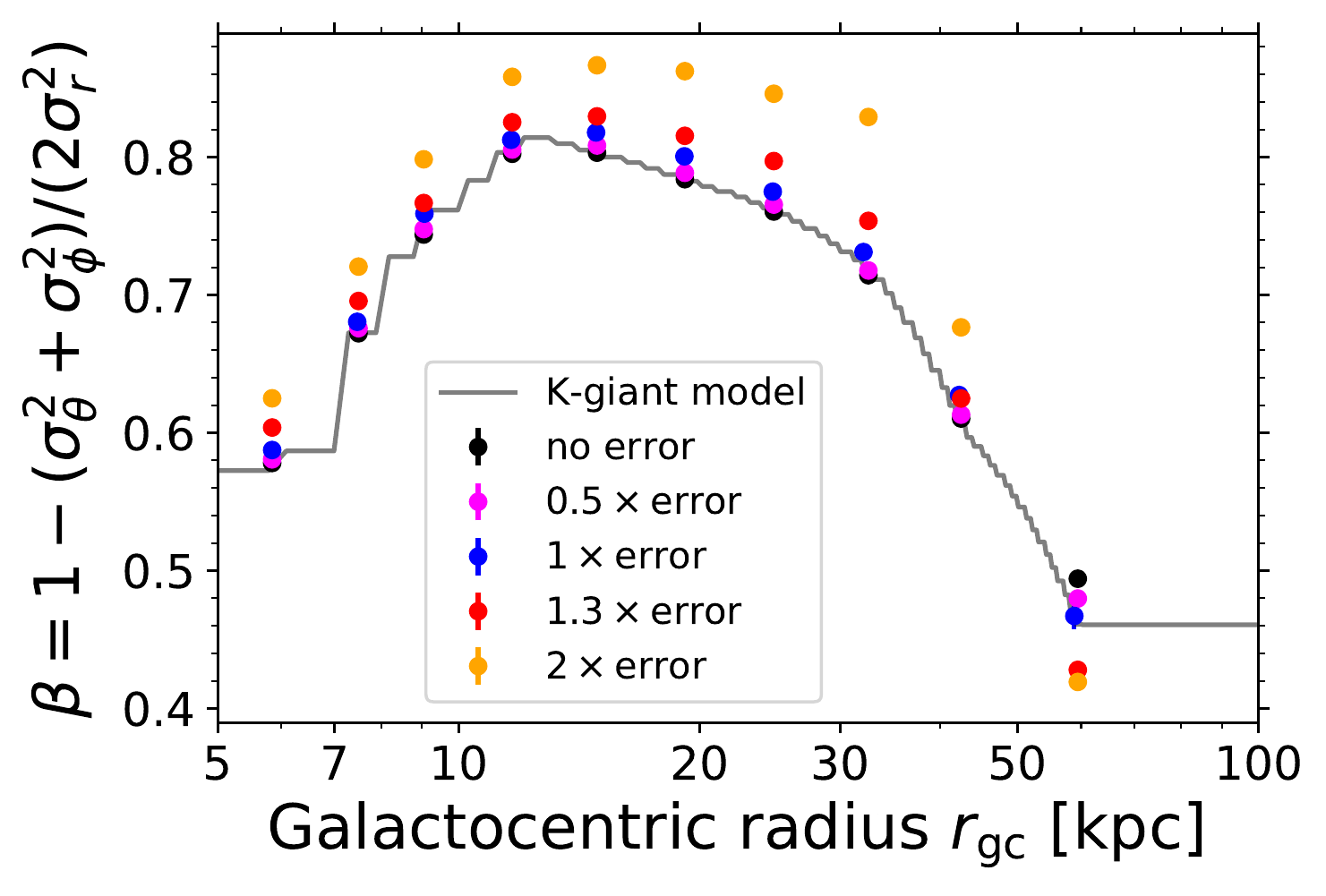}
\end{tabular}
\caption{Anisotropy profiles of the tests for biases due to increasing uncertainties. Lines indicate the input model for the mock stars (isothermal $\beta=0$ model and K-giant models are shown in the left and right panels, respectively). Black markers are the recovered profiles from the mocks before the distances, line-of-sight velocities, and proper motions are scattered by the corresponding uncertainties. Magenta, blue, red, and gold markers are the recovered profiles after scattering by $0.5\times$, $1.0\times$, $1.3\times$, and $2.0\times$ the uncertainties, respectively. All mocks have $100\times$ the number of sample K giants. Systematic bias is introduced which grows with increasing uncertainty, but even with errors twice as large, at most the bias is around $\Delta\beta\sim0.1$.
}
\label{fig:beta_13err}
\end{figure}

We further validate the analysis codes of \citet{Bird2019beta} via mock data sets as was done in \citet{Bird2019beta} using standard deviation with sigma clipping and median velocity uncertainties subtracted in quadrature to estimate $(\sigma_r,\sigma_\theta,\sigma_\phi)$ for the sample.
The mocks are
based on our LAMOST/SDSS halo K-giant and SDSS BHB samples. 

The mocks are made using the sky positions and distances of the real data and assigning velocities $V_r, V_\theta, V_\phi$
to each star according to a range of test velocity dispersion profiles. Once the mocks are made, we scatter their distances, line-of-sight velocities, and proper motions by the uncertainties $\delta d_\mathrm{helio}$, $\delta v_\mathrm{los}$, and $\delta\mu$ from the real data.
We make several runs of mocks with $1\times$, $10\times$, and $100\times$ the number of K giants and BHB stars.
Having imposed a particular functional form
for $\beta$ on the data set stars, we test our ability to recover it
using the methods applied to the actual data (we test the mocks both before and after scattering by the uncertainties). 
We tested a range of
models for $\beta$: isothermal ($\beta = 0$), 
radially 
anisotropic ($\beta = 0.5$), 
models for which $\beta$ is a function of $r_\mathrm{gc}$
(e.g., an increasingly large radially anisotropic model as frequently seen in simulations, a model based on the current halo K giant anisotropy profile). The velocity dispersion and anisotropy profiles for the radially rising $\beta$ model and LAMOST/SDSS K giant model are shown in Figures \ref{fig:veldisp100} and \ref{fig:beta_compare} (solid lines) for the K-giant mock stars. The markers in these figures are the recovered profiles from various mock tests with $1\times$, $10\times$, and $100\times$ the number of K giants, both with and without scattering due to uncertainties (as indicated in the Figure legends and captions).

We find that even with
the quite large errors that develop in the tangential velocities at
large distance (of order the velocity dispersion of the stars) seen in
Figure \ref{fig:profs}, we are able to recover $\beta$
correctly with systematic errors of $\Delta\beta<0.1$.

As a secondary test, we multiplied all uncertainties by a factor of 0.5, 1.3, and 2.0 to check for biases introduced due to a range of measurement uncertainties. The mock samples we test are 100$\times$ as many stars in our K-giant sample. Figure \ref{fig:beta_13err} displays the results for the isothermal model and the K-giant model. We find a systematic bias is introduced which grows with increasing uncertainty, but even with errors twice as large, at most the bias is around $\Delta\beta\sim0.1$.

These tests confirm that the dominant source of uncertainty is Poisson statistics in the bins for our particular halo star samples.

Our mock tests show that we recover the underlying models well with little bias. The similarity in scatter between recovered models with and without uncertainties in distances, line-of-sight velocities, and proper motions included signifies that we are properly able to subtract the velocity uncertainties from our kinematic dispersions. 

The scatter seen is largely due to Poisson uncertainties. The tests with sample sizes 10 and 100 times larger show that with small Poisson uncertainty, we recover the model profiles with negligible bias introduced due to the method. With the low Poisson noise of the large mock data sets, we find a systematic bias which grows as the uncertainties are multiplied by increasingly larger factors; but even with uncertainties twice as large, the difference in anisotropy is small ($\Delta\beta\sim0.1$). These test show the success in recovering the kinematic statistics of our halo sample with negligible bias using standard deviation with sigma clipping and median velocity uncertainties subtracted in quadrature (the method used in \citet{Bird2019beta}).

We have performed similar exercises with mocks based on our halo SDSS BHB sample and find similar results. The Poisson uncertainty is larger due to the smaller sample compared to the K giant sample, and the systematic bias introduced from the uncertainties of the sample is $\Delta\beta<0.1$. 

\begin{figure}[htb]
\begin{tabular}{cc}
  \includegraphics[width=.5\columnwidth]{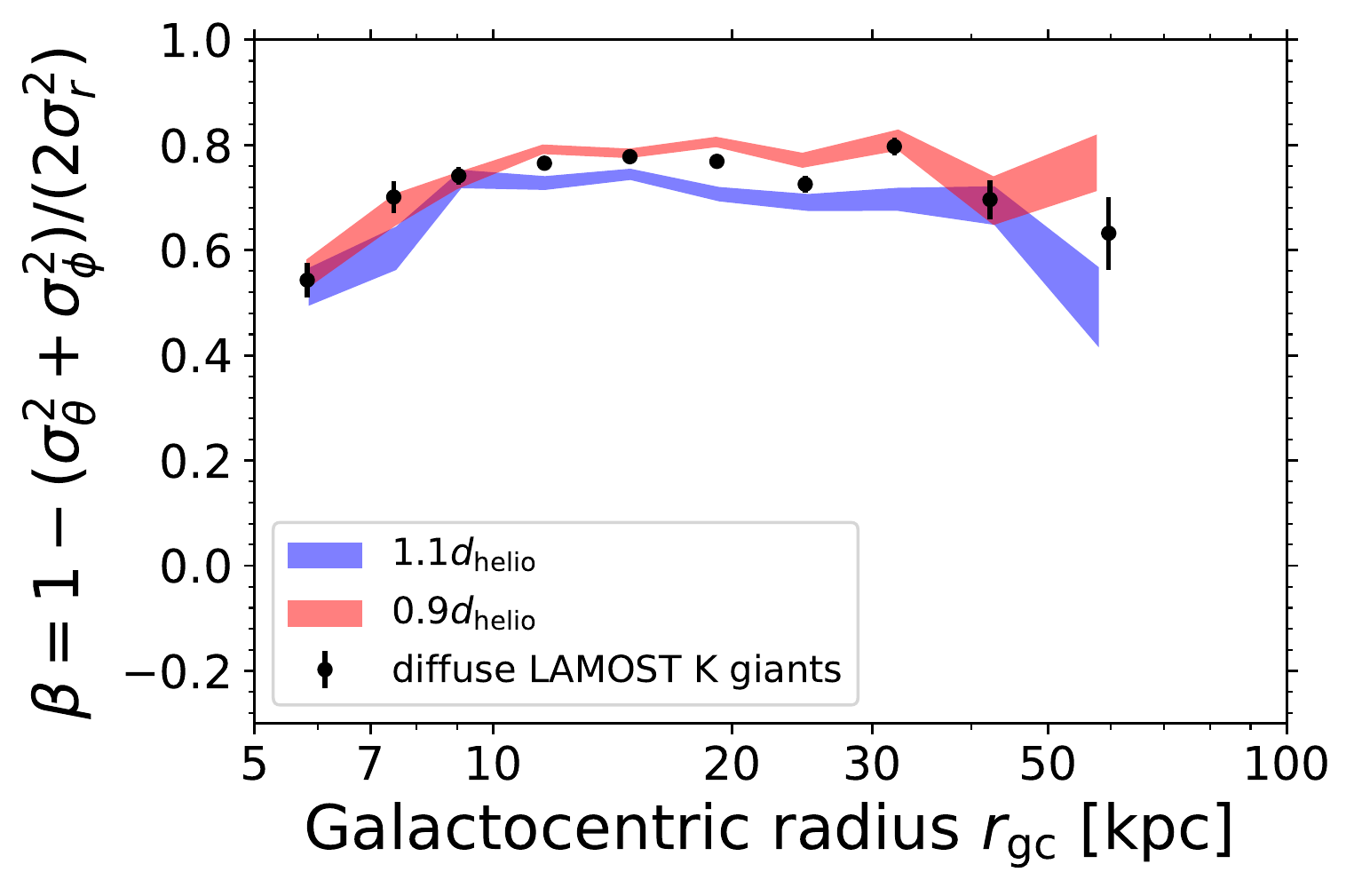}&
  \includegraphics[width=.5\columnwidth]{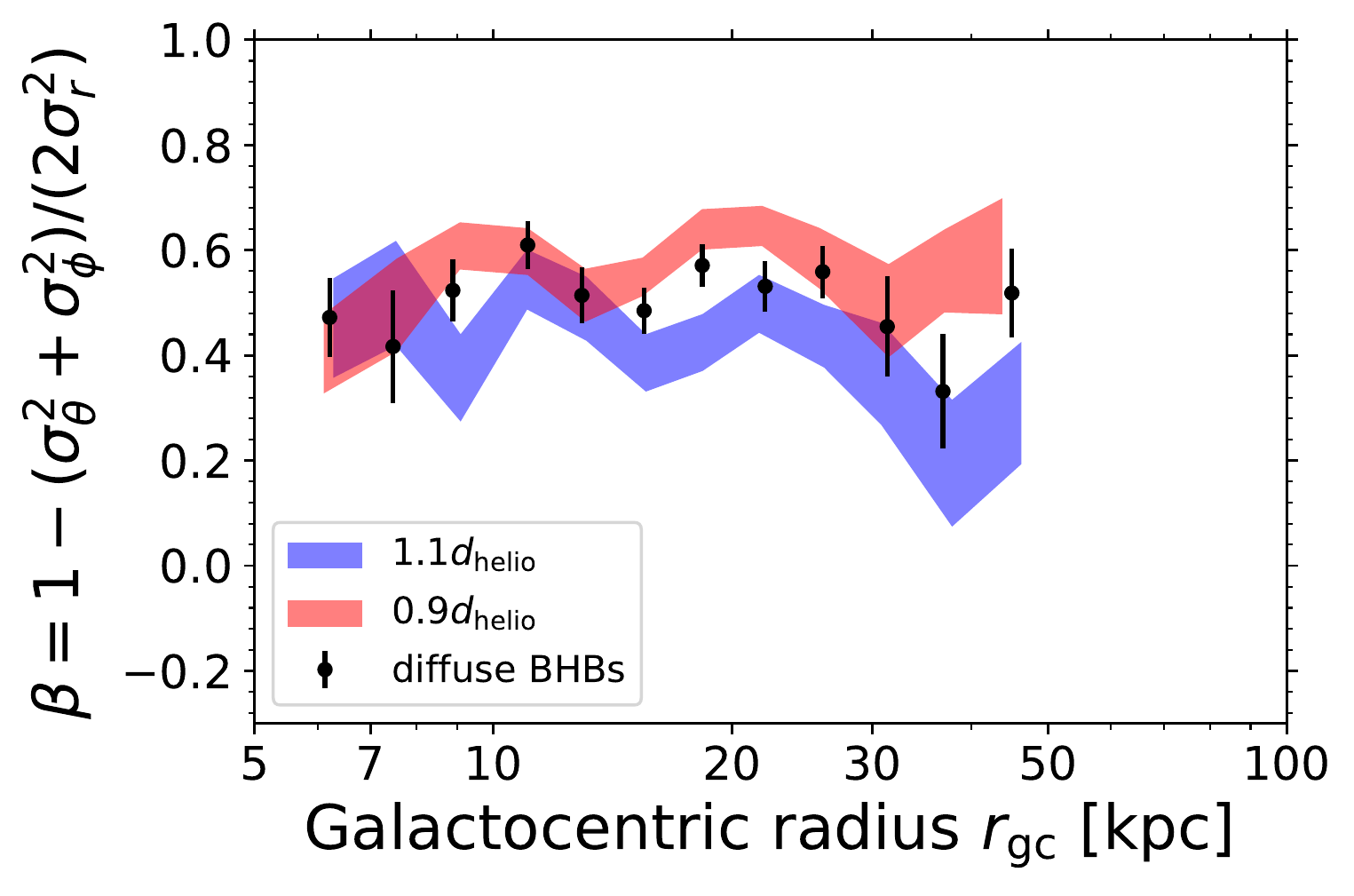}
\end{tabular}
  \caption{Test of distance uncertainties on the estimates of the anisotropy parameter $\beta$. 
We systematically change our distance scales
    for halo LAMOST K giants and SDSS BHB stars by $\pm10$\% to check the effect on $\beta$.
Black symbols show the $\beta$ profile for the smooth, diffuse halo LAMOST K giants and SDSS BHB stars, as the red and blue symbols, respectively, in Figure \ref{fig:rgc-beta-multi}.
Shaded regions show the results of the test for under-estimated $\beta$ (blue, due to overestimating distances by $1.1d_\mathrm{helio}$) and over-estimated $\beta$ (red, due to underestimating distances by $0.9d_\mathrm{helio}$). The width of the shaded regions are based on the estimated uncertainties for $\beta$.
Each marker represents the median radius of the stars within our selected radial bins.
This test shows that the main result of the paper, that the bulk of the K giant and BHB halo stars are on highly radial orbits,
is very robust to systematic error in the distance scale of this order.
  }
  \label{fig:distance_systematics_beta}
\end{figure}

The adopted distance scale for our K giants was
    cross-checked in \citet{Bird2019beta} and for our K giants and BHB stars in \citet{Yang2019b}. In Section 2 of \citet{Bird2019beta}, we compared
    our halo K-giant distance scale to that of
    \citet{Bailer-Jones2018}, for halo K giants within approximately 4
    kpc of the Sun. This test showed evidence for our distances being
    closer on average than the \citet{Bailer-Jones2018} distances by
    approximately 10\%
    (See Figure 2 of \citet{Bird2019beta}). \citet{Bailer-Jones2018} use a
    model of a ``lengthscale'' at any given $(l,b)$ position, to
    derive improved distance estimates of stars compared to simply
    inverting the parallax. The modeling is validated using stars in
    open clusters, i.e., on metal rich stars. The \citet{Xue2014}
    distances have been calibrated using K giants in globular
    clusters, stars very similar to our halo K giants, which gives us confidence in using these distances in regions within 4 kpc where we sample halo stars. \citet{Yang2019b} select stars with high quality parallaxes within 4 kpc and compare the K giants and BHB stars with inverted parallax. They also find the \citet{Xue2014} K giant distances are closer on average by $\sim15$\% and BHB distances are unbiased. \citet{Yang2019b} find the bias of the \citet{Xue2014} K giant distances decreases with fainter $G$ magnitudes. As the large bulk of our K giant halo stars are at fainter magnitudes compared to the distance bias with inverted parallax, we 
also prefer to use the \citet{Xue2014} distance method. 

A systematic
    distance correction (increasing the K giant distances by $10-15$\%) for our sample stars will have the
    effect of increasing the estimated tangential velocity dispersions
    (which are primarily sensitive to the distance scale via proper
    motions), while leaving the Galactocentric radial velocity
    dispersions relatively unaffected (as these are typically
    dominated by line-of-sight velocities). Consequently, 
    such a distance correction tends to reduce the estimates of the
    anisotropy $\beta$ slightly.

We test the influence of systematic distance uncertainties (bias and Poisson noise)
and compare the resulting $\beta$ profiles in
    Figure \ref{fig:distance_systematics_beta} 
using our smooth, diffuse halo LAMOST/SDSS K giants and SDSS BHB stars after stream removal. 
The anisotropy $\beta$ decreases/increases very
    slightly 
as a result of
    increasing/decreasing the distances by 10\%. 
These results are similar to the mock tests presented in \citet[Figure 7]{Bird2019beta} for LAMOST K giants. 
Figure \ref{fig:distance_systematics_beta} shows that the main result of the paper, that the bulk
    of the K giant and BHB halo stars are on highly radial orbits, is very robust
    to systematic error in the distance scale of this order. 


\end{document}